\documentclass[useAMS,usenatbib]{mn2e}

\usepackage{threeparttable}
\usepackage{graphics,graphicx,epsfig,ulem}
\usepackage{multirow}
\usepackage{amsmath}
\usepackage{xcolor} 
\usepackage{soul} 

\def\la{\mathrel{\hbox{\rlap{\hbox{\lower4pt\hbox{$\sim$}}}{\raise2pt\hbox{$<$}}}}}
\def\ga{\mathrel{\hbox{\rlap{\hbox{\lower4pt\hbox{$\sim$}}}{\raise2pt\hbox{$>$}}}}}

\title[A deficit of ULXs in LIRGs]{A deficit of ultraluminous X-ray sources in luminous infrared galaxies}
\author[W. Luangtip et al.]{W. Luangtip$^{1}$\thanks{E-mail:wasutep.luangtip@durham.ac.uk(WL); t.p.roberts@durham.ac.uk (TPR)}, T.\,P. Roberts$^{1}\footnotemark[1]$, S. Mineo$^{2,1}$, B.\,D. Lehmer$^{3,4}$, D.\,M. Alexander$^{1}$, \and F.\,E. Jackson$^{5}$, A.\,D. Goulding$^{2}$ and J.\,L. Fischer$^{6}$\\
$^{1}$Department of Physics, University of Durham, South Road, Durham, DH1 3LE, UK\\
$^{2}$Harvard-Smithsonian Center for Astrophysics, 60 Garden Street Cambridge, MA 02138, USA\\
$^{3}$The Johns Hopkins University, Homewood Campus, Baltimore, MD 21218, USA\\
$^{4}$NASA Goddard Space Flight Center, Code 662, Greenbelt, MD 20771, USA\\
$^{5}$Department of Physics and Astronomy, University of Toledo, Toledo, OH 43606, USA\\
$^{6}$Department of Physics and Astronomy, University of Pennsylvania, 209 S. 33rd St., Philadelphia, PA 19104, USA}

\begin{document}

\date{\today}

\pagerange{\pageref{firstpage}--\pageref{lastpage}} \pubyear{2014}

\maketitle

\label{firstpage}

\begin{abstract}
We present results from a \textit{Chandra} study of ultraluminous X-ray sources (ULXs) in a sample of 17 nearby ($D_{\rm L}$ $<$ 60 Mpc) luminous infrared galaxies (LIRGs), selected to have star formation rates (SFRs) in excess of 7 $M_{\odot}$ yr$^{-1}$ and low foreground Galactic column densities ($N_{H} \la 5 \times 10^{20}$ cm$^{-2}$).  A total of 53 ULXs were detected and we confirm that this is a complete catalogue of ULXs for the LIRG sample.  We examine the evolution of ULX spectra with luminosity in these galaxies by stacking the spectra of individual objects in three luminosity bins, finding a distinct change in spectral index at luminosity $\sim 2 \times 10^{39} \rm ~erg~s^{-1}$.  This may be a change in spectrum as $10 ~M_{\odot}$ black holes transit from a $\sim$ Eddington to a super-Eddington accretion regime, and is supported by a plausible detection of partially-ionised absorption imprinted on the spectrum of the luminous ULX ($L_{\rm X} \approx 5 \times 10^{39} \rm ~erg~s^{-1}$) CXOU~J024238.9-000055 in NGC 1068, consistent with the highly ionised massive wind that we would expect to see driven by a super-Eddington accretion flow. This sample shows a large deficit in the number of ULXs detected per unit SFR (0.2 versus 2 ULXs, per $M_{\odot} \rm ~yr^{-1}$) compared to the detection rate in nearby ($D_{\rm L}$ $<$ 14.5 Mpc) normal star forming galaxies.  This deficit also manifests itself as a lower differential X-ray luminosity function normalisation for the LIRG sample than for samples of other star forming galaxies.  We show that it is unlikely that this deficit is a purely observational effect.  Part of this deficit might be attributable to the high metallicity of the LIRGs impeding the production efficiency of ULXs and/or a lag between the star formation starting and the production of ULXs; however, we argue that the evidence -- including very low $N_{\rm ULX}/L_{\rm FIR}$, and an even lower ULX incidence in the central regions of the LIRGs -- shows that the main culprit for this deficit is likely to be the high column of gas and dust in these galaxies, that fuels the high SFR but also acts to obscure many ULXs from our view.  

\end{abstract}


\section{Introduction}
\label{sec:intro}

Ultraluminous X-ray sources (ULXs; see \citealt{feng2011} for a recent review) are extra-galactic X-ray sources for which the observed luminosity in the $0.3 - 10$ keV band reaches or exceeds the Eddington limit for a $\sim$10$~M_{\odot}$ black hole ($L_{\rm X} \geq10^{39}$ erg s$^{-1}$).  Indeed, it is now generally accepted that ULXs are accreting black holes (BHs); however, the class of BHs powering ULXs is still a topic of much debate due to their extreme luminosities.  Given that they are defined as non-nuclear point-like sources they cannot be powered by accretion onto the central supermassive black hole (SMBH) of their host galaxy.  This then leaves two main possibilities, the first being that ULXs contain intermediate mass black holes (IMBHs; 10$^{2}$$M_{\odot}$ $\la$ $M_{\rm BH}$ $\la$ 10$^{4}$$M_{\odot}$), which would be accreting material at sub-Eddington rates \citep{colbert1999}. Alternatively, ULXs could be stellar mass BHs ($M_{\rm BH} \la 20 M_{\odot}$) that, somehow, are accreting material at or above the Eddington rate.

In fact, although some of the very brightest ULXs remain good candidates for sub-Eddington accretion onto IMBHs, for instance HLX-1 in ESO 243-49 \citep{farrell2009} and a sample of extreme ULXs discussed by \citet{sutton2012}, there are many reasons to doubt this interpretation for the bulk of the ULX population.  For example, it is difficult to reconcile the presence of a break in the X-ray luminosity function (XLF) of point-like X-ray sources in nearby galaxies at $\sim 1 \times 10^{40} \rm ~erg~s^{-1}$ (see e.g. \citeauthor{mineo2012a} 2012a) with a population of IMBHs dominating in the ULX regime.  Similarly, the association of ULXs with ongoing star formation (see below) is difficult to interpret solely using IMBHs \citep{king2004}.  Moreover, the high quality X-ray spectra of multiple ULXs are inconsistent with the sub-Eddington accretion states displayed by Galactic BH binaries (\citealt{stobbart2006}; \citealt{roberts2007}; \citealt{bachetti2013}). The spectral characteristics and high luminosities of ULXs suggest that they occupy a new, super-Eddington accretion state, named the {\it ultraluminous state\/}  (\citealt{gladstone2009}; see also \citeauthor{sutton2013b} 2013b).   However, some ULXs might still be massive stellar BHs (MsBHs, in the 20$M_{\odot}$ $\la$ $M_{\rm BH}$ $\la$ 100$M_{\odot}$ regime) that could plausibly form in metal-poor environments \citep{zampieri2009,mapelli2010,belczynski2010}, which would be accreting matter at about or slightly higher than the Eddington rate. 

If stellar remnant BHs can produce such high luminosities, then there must be physical mechanisms by which they are able to accrete material at rates higher than the classical Eddington rate.  Much recent theoretical work has sought to establish and/or explain these mechanisms (see e.g. \citealt{poutanen2007}; \citealt{dotan2011}; \citealt{kawashima2012}), with the notion that super-Eddington accretion rates lead to large scale-height discs and massive outflowing, radiatively-driven winds providing a plausible scenario to explain the X-ray emission characteristics.  These predicted properties of super-Eddington accretion flows are now being substantiated by observations.  The highest quality ULX X-ray spectra, investigated by \cite{gladstone2009}, demonstrated that ULXs appear disc-like at close to the Eddington limit (albeit the emission is broadened beyond that seen in classic disc models); at higher luminosities the spectra become two component, with a soft excess well-modelled by a cool accretion disc ($kT \sim 0.2$ keV) and a harder component modelled by an optically-thick corona ($kT_{\rm e} \sim 2$ keV; $\tau \sim 10$).  Subsequent work has explained these spectra by attributing the cool disc component to the optically thick, massive outflowing wind that thermalises and downscatters much of the emission within it (e.g. \citealt{kajava2009}; although see \citealt{miller2013}), whereas the optically thick coronal component could be physically attributed to the hot inner part of the disc (e.g. \citeauthor{middleton2011a} 2011a; \citealt{kajava2012}).  The broadened disc-like spectra might then be composed of the sum of these two components, emerging as the ULX begins to exceed the Eddington limit (\citeauthor{middleton2011b} 2011b).  Furthermore, new work considering both spectral and timing characteristics of ULXs shows that the balance between the two components in ULX spectra appears dependent upon two factors: accretion rate and viewing angle (\citeauthor{sutton2013b} 2013b; Middleton et al. submitted).

ULXs are found in all types of galaxy.  In elliptical galaxies there is an average $\sim 1$ ULX per 10$^{11} M_{\odot}$ (\citealt{feng2011} and references therein; also \citealt{plotkin2014}), a population of ULXs that must be related to the low mass X-ray binary (LMXB) populations found in the old stellar populations of these systems.  In contrast, the number of ULXs per unit stellar mass in star forming galaxies is much higher at $\sim 1$ ULX per 10$^{10} $ $M_{\odot}$, and this ratio tends to increase for the least massive galaxies (\citealt{feng2011}; also \citealt{swartz2008}, \citealt{walton2011}).  This suggests that most ULXs are related to the ongoing star formation hosted in younger systems, and indeed a connection between ULXs and star formation is well established in the literature. For example, \citet{fabbiano2001} and \citet{gao2003} found that high numbers of ULXs were detected in individual starburst galaxies; \citet{swartz2009} and \citet{mineo2013} find a direct spatial correlation between ULXs and star forming regions; and \citeauthor{mineo2012a} (2012a) studied the population of high mass X-ray binaries (HMXBs, including ULXs) in a number of star forming galaxies via their X-ray luminosity functions (XLFs), and found that the number of detected HMXBs is proportional to the star formation rate (SFR).  In fact, the global average number of ULX detections is $\sim 2$ per SFR of 1~$M_{\odot}$ yr$^{-1}$ in nearby galaxies \citep{mapelli2010,swartz2011}.  Given this relationship we would therefore expect to detect relatively large numbers of ULXs in high SFR galaxies, and the number of ULXs in these systems should increase proportionally with the SFR. 

\begin{table*}
      \centering
      \caption{Properties of the LIRG sample}\label{tab:LIRGsample}
      \smallskip
      \begin{threeparttable}
          \begin{tabular}{lcccccccccc}
             \hline
						 
 & &  & \multicolumn{3}{c}{$R_{20}$ ellipse parameters} & & & & \\
 
\cline {4-6} 
 
Galaxy name	&	RA	&	Dec	&	\textit{a}	&	\textit{b}	&	P.A.	&	$D_{\rm L}$	&	SFR	&	$Z$	&	AGN?	\\
	&	&	&	(arcmin)	&	(arcmin)	&	(deg)	&	(Mpc)	&	($M_{\odot}$ yr$^{-1}$)	&		&		\\
 ~~~~~~(1)	&	(2)	&	(3)	&	(4)	&	(5)	&	(6)	&	(7)	&	(8)	&	(9)	&	(10)	\\

\hline
NGC 1068	&	02 42 41	&	-- 00 00 48	&	2.42	&	1.99	&	35	&	13.8	&	7.2	&	9.07\tnote{1}	&	Y	\\
NGC 1365	&	03 33 36	&	-- 36 08 25	&	4.60	&	3.45	&	50	&	18.0	&	9.1	&	9.01\tnote{1}	&	Y	\\
NGC 7552	&	23 16 11	&	-- 42 35 05	&	2.21	&	1.30	&	95	&	21.6	&	9.5	&	9.16\tnote{2}	&	N	\\
NGC 4418	&	12 26 55	&	-- 00 52 39	&	0.83	&	0.45	&	55	&	32.1	&	10.3	&	-	&	Y	\\
NGC 4194	&	12 14 09	&	+54 31 37	&	0.64	&	0.44	&	170	&	40.7	&	10.9	&	-	&	Y	\\
IC 5179	&	22 16 09	&	-- 36 50 37	&	1.48	&	0.56	&	55	&	47.2	&	13.5	&	8.90\tnote{3}	&	N	\\
ESO 420-G013	&	04 13 50	&	-- 32 00 25	&	0.61	&	0.58	&	110	&	48.2	&	8.5	&	-	&	Y	\\
Arp 299	&	11 28 30	&	+58 34 10	&	1.42	&	1.25	&	28	&	48.2	&	73.2	&	8.80\tnote{3}	&	Y	\\
NGC 838	&	02 09 39	&	-- 10 08 46	&	0.62	&	0.43	&	95	&	50.8	&	8.5	&	-	&	N	\\
NGC 5135	&	13 25 44	&	-- 29 50 01	&	1.75	&	0.86	&	125	&	52.9	&	14.5	&	8.70\tnote{3}	&	Y	\\
NGC 5395	&	13 58 38	&	+37 25 28	&	1.53	&	0.80	&	5	&	54.0	&	11.3	&	-	&	Y	\\
NGC 5653	&	14 30 10	&	+31 12 56	&	0.76	&	0.70	&	75	&	55.5	&	11.0	&	-	&	N	\\
NGC 7771	&	23 51 25	& +20 06 43	&	1.52	&	0.62	&	75	&	57.9	&	20.6	&	8.80\tnote{3}	&	Y	\\
NGC 3221	&	10 22 20	&	+21 34 11	&	1.62	&	0.34	&	167	&	59.5	&	11.1	&	-	&	N	\\
CGCG 049-057	&	15 13 13	&	+07 13 32	&	0.45	&	0.23	&	20	&	59.8	&	20.4	&	-	&	N	\\
IC 860	&	13 15 03	&	+24 37 08	&	0.56	&	0.32	&	20	&	59.9	&	15.1	&	-	&	Y	\\
NGC 23	&	00 09 53	&	+25 55 26	&	1.13	&	0.54	&	155	&	60.5	&	9.7	&	-	&	N	\\

 \hline  
 
         \end{tabular}
         \begin{tablenotes}
         \item \textbf{Notes.} Basic properties of the LIRG sample, ordered by distance (based on table~1 of \citealt{lehmer2010}). Column 1: common galaxy name. Columns 2 and 3: right ascension and declination, at epoch J2000, respectively. Columns 4--6: parameters of the $R_{20}$ region for each galaxy: semi-major axis, semi-minor axis and position angle, respectively. Column 7: luminosity distance. The values in columns 1--7 were given in \citet{lehmer2010}. Column 8: star formation rate, corrected for an AGN contribution in the calculation (see Section~\ref{sec:ULX number and host galaxy}).  Column 9: metallicity, in the form of the oxygen abundance (i.e. 12+log(O/H)). The references are: $^{1}$\citet{zaritsky1994}, $^{2}$\citet{moustakas2010} and $^{3}$\citet{relano2007}.
Column 10: Indication of whether the galaxies possess an AGN (see text for details).

         \end{tablenotes}
      \end{threeparttable}
    \end{table*}


However, another environmental factor might also affect the production efficiency of ULXs in galaxies: metallicity. It has been suggested that ULXs should be formed in higher numbers in low metallicity regions, as it is easier to evolve binaries to produce Roche lobe overflow onto a compact object in such environments \citep{linden2010}.  In addition, these environments should also lead to more efficient production of MsBHs via direct stellar collapse, due to reduced metal line-driven wind loss from massive stars (\citealt{zampieri2009,mapelli2013}; \citeauthor{fragos2013a} 2013a, 2\citeyear{fragos2013b}). Observational support for these hypotheses is beginning to emerge. For instance, \textit{Chandra} observations of extremely metal poor galaxies \citep{prestwich2013} show that the number of ULXs normalised to SFR in these galaxies is high when compared to that of metal rich galaxies, although this is only at a formal significance of 2.3$\sigma$.   Similarly, a marginally significant anticorrelation between the number of ULXs per unit SFR and metallicity was reported by \citet{mapelli2011}.  In addition, \mbox{\citeauthor{basu-zych2013a} 2013a} demonstrated that the X-ray luminosity per unit SFR of Lyman break galaxies is elevated compared to local galaxies; also a further result from a study of Lyman break galaxy analogue sample (\mbox{\citeauthor{basu-zych2013b} 2013b}) showed the anti-correlation between the X-ray luminosity per unit SFR and metallicity. They conclude that this is attributable to an enhanced population of X-ray binaries (XRBs) in the low metallicity environment of the galaxies, that must have a very strong ULX contribution.

Given the strong relationship between ULXs and star formation an obvious place to look for a large sample of ULXs is in the most active local star forming systems.  In this paper we study the population of ULXs in such an environment -- a sample of the nearest luminous infrared galaxies (LIRGs).  Interestingly, such galaxies are typically relatively abundant in metals (\textit{Z} $\ga$$Z_{\odot}$), providing a sample of high-metallicity host environment ULXs to compare with other samples.  The paper is laid out as follows.  In Section 2 we describe the sample of galaxies, and how the X-ray point source catalogue for these galaxies was constructed.  We detail our analyses of the ULXs, in terms of their X-ray spectra, X-ray luminosity function and aggregated properties in Section 3, and discuss these in light of other results in Section 4.  Our findings are summarised in Section 5.

\section{Observations and data reduction}
\label{sec:Observation and Data reduction}

\subsection{Observations and initial analyses}
\label{sec:Observation and initial analyses}

In this paper we study the ULX population hosted by the sample of 17 LIRGs studied  in \citet{lehmer2010}, using \textit{Chandra} X-ray observations.  In brief, this sample was selected as all LIRGs within a luminosity distance ($D_{\rm L}$) of 60 Mpc with a foreground Galactic column density ($N_{\rm H}$) $< 5\times10^{20}$ cm$^{-2}$.  To qualify as LIRGs all galaxies have total infrared luminosity (8~--~1000~$\mu$m band; $L_{\rm IR}$) $> 1 \times 10^{11} L_{\odot}$ (where $L_{\odot}$ is the bolometric luminosity of the Sun), with the most luminous LIRG in the sample -- Arp~299 -- as bright as $\sim 8 \times 10^{11} L_{\odot}$.  These luminosities correspond to SFRs $\ga$ 7 $M_{\odot}$ yr$^{-1}$ for each galaxy in the sample (see Section~\ref{sec:ULX number and host galaxy} for the calculation of SFR).  The general properties of each galaxy are shown in Table~\ref{tab:LIRGsample}.  For the purposes of this paper we consider all galaxies for which \citet{lehmer2010} present evidence of possessing an AGN in their Tables~1~and~3 (including the two objects classified as harbouring LINER/Sy 2 nuclei) as AGN hosts. The effect of an AGN contributing to the host galaxy IR emission and hence SFR will be taken into account in the following analyses.

The observational data for these LIRGs were extracted from the \textit{Chandra} data archive.\footnote{\texttt{http://cxc.harvard.edu/cda/}} For the few galaxies that have multiple archival datasets, we selected the observation with the longest exposure time in order to obtain the most complete snapshot of the ULX population, and the best quality data for each individual object. The exception to this was NGC~1365, for which we selected an observation with shorter exposure time as it covered a larger fraction of the galaxy's area, as defined by the $R_{20}$ ellipse,\footnote{The elliptical isophote equivalent to a level of 20 mag per arcsecond$^2$ in the K$_{\rm s}$ band, see \citet{lehmer2010} and references therein.} than the longest exposure, but still permitted us a statistically complete vista of the ULXs within its field-of-view (see Section 2.2.3). We decided not to merge the observations together to avoid an artificial boost in the number of ULXs compared to a single epoch observation, via the inclusion of multiple transient objects. Table~\ref{tab:observation} lists the observational datasets we selected for our analysis; more details on the instrument and set-up for each observation are provided in Table 2 of \cite{lehmer2010}.

The data were reduced using version 4.4 of the Chandra Interactive Analysis of Observation ({\sc ciao})\footnote{\texttt{http://cxc.harvard.edu/ciao/}} tools. We first used the {\sc chandra\_repro} script to create a new level 2 event file and a new bad pixel file, in order to utilise the latest calibration data. Then background flares were inspected and the exposure ranges that have a background level $> 3\sigma$ above the mean level were removed by the {\sc deflare} script.  The level 2 event files created by these steps were then used as the basis for further analysis.

\begin{table} 
      \centering
      \caption{\textit{Chandra} observational data}\label{tab:observation}
      \smallskip
      \begin{threeparttable}
          \begin{tabular}{lccc}
          
             \hline

Galaxy	&	Obs. ID$^{a}$	&	Exp. time$^{b}$	& Refs.$^{c}$\\
        &         &   (ks)    &     \\

\hline
NGC 1068	&	344\tnote{d}	&	47.05	&	1,2,3,4,5,6	\\
NGC 1365	&	3554\tnote{d}	&	13.42	&	7,8,9	\\
NGC 7552	&	7848	&	5.08	&	10	\\
NGC 4418	&	4060\tnote{d}	&	19.62	&	-	\\
NGC 4194	&	7071	&	35.30	&	11,12	\\
IC 5179	&	10392	&	11.96	&	-	\\
ESO 420-G013	&	10393	&	12.42	&	-	\\
Arp 299	&	1641	&	24.10	&	13,14,15,16	\\
NGC 838	&	10394	&	13.79	&	-	\\
NGC 5135	&	2187	&	26.41	&	-	\\
NGC 5395	&	10395	&	15.65	&	17	\\
NGC 5653	&	10396	&	16.52	&	-	\\
NGC 7771	&	10397	&	16.71	&	-	\\
NGC 3221	&	10398	&	18.96	&	-	\\
CGCG 049-057	&	10399	&	19.06	&	-	\\
IC 860	&	10400	&	19.15	&	-	\\
NGC 23	&	10401	&	19.06	&	-	\\

               \hline
         \end{tabular}
         \begin{tablenotes}
         \item \textbf{Notes.} $^a$The {\it Chandra\/} observation identifier.  $^{b}$The net exposure time after the removal of high background periods (see Section~\ref{sec:Observation and initial analyses}). $^{c}$References to other analyses of the listed datasets, all of which were used by \citet{lehmer2010}.  These are: (1) \citet{smith2003}; (2) \citet{ptak2006}; (3) \citet{Gultekin2009}; (4) \citet{swartz2004}; (5) \citet{swartz2011}; (6) \citet{peterson2006}; (7) \citet{strateva2009}; (8) \citet{wang2009}; (9) \citet{risaliti2005}; (10) \citet{grier2011}; (11) \citet{kaaret2008}; (12) \citeauthor{mineo2012a} (2012a); (13) \citet{huo2004}; (14) \citet{nelemans2010}; (15) \citet{gonz2006}; (16) \citet{zezas2003}; (17) \citet{smith2012}.  $^{d}$Galaxies with multiple observations; further details of these can be found in \citet{lehmer2010}.

         \end{tablenotes}
      \end{threeparttable}
    \end{table}


\subsection{The source catalogue}

\subsubsection{Source detection} \label{sec:Source detection}

We began by searching for the X-ray point source candidates within the $R_{20}$ ellipse region of each galaxy, using the {\sc ciao wavdetect} algorithm, that applies a Mexican Hat wavelet function with radii of 1, 2, 4, 8 and 16 pixels to images created from the level 2 data.  A detection threshold level of $10^{-6}$ was used, corresponding approximately to a false detection probability of 1 per $10^6$ spatial resolution elements.  In addition, for the galaxies in which a part of the $R_{20}$ ellipse was on (or near) the CCD edge, i.e. NGC~1365, NGC~7552 and Arp~299, we also supplied an exposure map at 1.5 keV to the {\sc wavdetect} algorithm, created using the {\sc fluximage} tool, in order to increase the accuracy of the source detection in these regions.  {\sc wavdetect} was run over data segregated into images in three separate energy bands per observation: soft (0.3--2\,keV), hard (2--10\,keV) and full (0.3--10\,keV).  Subsequently, the sources detected in each energy band were cross-correlated with source lists from the other bands, with individual objects detected in multiple bands identified using a matching radius of 3 arcseconds.  Each individual object was listed as one detection in the resulting combined source lists (one list per galaxy), with its position adopted from its full, soft or hard band detection, in that order of priority. There were a total of 214 candidate X-ray sources at this stage.

The {\sc acis extract} package (\citealt{broos2010,broos2012}; hereafter AE\footnote{\texttt{http://www.astro.psu.edu/xray/acis/acis\_analysis.html}}) was then used to finalise a catalogue composed of the most reliable point-like sources, and to characterise these objects.  Firstly, two low-significance sources -- defined as detections which have AE binomial no-source probability $> 0.01$ -- were removed.  After that, AE compared the source radial profile with the point spread function (PSF) for a point-like object at  the source position and removed the sources that were not consistent with the PSF according to a Kolmogorov-Smirnov test (AE K-S probability $< 0.01$; i.e. extended sources).  We also picked out any remaining objects by eye that might still be extended, and used the {\sc srcextent}\footnote{{\tt http://cxc.harvard.edu/ciao/threads/srcextent/}} script to test whether to reject them, excluding ones shown to be extended objects. Briefly, the tool derived the elliptical Gaussian parameters which are most strongly correlated with the source image. The same algorithm was also applied to derive the elliptical Gaussian parameters for the PSF of a point-like object at the source position. Then these two apparent ellipses were tested for consistency given the image statistics, and the algorithm reported whether the source was extended within 90\% confidence intervals. In total, 61 extended sources ($\sim$ 30$\%$ of all detected sources) were removed by the AE K-S test and {\sc srcextent} tools. As the final step we compared our point source catalogue with NASA/IPAC Extragalactic Database (NED)\footnote{\texttt{http://ned.ipac.caltech.edu}} and SIMBAD astronomical database (hereafter
SIMBAD)\footnote{\texttt{http://simbad.u-strasbg.fr}}, and identified and removed the X-ray sources within 5 arcseconds of each galaxy centre, to exclude all possible AGN contamination from the host galaxy; 8 sources were removed in this way.  We also used these databases to search for foreground and/or background source contamination and 4 objects were identified, i.e. two stars, one QSO and one AGN, and removed. At the conclusion of these steps we were left with the final catalogue of X-ray point sources for each galaxy, that we split into two source lists in Appendix~\ref{sec:appendixA}.  Overall, 139 X-ray point sources were detected in 14 LIRGs; no point sources were detected in ESO~420-G013, CGCG~049-057 or IC~860.  More than 50\% of the point sources were detected in the two nearest galaxies, NGC~1068 and NGC~1365, due to a combination of their proximity and exposure times yielding far greater sensitivity to faint point sources than in the other galaxies in the sample.

\subsubsection{Source photometry}
\label{sec:source photometry}

The net photon count rate detected from each point source was calculated in the 0.3--10\,keV band by AE, using a polygonal aperture set to the 90$\%$ encircled energy of the source PSF at 1.5 keV.  The background level was set using an annulus around the source or, for the cases in which the source resided in densely populated regions or areas of heightened diffuse emission, we used a circular aperture located in a source-free region near to the source position, with a similar level of diffuse emission. The background area was set to be at least 4 times bigger than the source region and to encompass at least 100 counts.  The web interface version of the Portable Interactive Multi-Mission Simulator ({\sc webpimms})\footnote{\texttt{http://heasarc.nasa.gov/Tools/w3pimms.html}} was applied to convert the calculated count rates into a 0.3--10\,keV source flux, using an absorbed power-law continuum model. A typical value of ULX photon index ($\Gamma$) and column density ($N_{\rm H}$)\footnote{Here this is a combination of both foreground Galactic and additional, extragalactic column density.} were used in the model: $\Gamma$ = 2 and $N_{\rm H}$ = 1.5$\times$10$^{21}$ cm$^{-2}$ (see e.g. \citealt{swartz2004,gladstone2009,sutton2012}). Finally, the luminosity of each source was calculated using the luminosity distance of its host galaxy. All source photometry is provided in Tables~\ref{tab:ULXcandidates} and~\ref{tab:other_xray} in Appendix~\ref{sec:appendixA}, where we separate the detections into two lists according to whether their calculated luminosity is above or below the 0.3--10\,keV luminosity threshold of $10^{39} ~\rm erg~s^{-1}$  used to demarcate a ULX.

\subsubsection{ULX catalogue and completeness}
\label{sec:completeness}

\begin{figure*}
\begin{center}

\includegraphics[width=4cm]{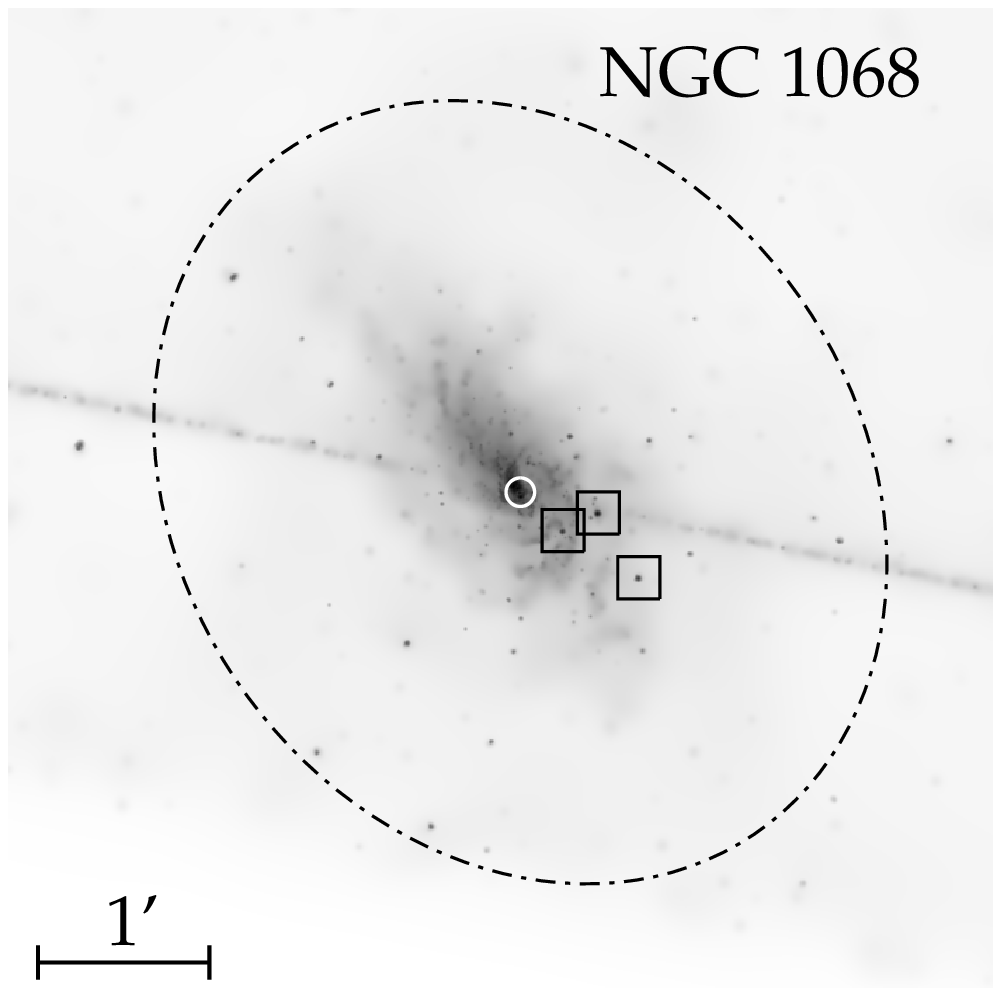}\includegraphics[width=4cm]{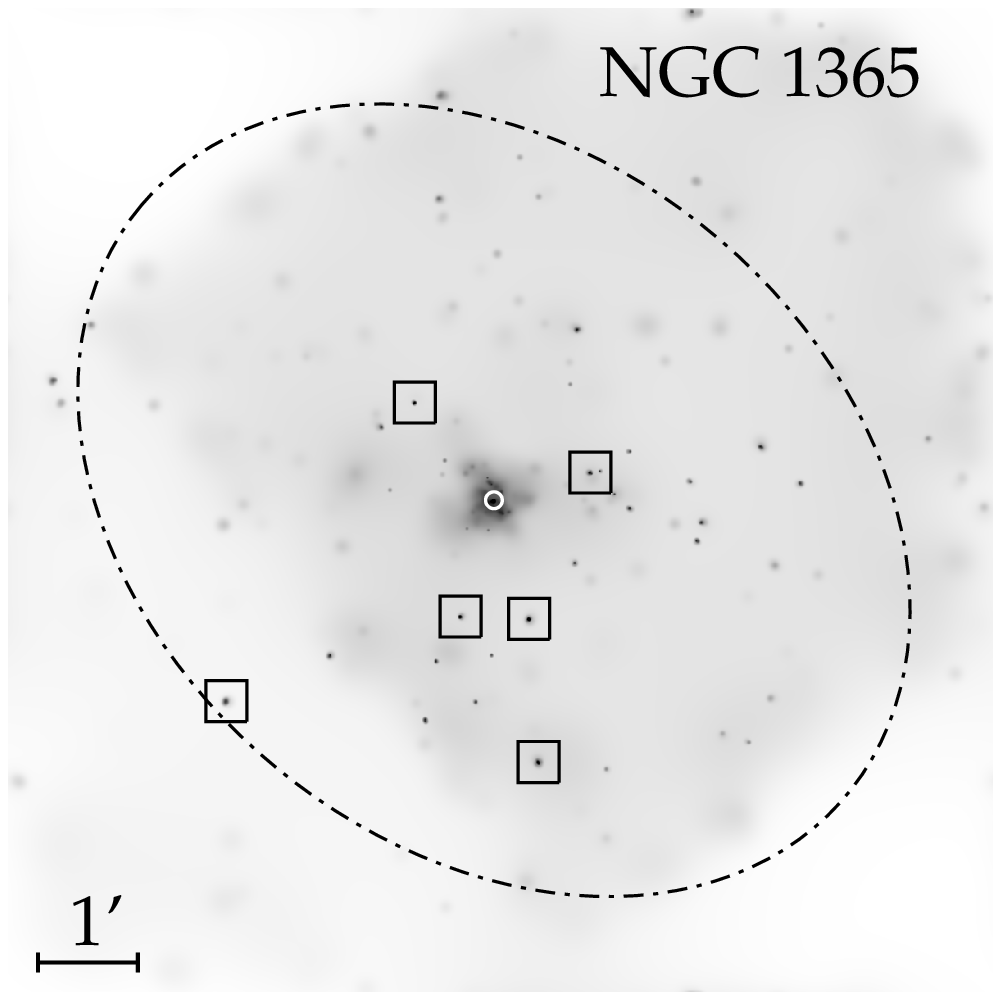}\includegraphics[width=4cm]{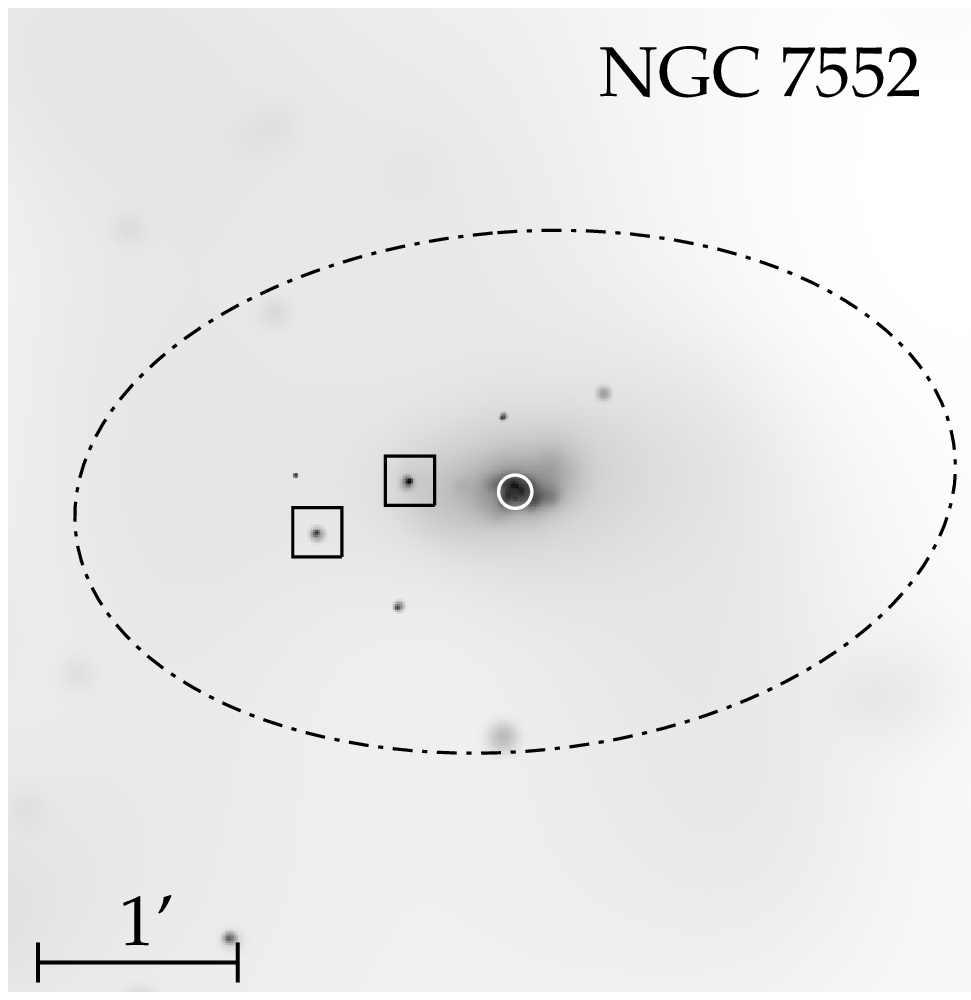}\includegraphics[width=4cm]{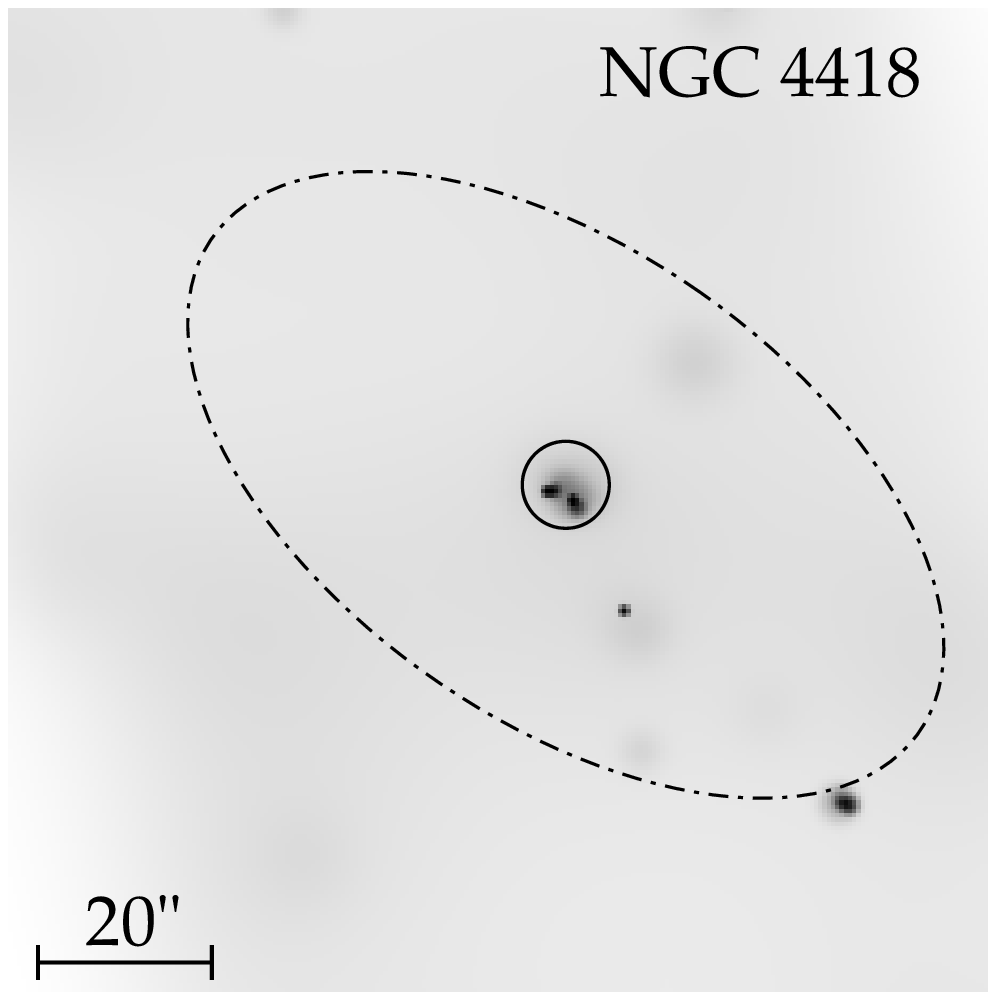}
\includegraphics[width=4cm]{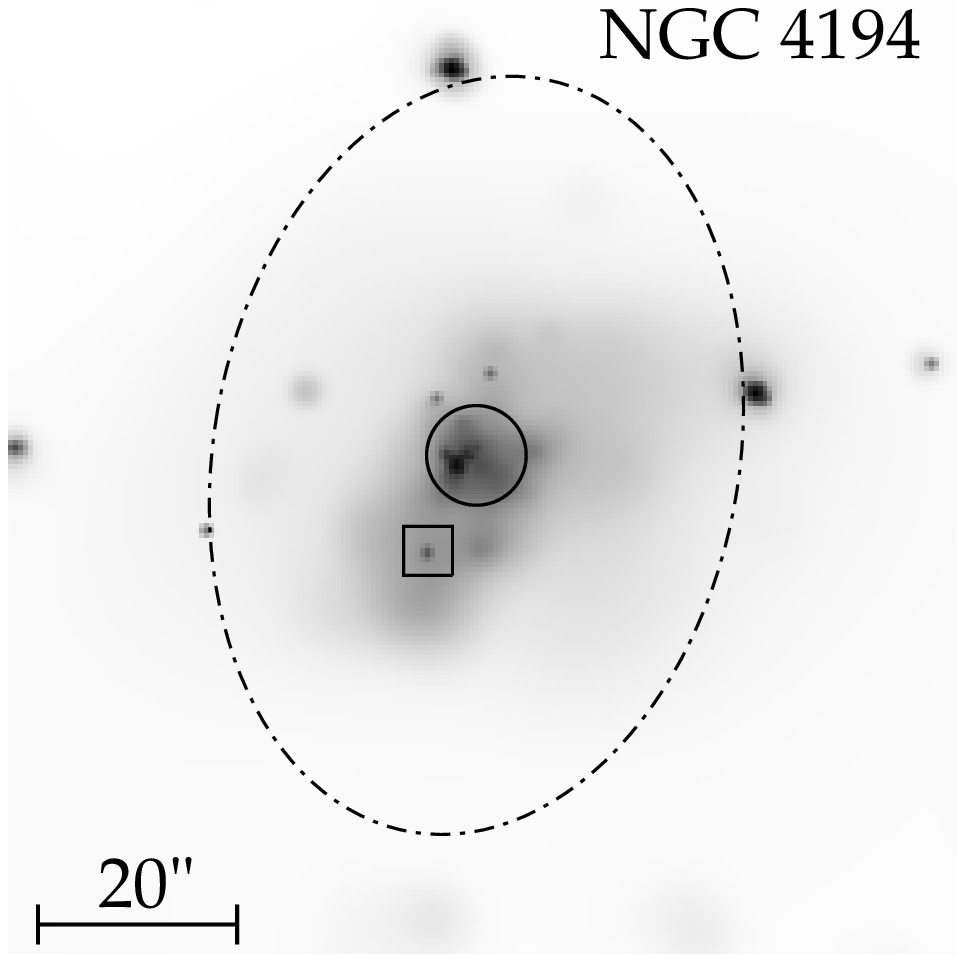}\includegraphics[width=4cm]{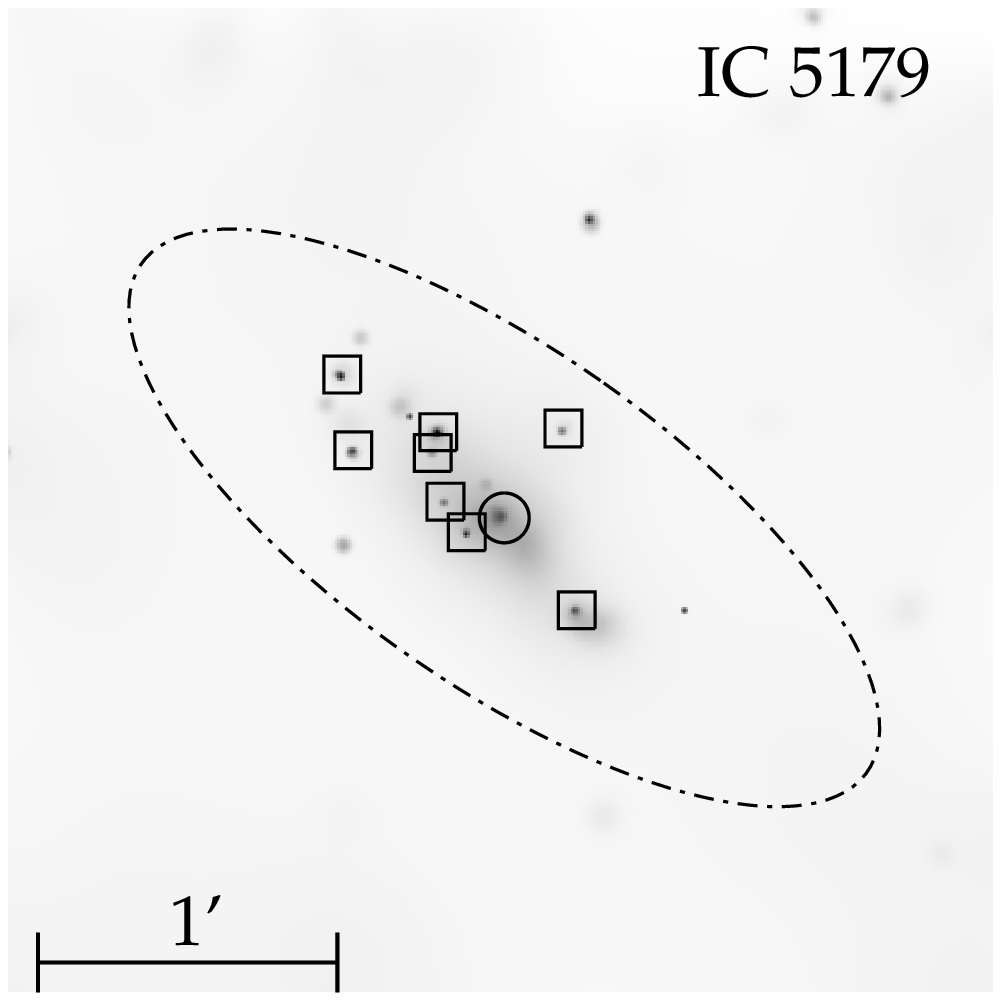}\includegraphics[width=4cm]{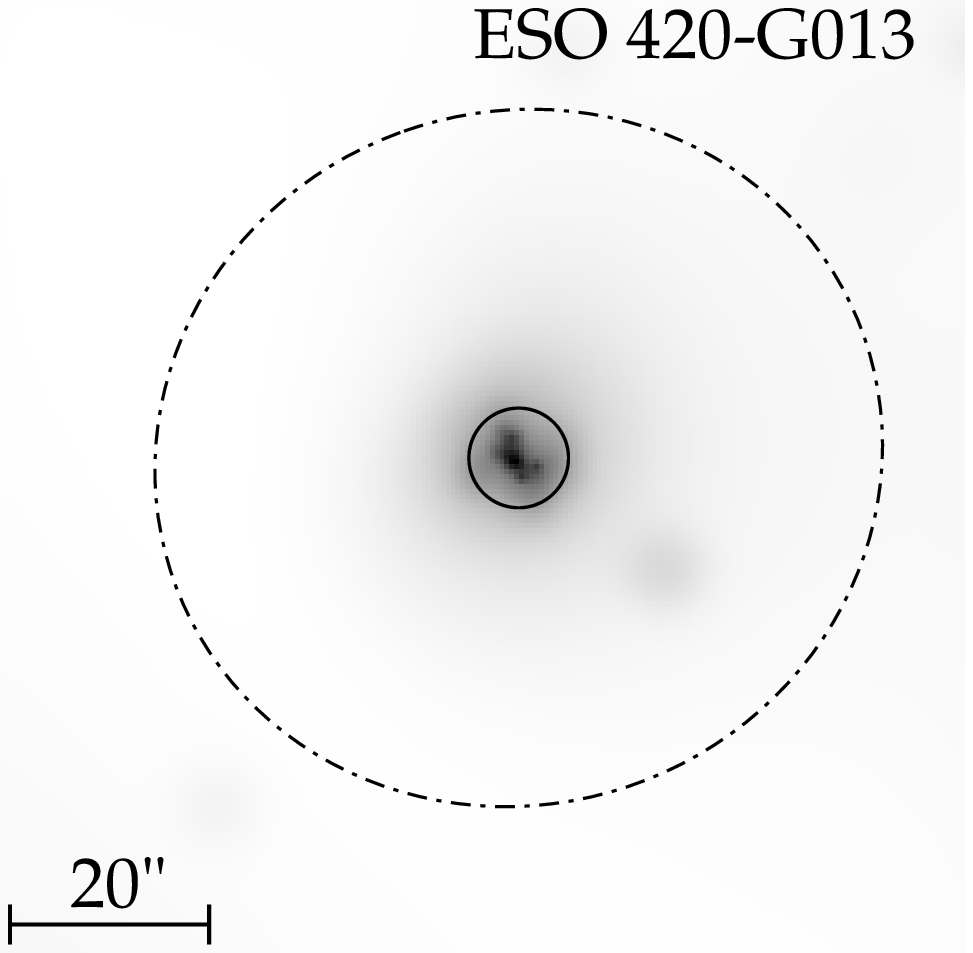}\includegraphics[width=4cm]{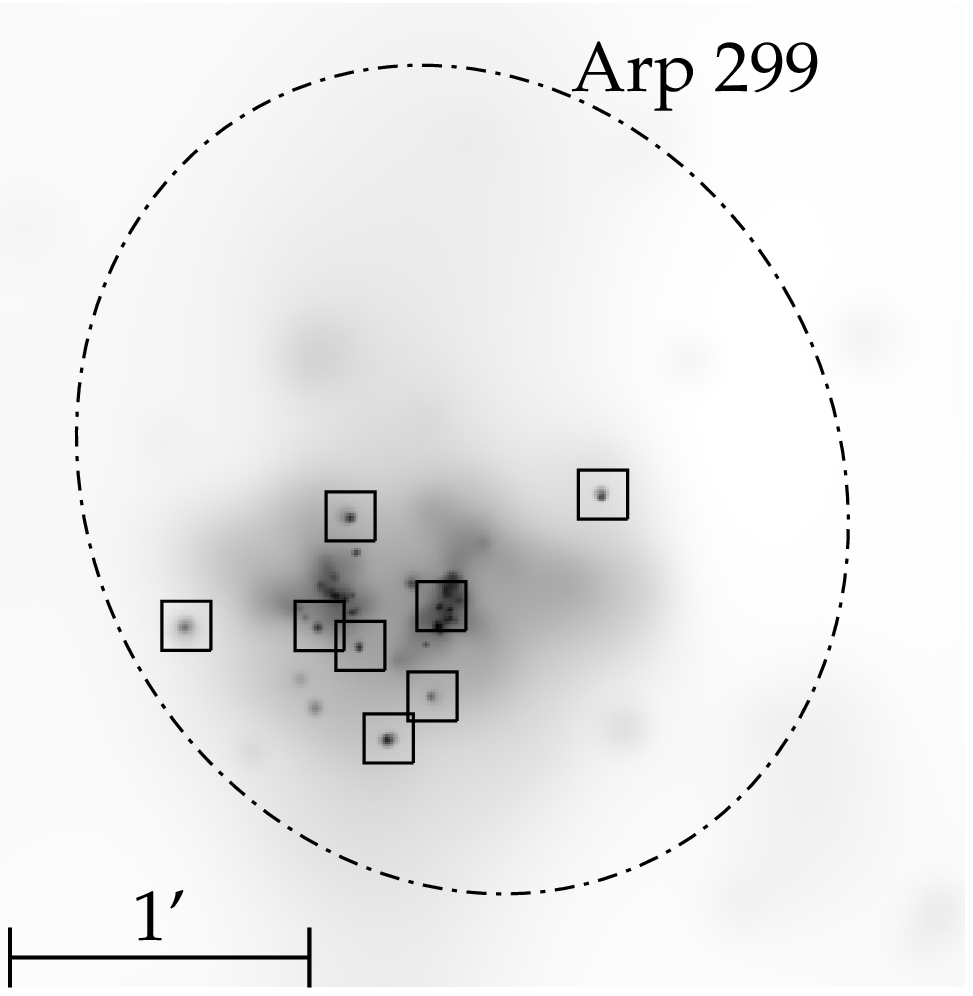}
\includegraphics[width=4cm]{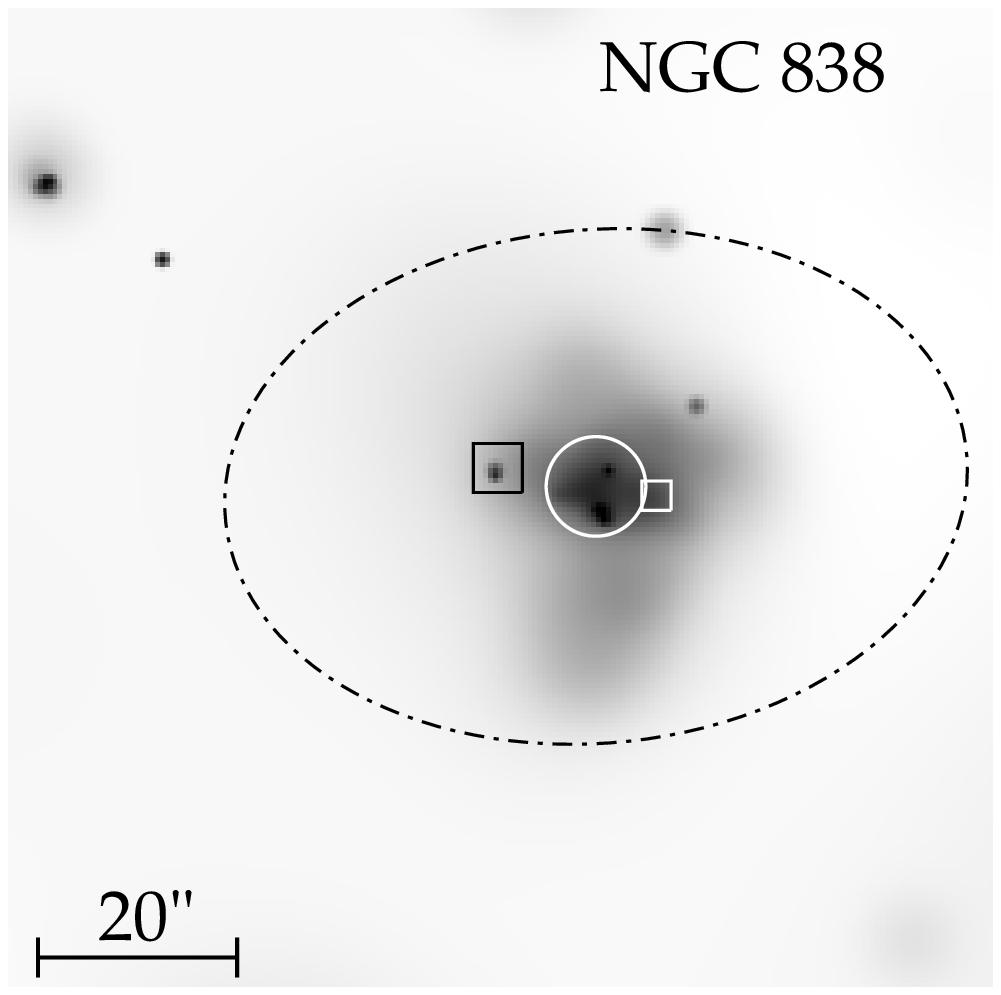}\includegraphics[width=4cm]{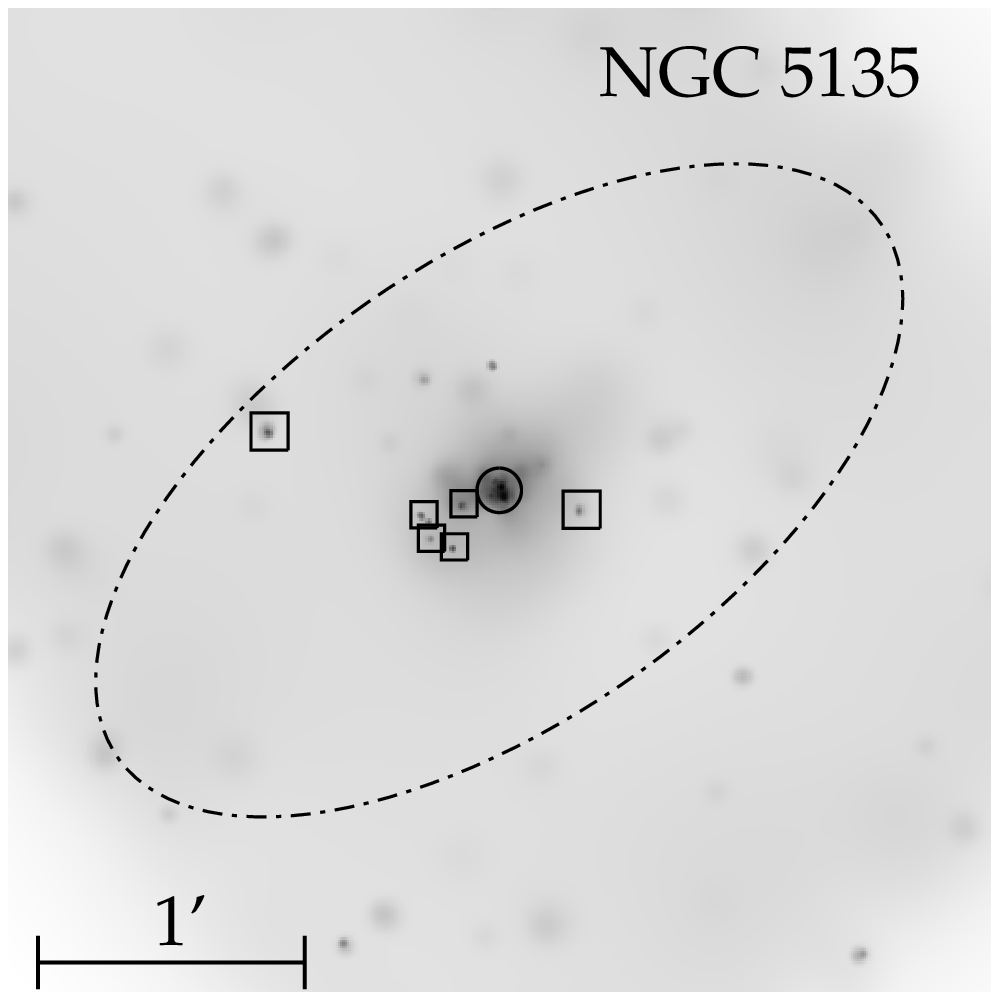}\includegraphics[width=4cm]{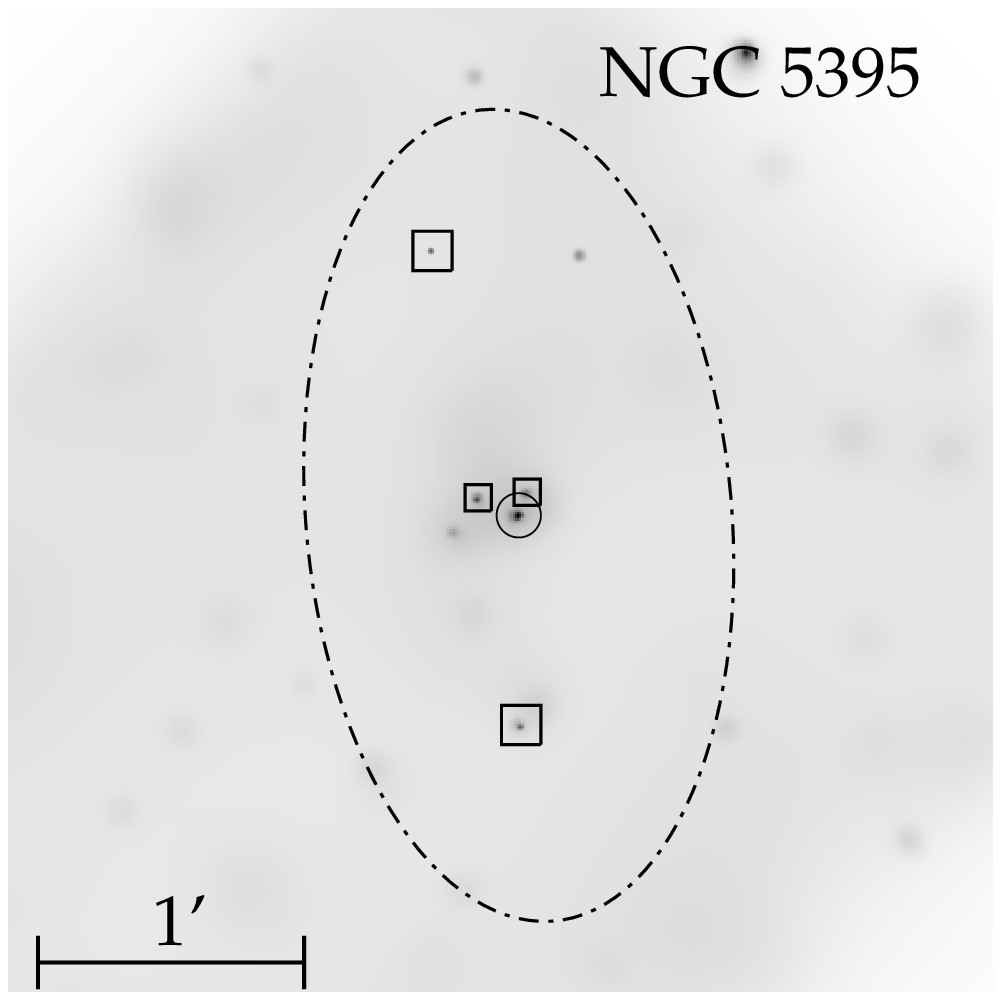}\includegraphics[width=4cm]{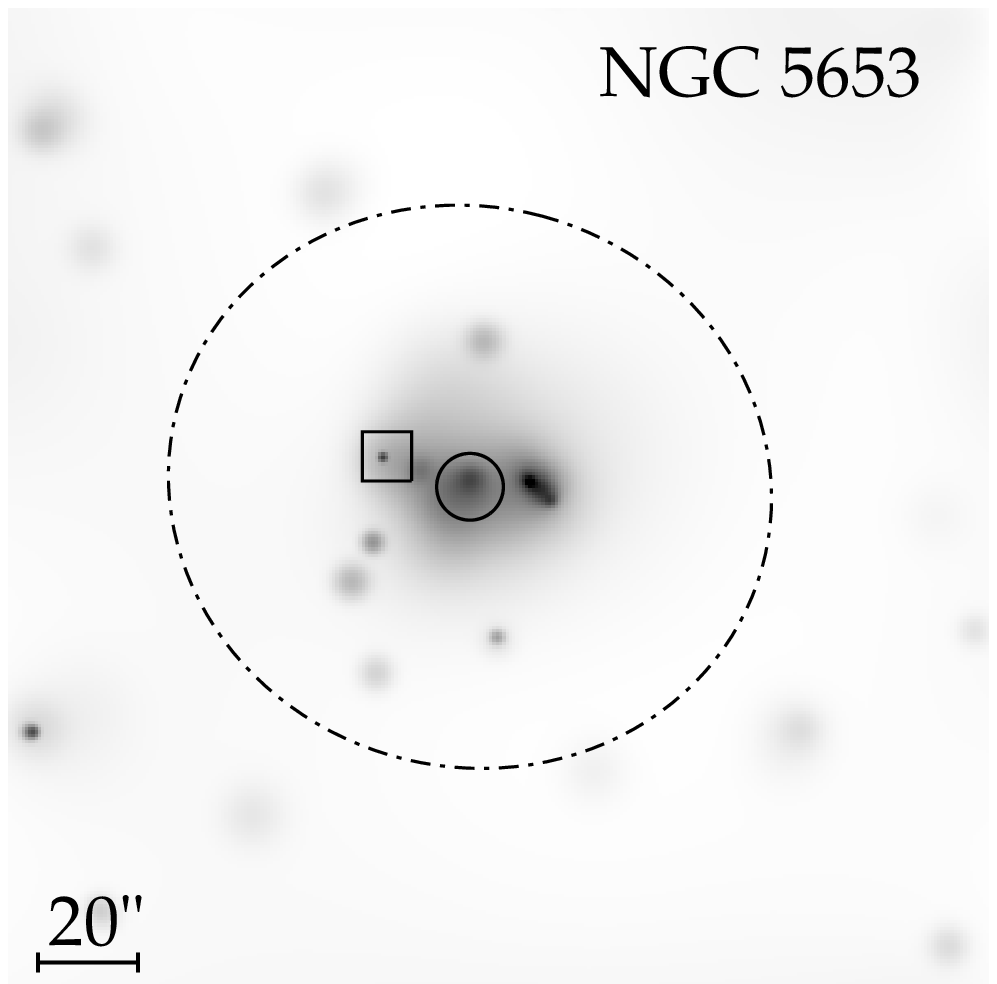}
\includegraphics[width=4cm]{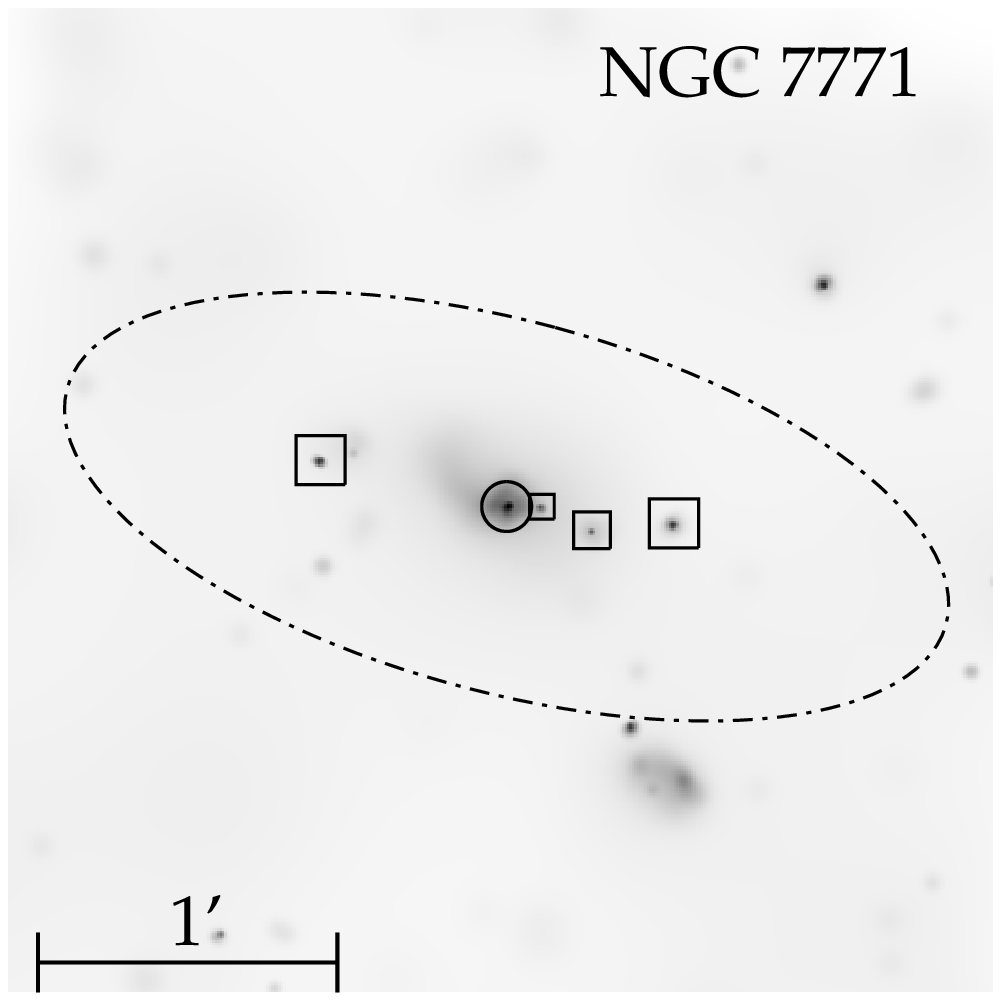}\includegraphics[width=4cm]{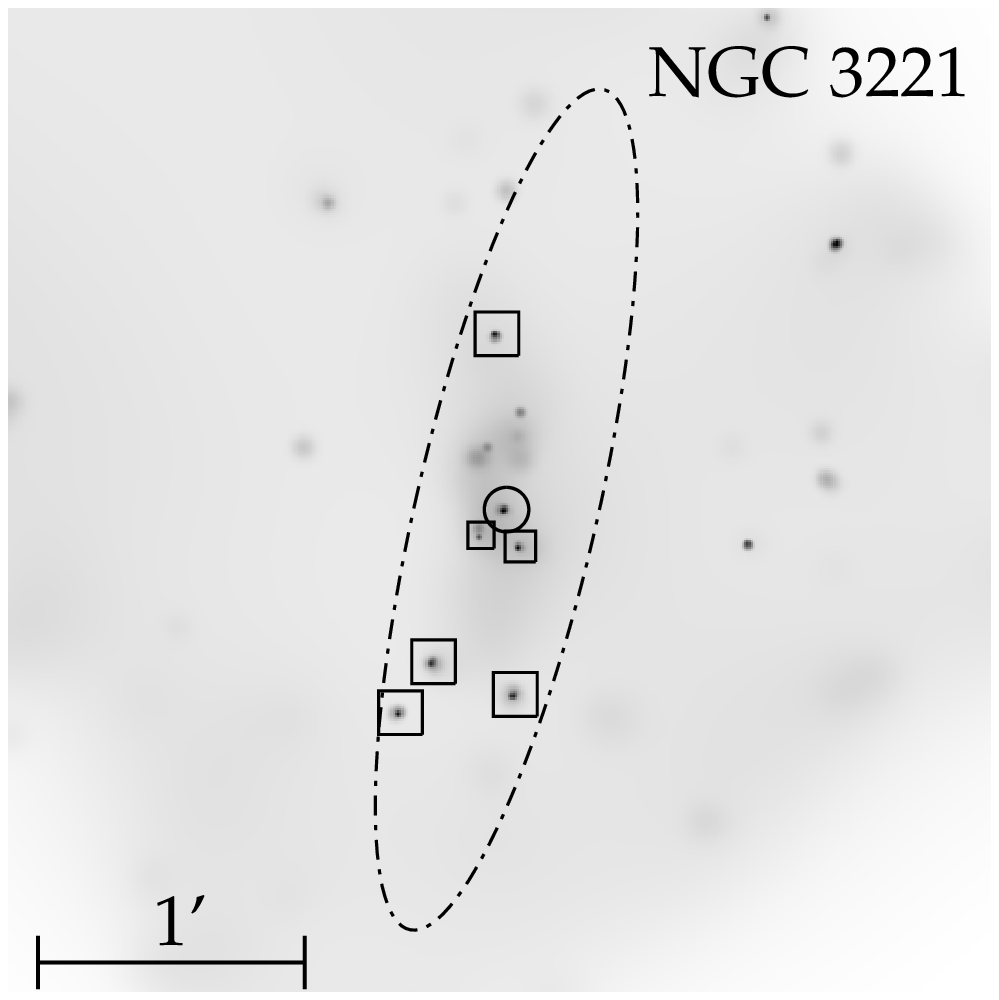}\includegraphics[width=4cm]{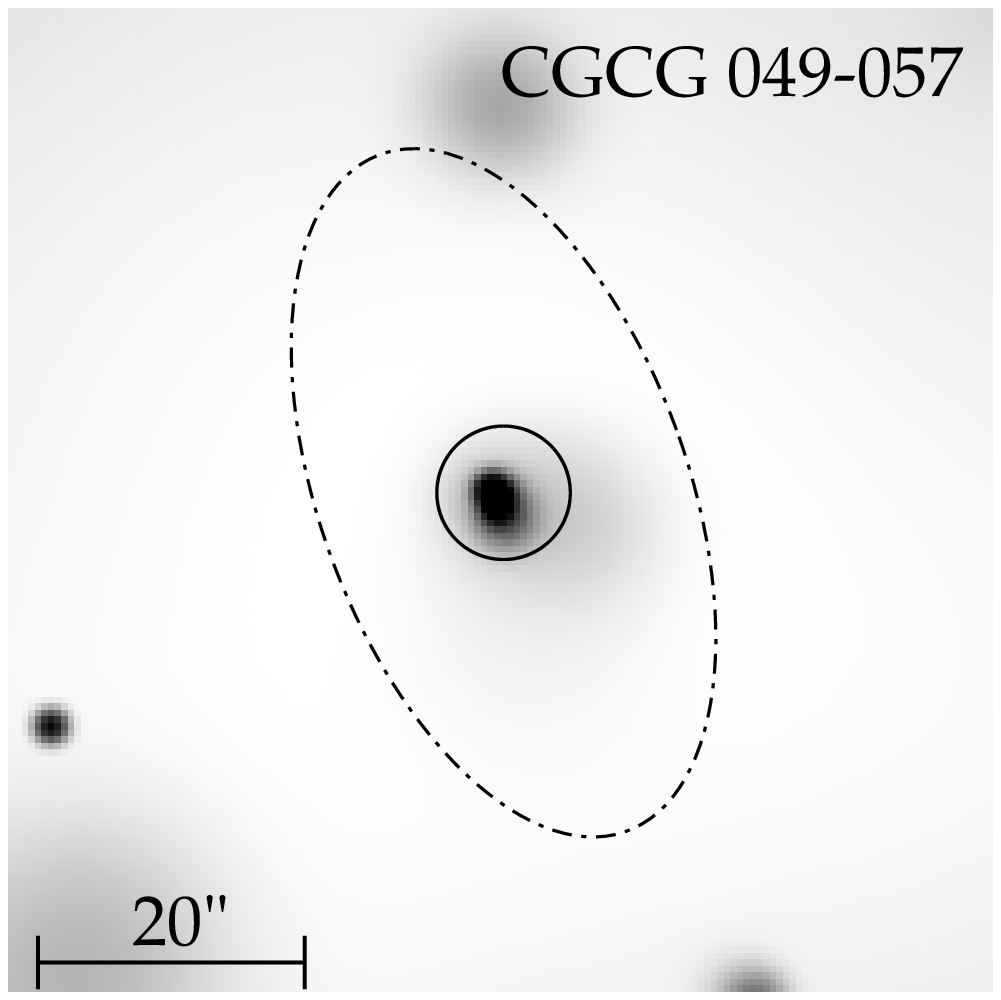}\includegraphics[width=4cm]{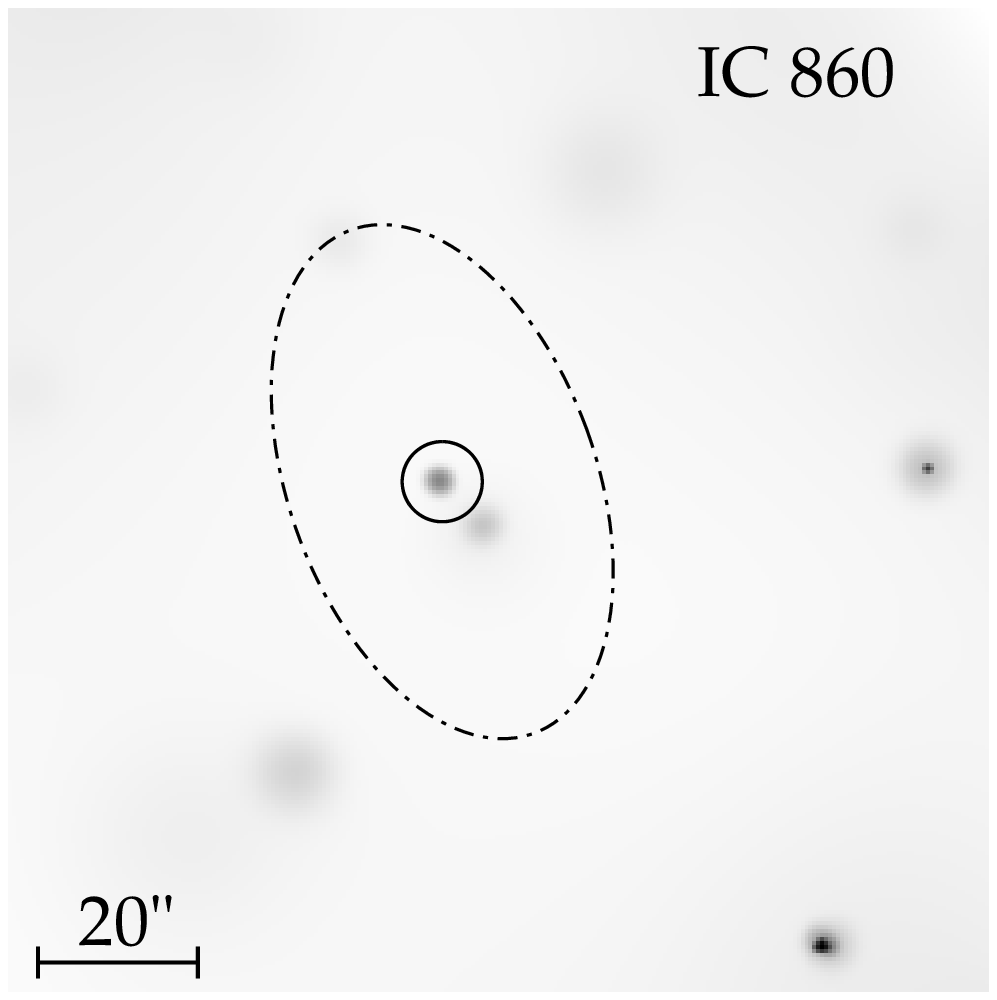}
\includegraphics[width=4cm]{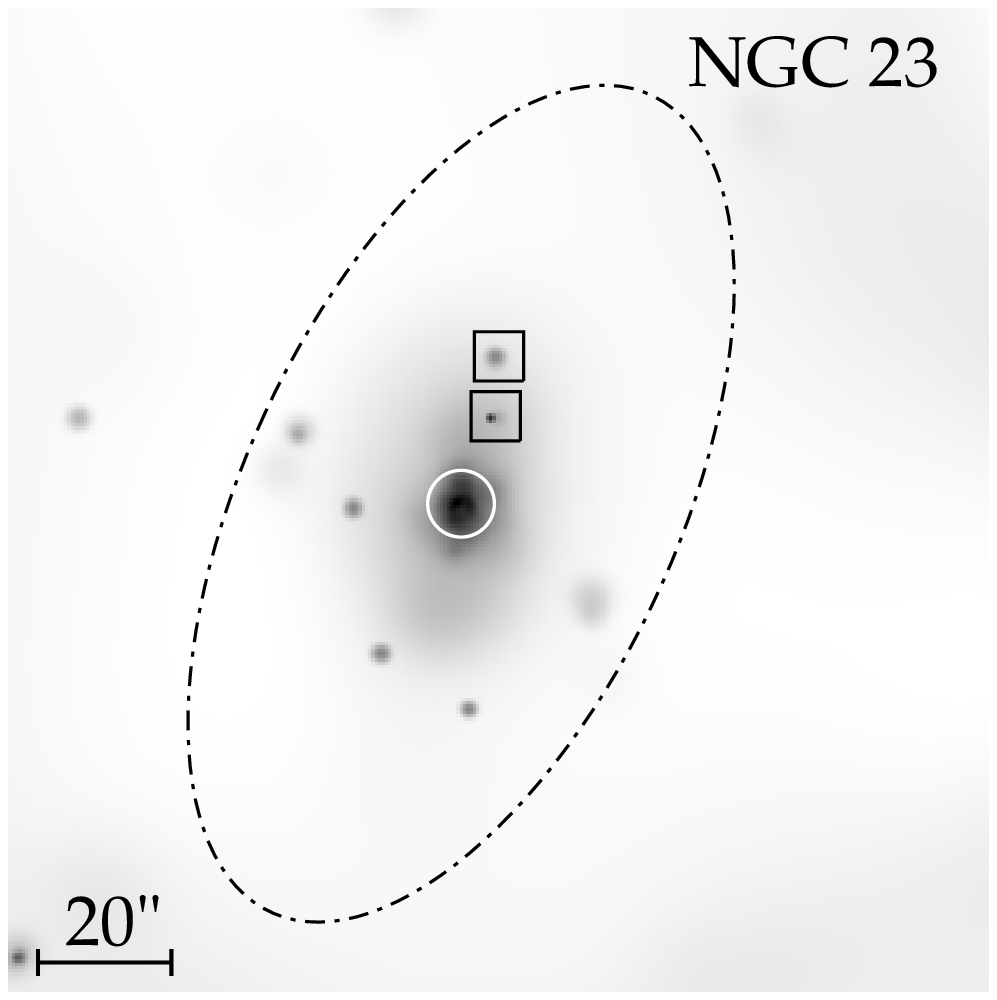}

\caption[]{The ULX detections (boxes), presented on the 0.3--10\,keV {\it Chandra\/} X-ray images of their host galaxies.  The images were adaptively smoothed using the {\sc csmooth} algorithm with a S/N ratio set between 2 -- 3 and a minimum smoothing scale of 0.5 pixels for a Gaussian kernel, and are displayed with arbitrary logarithmic scaling intended in each case to permit the ULXs to be visible.  The dash-dot ellipses define the $R_{20}$ region of each galaxy, and the circles indicate the 5 arcsecond radius regions excluded from the point source analysis due to their proximity to the nucleus.  Note that no exclusion was made in the case of Arp 299 where the irregularity of this galaxy merger makes the definition of a single nuclear position difficult.}
\label{fig:detectedULX}
\end{center}
\end{figure*}

The ULX candidates are tabulated in Table~\ref{tab:ULXcandidates}.  The number of ULXs detected in each galaxy is shown in Table~\ref{tab:detectedULX}, and the position of the ULXs within their host galaxies are shown in Fig.~\ref{fig:detectedULX}.  In total, 53 ULXs were detected in 13 LIRGs; no ULXs were detected in NGC 4418, ESO 420-G013, CGCG 049-057 or IC 860.


      \begin{table}
      \centering
      \caption{ULX detections and completeness per galaxy}\label{tab:detectedULX}
      \smallskip
      \begin{threeparttable}
          \begin{tabular}{lccc}
          
             \hline

\multirow{2}{*}{~~~Galaxy}	&	Number of  & \multirow{2}{*}{log $L_{\text{comp}}$$^{b}$}& \multirow{2}{*}{\textsl{K(L$_{\rm ULX}$)}$^{c}$}\\
 &	ULXs$^{a}$	&	&	\\

\hline
NGC 1068	& $	3	\pm 0 $ & 37.53& 1.00\\
NGC 1365	& $	6	_{-	1	}^{+	2	} $ & 38.11& 1.00\\
NGC 7552	& $	2	_{-	1	}^{+	3	} $ & 38.47& 1.00\\
NGC 4418	& $	0	\pm 0 $ & 38.31& 1.00\\
NGC 4194	& $	1	\pm	1	$ & 38.55& 1.00\\
IC 5179	& $	8	_{-	3	}^{+	0	} $ & 38.88& 0.98\\
ESO 420-G013	& $	0	\pm 0 $ & 38.87& 0.96\\
Arp 299	& $	8	_{-	0	}^{+	1	}$ & 38.96 & 0.93\\
NGC 838	& $	2	_{-	0	}^{+	1	} $  & 39.01& 0.89\\
NGC 5135	& $	6	_{-	2	}^{+	1	} $  & 38.70& 0.99\\
NGC 5395	& $	4	_{-	2	}^{+	1	} $  & 38.85& 1.00\\
NGC 5653	& $	1	\pm 1 $  & 38.90& 0.95\\
NGC 7771	& $	4	_{-	0	}^{+	4	} $   & 38.92& 0.97\\
NGC 3221	& $	6	_{-	1	}^{+	0	} $  & 38.90& 0.99\\
CGCG 049-057	& $	0	\pm 0 $  & 38.87& 1.00\\
IC 860	& $	0	\pm 0 $ &  38.86& 1.00\\
NGC 23	& $	2	\pm2 $  & 38.92& 0.96\\
\hline

Total & $	53	_{-	13	}^{+	16	} $  & & \\
\hline

         \end{tabular}
         \begin{tablenotes}
         \item \textbf{Notes.} $^{a}$The number of ULXs detected in each galaxy. The upper and lower limits were calculated from the number of sources whose 1$\sigma$ luminosity upper limit lies within the ULX regime, and the number of ULXs with lower luminosity limit below the ULX regime, respectively. $^{b}$The completeness luminosity in decimal logarithmic units, defined as the luminosity at which the completeness function, $K(L)$ = 0.9. $^{c}$The value of the completeness function at the minimum ULX luminosity
of 10$^{39}$ erg s$^{-1}$.
                        
         \end{tablenotes}
      \end{threeparttable}
    \end{table}


A pertinent question is whether we have detected all of the ULXs in each galaxy? To address this question, we calculated a completeness function for each galaxy following the method of \citet{voss2006}.  We first created sensitivity maps in the full band (0.3 -- 10 keV) for each galaxy, assuming source spectra with power-law continuum forms (we still used typical ULX model parameters: $\Gamma = 2$ and $N_{\rm H} = 1.5\times10^{21}$ cm$^{-2}$, as this is appropriate for ULXs; cf. \citealt{gladstone2009,sutton2012}). We then computed the completeness function, $K(L)$, within the $R_{20}$ region of each galaxy as per \citet{voss2006}, assuming that the fraction of cosmic X-ray background sources contributing to the sample is small and can be neglected (this is justified in Section \ref{sec:CXBcontribution}). Once the completeness functions were constructed, we defined the completeness luminosity ($L_{\text{comp}}$) as the luminosity at which the completeness function, $K(L)$ = 0.9; i.e. no more than 10 per cent of point sources are missing in the detection at this luminosity. We found that all galaxies have $L_{\text{comp}}$ $\la$ 10$^{39}$ erg s$^{-1}$ (see the third column of Table~\ref{tab:detectedULX}).  Indeed, as we are interested in the completeness of the ULX detection, we defined another parameter, \textit{K}($L_{\rm ULX}$), as the completeness of the point source detection at minimum ULX luminosity of 10$^{39}$ erg s$^{-1}$ (the fourth column of Table~\ref{tab:detectedULX}). The analysis showed that the observations are sufficiently sensitive for 100 per cent of the ULXs to be detected in 8 of the sample galaxies (i.e. $K(L_{\rm ULX}) = 1$).  In addition, the remainder of the galaxies are complete to the $\ga 90$ per cent level.  Thus we are satisfied that the ULX catalogue is complete in terms of ULX detection.

\subsubsection{Contamination from the cosmic X-ray background}
\label{sec:CXBcontribution}

An issue that must always be considered when examining source populations in external galaxies is the extent to which these populations are contaminated by foreground and/or background interlopers.  While the majority of foreground objects, particularly at low Galactic latitudes, are bright stars that are readily identified in optical images, at the high Galactic latitudes of the LIRG sample the predominant contamination comes from the background QSOs that constitute the cosmic X-ray background (CXB).  Fortunately, this contamination can be taken into account statistically from the known CXB source counts.

The number of CXB sources contributing to the ULX catalogue was estimated using the $logN-logS$ relationships of \citet{georgakakis2008}. We firstly calculated the flux in the 0.5--10\,keV band of a source at the minimum ULX luminosity of 10$^{39}$~erg~s$^{-1}$ at the distance of each galaxy. Then {\sc webpimms} was used to divide this minimum ULX flux between the 0.5--2\,keV and 2--10\,keV bands, the soft and hard bands used in \citet{georgakakis2008}, respectively. In this calculation we used a CXB AGN-like power-law spectrum to convert the flux between energy bands; a Galactic column density of 5$\times$10$^{20}$ cm$^{-2}$ (the upper limit on the foreground Galactic column density of the LIRGs) and a photon index of 1.7.  We then converted these minimum fluxes to a maximum CXB contribution per unit area using Equation 2 of \citet{georgakakis2008}, before normalising the CXB contamination estimate to the $R_{20}$ area of every LIRG in both the soft and hard bands.  Finally, we summed these contamination estimates together to calculate a total contamination for the sample.

The calculation predicts a total of $\sim 10$ soft band contaminants, and $\sim 14$ contaminants in the hard band.  This compares to 46 soft band detections\footnote{Given the low effective area of ACIS-S below 0.5 keV, we regard our 0.3--2\,keV band as equivalent to the 0.5--2\,keV band to first order.}, and 27 hard band detections, leading to contamination estimates of $\sim 22$ per cent and $\sim 52$ per cent respectively.  However, we regard these numbers as upper limits, for the following reasons.  Firstly, the soft band numbers do not account for the absorption of the background CXB source photons by the cold, neutral interstellar medium of the LIRGs.  This will act to reduce the detectability of the background objects, and so diminish the numbers of contaminants.  Secondly, the calculation is based on detectability in the full band; this does not guarantee a detection in the hard band.  Indeed, in many galaxies a full band detection of a ULX is at the limit of the capabilities for the given exposure times (cf. the completeness analysis in Section~\ref{sec:completeness}).  So, given that $< 30$ per cent of the photons detected by the {\it Chandra\/} ACIS-S detector for the AGN-like spectral model used here are above 2 keV, these would not result in hard band detections for the ULX flux limit (although, conversely, given that most of the photon flux is below 2 keV the situation is not as bad for that band -- as is indeed shown by the relative soft/hard detection statistics given above).  Indeed, a factor $\sim 4$ increase in flux would be required to turn these full band detections into hard band detections; this means that the current estimate of contamination in this band is very likely a grossly inflated overestimation.  We therefore disregard it, and quote only the soft band upper limit of $\la 22$ per cent contamination for the sample.  We note this is relatively low (thus permitting the analysis in Section~\ref{sec:completeness}), so we are confident our sample is dominated by {\it bona fide\/} ULX detections.  It is also consistent with the contamination fractions in previous similar ULX sample studies, with $\sim 15$ per cent of the ULXs detected in the spiral galaxy samples of \citet{swartz2004} and \citet{walton2011} estimated to be background objects via similar calculations.

\section{Data analysis \& results}

\subsection{Spectral analyses}

The selection of a large sample of ULXs permits us to perform several experiments to investigate their properties, and how these properties relate to that of their host galaxies.  One such experiment is to examine how their spectra evolve as a function of luminosity, where our sample can act as an independent control to investigate the suggested evolution from the study of the nearest and brightest objects, such as presented in \cite{gladstone2009}.  However, given the low number of counts available for most of the ULX detections in our sample (cf. Table~\ref{tab:ULXcandidates}) it is clear that we cannot do this for each individual ULX. Hence, we simply split the ULXs by luminosity, segregating into three equal logarithmical luminosity bins: low luminosity ($39 \leq$ log $ L_{\rm X}<39.33$, 25 sources); medium luminosity ($39.33 \leq $ log $ L_{\rm X}<39.67$, 21 sources); and high luminosity ($39.67 \leq $ log $ L_{\rm X} < 40$, 6 sources), and proceeding to study their stacked spectra. These luminosity boundaries are broadly supported by evidence that ULXs do change in their spectral shape, from a disc-dominated to a two-component ultraluminous regime, at a luminosity $\sim$3 $\times$ 10$^{39}$ erg s$^{-1}$ (cf. \citealt{gladstone2009}; \citeauthor{sutton2013b} 2013b). In one case -- CXOU~J024238.9-000055 -- the high number of counts detected would have dominated the stacked spectrum of the high luminosity bin, and so we study it individually below.

We extracted the spectrum of each ULX using the {\sc ciao specextract} script.  The same source and background regions as were used to extract source photometry in Section \ref{sec:source photometry} were also used here as data extraction apertures.  Once extracted, the spectra were stacked by the {\sc combine\_spectra} script using the {\it summation\/} method, to create a single spectrum for each luminosity bin, and grouped to a minimum of 25 counts per bin to permit $\chi^2$ fitting. In the case of CXOU~J024238.9-000055, the individual spectrum was also grouped to a minimum of 25 counts per bin. The spectra were then fitted in {\sc xspec}\footnote{\texttt{http://heasarc.gsfc.nasa.gov/xanadu/xspec/}} version 12 over the energy range 0.3--10\,keV.  In the following fits the absorption is modelled using the {\sc tbabs} model with interstellar abundances set to the values of \cite{wilms2000}; we do not explicitly account for foreground Galactic column in the fits as the stacked spectra will contain data from objects with differing values of this parameter, although we note that the initial selection criteria of the LIRG sample ($N_{H}$ $< 5 \times 10^{20}$ cm$^{-2}$) ensures this is relatively low.  All best fitting spectral parameter values are quoted with 90 per cent confidence intervals \citep{avni1976} unless otherwise specified.

\subsubsection{Combined spectral fits}\label{sec:stack}

\begin{figure}
\begin{center}

\includegraphics[width=8.5cm]{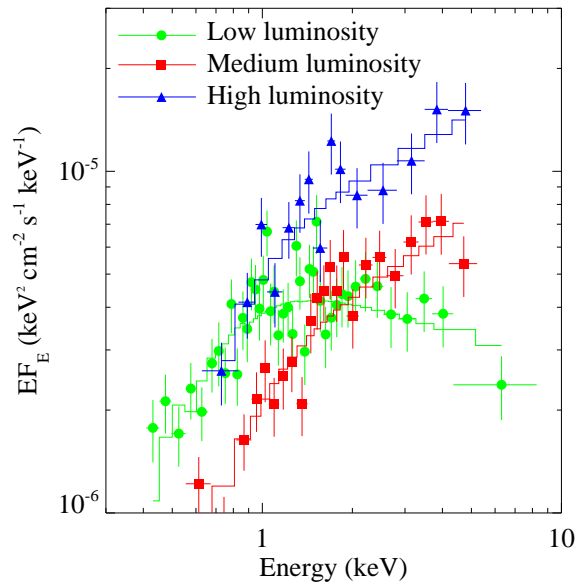}

\caption[]{The stacked spectra of the low ($39 \leq$ log $ L_{\rm X}<39.33$, green circle), medium ($39.33 \leq $ log $ L_{\rm X}<39.67$, red square) and high ($39.67 \leq $ log $ L_{\rm X} < 40$, blue triangle) luminosity groups, unfolded from the detector response.  We also show the best-fitting power-law continuum model as the similarly-coloured solid line in each case.}
\label{fig:grouped spectra}
\end{center}
\end{figure}

We began by fitting the stacked spectra with a simple empirical model, namely an absorbed power-law. The results are shown in Fig.~\ref{fig:grouped spectra} and the model fitting parameters are reported in Table~\ref{tab:fittingresult}.  Two main results are apparent.  Firstly, the absorption for this model is consistent between all three luminosity groups at $\sim 2 \times 10^{21}$ cm$^{-2}$; this is significantly ($\sim$4 times) higher than the foreground columns for any of the LIRGs, indicating that absorption from cold material is present in these galaxies.  Secondly, a significant change in the stacked spectrum is seen between the low and medium luminosity groups, with the spectrum hardening from $\Gamma \sim 2.3$ to $\sim 1.5$ respectively.  However, the spectrum then remains constant (within errors) between the medium and high spectra: there is no statistical distinction in power-law photon index between these 2 luminosity bins.

Similar results were seen when an absorbed multicolour disc blackbody model (MCD; {\sc diskbb} in {\sc xspec}; \citealt {mitsuda1984}) was used instead.  In this case the absorption columns ($N_{H}$) of the three luminosity bins were also formally consistent with each other, and with no absorption, although the column for the low luminosity group had a much lower upper limit than the more luminous bins.  But again, a significant hardening is seen between the low and medium luminosity bins, with the medium and high luminosity stacked spectra very consistent.  While it would not be prudent to extract physical meaning from the results of this MCD fit given the likely wide differences in underlying spectra of the contributing sources, the similarity of the main results with the power-law fit -- namely an apparent change in spectrum for objects at luminosities above and below $\sim 2 \times 10^{39} \rm ~erg~s^{-1}$ -- points at a real change in underlying spectrum.  

However, one must be extremely careful when stacking spectra not to induce any characteristics into the spectra by the choice of bins, or the relative contributions of sources to the bins.  In this case, we note that all the source detections with very low total counts ($\la 10$ counts) lie in the lowest luminosity bin; given the predominantly soft response of {\it Chandra\/} this raises the concern that these objects may not be contributing many counts above 2 keV, where the ACIS-S detector sensitivity falls off, thus artificially softening the stacked spectrum in this bin.  However, we note two things: firstly, that sources would need to be very soft not to contribute counts above 2 keV in the ACIS-S -- for example, an object with 10 detected counts and an intrinsic power-law slope of $\Gamma = 2.5$ and absorption $N_{\rm H} = 5 \times 10^{20} \rm ~cm^{-2}$ would contribute 1--2 counts above 2 keV.  Secondly, the 12 objects with low counts only contribute $\sim 7$ per cent of the total counts to this stacked spectrum.  We confirm that this is not an issue by re-stacking the spectra in the low and medium luminosity bins to exclude all sources with more than 70 counts; this leaves spectra with 371 and 405 counts respectively.  The power-law continuum fits to these spectra are consistent within errors with the full stacked spectrum fits to each bin, showing the same soft/hard spectral dichotomy (low luminosity bin:  $N_{\rm H} = 0.3 \pm 0.2 \times 10^{22} \rm ~cm^{-2}, \Gamma = 2.1 \pm 0.4$; medium luminosity bin: $N_{\rm H} = 0.4^{+0.3}_{-0.2} \times 10^{22} \rm ~cm^{-2}, \Gamma = 1.4 \pm 0.4$).  We therefore do not consider the low luminosity stacked spectrum to have been artificially softened by this effect.  

There is one more concern about the low luminosity bin, though; two sources (CXOU J024238-000118 and CXOU J024240-000101, both in NGC 1068) contribute roughly one third of the counts to the bin each.  We therefore examined their spectra (that were removed from the analysis reported in the above paragraph), and found that one object appears softer than the full stacked spectrum for this bin ($N_{\rm H} = 0.6 \pm 0.2 \times 10^{22} \rm ~cm^{-2}, \Gamma = 4.4 \pm 1.1$), and the other harder ($N_{\rm H} = 0.2 \pm 0.1 \times 10^{22} \rm ~cm^{-2}, \Gamma = 1.6 \pm 0.3$), cf. Table 4; hence their spectra average out across the bin.  Considering that the other objects in this bin have a stacked spectrum that is consistent with the full stack, we confirm that the stacked spectra reflect the average of the underlying low luminosity ULX properties.  The difference in spectra can therefore be regarded as a real effect, and we discuss it further in Section~\ref{accretion state in black holes}.
 
Finally, we attempted more complex spectral model fits, in particular an absorbed MCD plus power-law model.  However, in large part due to the moderate quality of the stacked spectra (a few hundred to $\sim 1000$ counts per bin), no statistical improvements over the single component models could be discerned.

\begin{table*}
      \centering
      \caption{Combined spectral fitting}\label{tab:fittingresult}
      \smallskip
      \begin{threeparttable}
      \begin{tabular}{ccccccccc}
 
             \hline
Luminosity&No. of& \multicolumn{3}{c}{Power-law} && \multicolumn{3}{c}{MCD}\\
\cline {3-5}  \cline {7-9}
bin&sources$^{a}$&&&&&&&\\ 
&&  $N_{\rm H}$$^{b}$  & $\Gamma$$^{c}$ & $\chi^{2}/d.o.f.$$^{d}$ && $N_{\rm H}$$^{e}$& $T_{\rm in}$$^{f}$& $\chi^{2}/d.o.f.$$^{d}$\\
              \hline
Low & \multirow{2}{*}{25} & \multirow{2}{*}{$	0.19	_{	-0.05	}^{+	0.04	}$} & \multirow{2}{*}{$	2.31	_{	-0.09	}^{+	0.11	}$} & \multirow{2}{*}{$	49.84	/	40	$} & \multirow{2}{*}{} & \multirow{2}{*}{$	< 0.01  $} & \multirow{2}{*}{$	0.84	_{	-0.06	}^{+	0.08	}$} & \multirow{2}{*}{$	60.82	/	40$}\\
\scriptsize($39\leq$log$L_{\rm X}<39.33$)&&&&&&&&\\

Medium & \multirow{2}{*}{21} & \multirow{2}{*}{$	0.25	_{	-0.09	}^{+	0.15	}$} & \multirow{2}{*}{$	1.48	_{	-0.14	}^{+	0.21	}$} & \multirow{2}{*}{$	26.6	/	20	$} & \multirow{2}{*}{} & \multirow{2}{*}{$ <0.17$} & \multirow{2}{*}{$	1.59	_{	-0.24	}^{+	0.44	}$} & \multirow{2}{*}{$	21.64	/	20	$}\\
\scriptsize($39.33\leq$log$L_{\rm X}<39.67$)&&&&&&&&\\

High  & \multirow{2}{*}{6} & \multirow{2}{*}{$	0.23	_{	-0.12	}^{+	0.20	}$} & \multirow{2}{*}{$	1.57 _{	-0.22	}^{+	0.21	}$} & \multirow{2}{*}{$	13.48	/	12	$} & \multirow{2}{*}{} & \multirow{2}{*}{$<0.13$} & \multirow{2}{*}{$	1.59	_{	-0.32	}^{+	0.63	}$} & \multirow{2}{*}{$	16.66	/	12	$} \\

\scriptsize($39.67\leq$log$L_{\rm X}\leq40$)&&&&&&&&\\
             
             \hline
         \end{tabular}
         \begin{tablenotes}
         \item \textbf{Notes.} $^{a}$the number of sources contributing to the luminosity bins. $^{b}$Total absorption column measured from power-law model (10$^{22}$ cm$^{-2}$).  $^{c}$Power-law photon index. $^{d}$$\chi^2$ fitting statistic, and number of degrees of freedom. $^{e}$Total absorption column measured from MCD model (10$^{22}$ cm$^{-2}$). $^{f}$Inner-disc temperature (keV). All best fitting parameter values are reported with 90 per cent confidence intervals.      
         
      \end{tablenotes}
      \end{threeparttable}
    \end{table*}


\subsubsection{CXOU~J024238.9-000055}\label{sec:srcspec}

CXOU~J024238.9-000055 is one of the brightest ULXs in the entire sample, at $L_{\rm X}\sim 5\times10^{39}$ erg s$^{-1}$, and given its location in the nearest galaxy in the sample (NGC 1068) and the relatively high exposure for this galaxy (46.9 ks) we have accumulated sufficient counts from this object for an analysis of its spectrum.  Previous work has shown it has a very hard spectrum: its power-law photon index is $\sim0.8-1$ \citep{smith2003,swartz2004,berghea2008}.  However, this is not due to excessive pile-up on the \textit{Chandra} detectors; its observed count rate of $\sim 0.03$ count s$^{-1}$ is no more than $\sim 5$ per cent piled-up\footnote{\texttt{http://cxc.harvard.edu/csc/memos/files/Davis\_pileup.pdf}}.  It is not due to out-of-time events from the bright AGN in NGC 1068 either; although the source is located along the readout streak\footnote{\texttt{http://cxc.harvard.edu/ciao/why/pileup\_intro.html}}, we have verified using on- and off-streak backgrounds that the spectrum is not significantly affected by this detector effect (indeed, we note that the adaptive smoothing algorithm used to create Fig.~\ref{fig:detectedULX} greatly exaggerates this effect compared to raw data).  Hence the interpretation of the unusual hardness of this ULX is currently unclear.  Here, we re-analyse the data to revisit this question.

\begin{figure}
\begin{center}

\includegraphics[width=8.5cm]{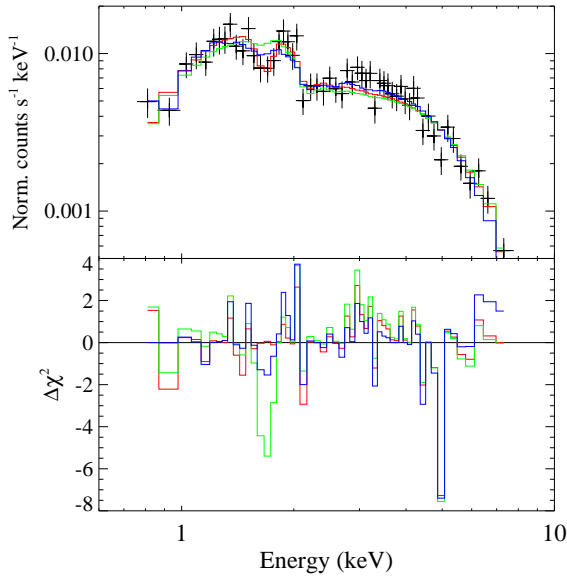}

\caption[]{The spectrum of CXOU~J024238.9-000055.  {\it Top panel:\/} The spectral data (in black), with the best fitting power-law model (green line), power-law model with Gaussian absorption component (red line) and power-law model with partial covering absorber (blue line).  {\it Bottom panel:\/} The residual differences between the data and the best fit, with the models plotted as above.}
\label{fig:src5 spectra}
\end{center}
\end{figure}



\begin{table}
      \centering
      \caption{Spectral fitting results for CXOU~J024238.9-000055}\label{tab:src5fittingresult}
      \smallskip
      \begin{threeparttable}
      \begin{tabular}{cc}
      
 \hline
 \multicolumn{2}{c}{\textbf{Single power-law model}} \\
 $N_{\rm H}$$^{a}$ & $0.67_{-0.14}^{+0.17}$ \\
 $\Gamma$$^{b}$ & $0.75_{-0.11}^{+0.12}$ \\
 $\chi^{2}/d.o.f.$ & 65.29/53\\
 
\hline
 \multicolumn{2}{c}{\textbf{Power-law model with Gaussian absorption}} \\    
 $N_{\rm H}$$^{a}$ & $0.75_{-0.07}^{+0.12}$ \\
 $\Gamma$$^{b}$ & $0.88_{-0.11}^{+0.16}$ \\   
  Line energy$^{c}$ & $1.68_{-0.08}^{+0.04}$\\
  Line width$^{d}$ & $0.08_{-0.04}^{+0.07}$\\
  $\chi^{2}/d.o.f.$ & 47.13/50\\ 
 
\hline
 \multicolumn{2}{c}{\textbf{Power-law model with partial covering absorption}} \\
 $N_{\rm H}$$^{a}$ & $0.68_{-0.19}^{+0.30}$ \\
 $\Gamma$$^{b}$ & $1.65_{-0.33}^{+0.77}$ \\
 $N_{\rm H, pa}$$^{e}$& $7.95_{-2.36}^{+6.08}$\\
 log $\xi$$^{f}$ & $2.20_{-0.19}^{+0.18}$\\
 Covering fraction$^{g}$ & $>0.83$\\
 $\chi^{2}/d.o.f.$ & 50.27/50\\

     \hline

         \end{tabular}
         \begin{tablenotes}
         \item \textbf{Notes.} A fixed Galactic column of $0.04 \times 10^{22}$ cm$^{-2}$ (the column density in the direction of NGC 1068) was added to each model. $^{a}$Absorption column intrinsic to the ULX and/or its host galaxy (10$^{22}$ cm$^{-2}$).  $^{b}$Power-law photon index. $^{c}$Central energy of the Gaussian absorption line (keV). $^{d}$Width of Gaussian absorption line (keV). $^{e}$Column density of partial covering absorber (10$^{22}$ cm$^{-2}$).  $^{f}$Logarithm of photo-ionisation parameter $\xi = L/(nR^2)$, where $L$ is the total luminosity, $n$ the density of the absorber and $R$ the distance of the material from the ionising source.  $^{g}$Covering fraction of partial covering absorber (where 0 = totally uncovered and 1 = full covering). All best fitting parameter values are reported with 90 per cent confidence intervals.
         
      \end{tablenotes}
      \end{threeparttable}
    \end{table}


We began by fitting the ULX spectrum with an absorbed power-law continuum model. A similar result to previous work was found: the power-law index was very hard, with $\Gamma = 0.75\pm0.1$ and $N_{H} = 0.67\pm0.2$ $\times$ $10^{22}$ cm$^{-2}$(see Table~\ref{tab:src5fittingresult}), and produced a statistically acceptable fit (null hypothesis probability of $\sim 12$ per cent).  However, the spectrum of this source does show residuals compared to the data, with the largest feature falling below the model level at an energy of $\sim 1.7$ keV (cf. Fig.~\ref{fig:src5 spectra}).  We therefore attempted to improve the fit to the data by adding a Gaussian absorption component to the power-law model.  This provided a reasonably significant improvement to the fit, with $\Delta\chi^2 = 18$ for 3 additional degrees of freedom for a broad absorption feature (width of $\sim 80$ eV) at 1.68 keV.  However, a single broad absorption feature at that energy is difficult to explain physically.

We therefore attempted a more physically meaningful fit to the data by modelling the residual using a partial covering, partially ionised absorption model, namely the {\sc zxipcf} model in {\sc xspec}, on top of an underlying power-law continuum model.  This produced a very similar improvement in the fit to the single Gaussian absorption line ($\Delta\chi^2 = 15$ for 3 additional degrees of freedom), and also largely modelled the feature at $\sim 1.7$ keV, using an absorber with a relatively high column of $\sim 8 \times 10^{22}$ cm$^{-2}$ and high covering factor ($> 0.83$) of moderately ionised material (log $\xi = 2.2$).  Notably, the addition of the partial covering, partially ionised absorber led to a change in the slope of the underlying power-law continuum, with it increasing to a more typical value for a ULX of $\Gamma \sim 1.7$.    This therefore appears to be a physically plausible interpretation of the spectral data for this ULX.  We summarise the analysis of CXOU~J024238.9-000055 in Table~\ref{tab:src5fittingresult} and Fig.~\ref{fig:src5 spectra}, and discuss it further in Section~\ref{sec:discussion}.

\subsection{The correlation between ULXs and star formation in LIRGs}
\label{sec:ULX number and host galaxy}

Whilst a relationship between the presence of ULXs and ongoing star formation is undisputed, the question of whether this relationship is universal, or whether it differs subtly in different environments is an open question, with evidence starting to suggest that differences may occur due to metallicity (see Section~\ref{sec:intro}).  Indeed, this sample itself presents an interesting puzzle -- according to \cite{swartz2011} we should see 2 ULXs per unit SFR; yet we only see 53 ULXs in a sample of galaxies with a summed SFR of $\sim 260~M_{\odot}$ yr$^{-1}$, i.e. 0.2 ULXs per unit SFR, a factor of 10 lower than expected\footnote{This is not the result of different SFR estimates -- although we use a different method to calculate SFR than \cite{swartz2011}, the summed SFR for our sample calculated using both methods differs by $< 2$ per cent.}.  It is therefore of great interest to examine the relationship between ULXs and star formation in our sample in more detail, and to contrast it with other samples of galaxies. 

Here two indicators are used to examine the relationship: the blue and far-infrared (FIR) luminosities of the host galaxies ($L_{\rm B}$ and $L_{\rm FIR}$). As reported by \citet{swartz2009} ULXs tend to reside in regions that have a bluer optical colour than other parts of galaxies; and it was found that the number of ULX per unit blue luminosity is likely to be enhanced in the bluer galaxies \citep{smith2012}, suggesting a connection between ULXs and recently formed, young stellar populations. Although the blue band luminosity is not a reliable indicator of star formation rate, here we derive it for direct comparison with previous work. The FIR luminosity was selected as a more accurate proxy of the SFR of the galaxies. We explain how we derived these luminosities below, and the values for each individual galaxy are given in Table~\ref{tab:galaxy luminosity}.

\textbf{\textit{$\bullet$ Host galaxy blue luminosity.}}  Here we define $L_{\rm B}$ to be the luminosity of the galaxy in the $B$ filter of the Johnson-Morgan $UBV$ system \citep{johnson1953}.  Most $L_{\rm B}$ values for the LIRGs were converted from $B$ magnitudes corrected for extinction and redshift, which were reported in Third Reference Catalog of Bright Galaxies (RC3 catalogue; \citealt{devaucouleurs1991}).  The standard photometric value for conversion was taken from \citet{zombeck1990}. For ESO~420-G013, we calculated $L_{\rm B}$ from the photographic B magnitude reported in the RC3 catalogue. In the case of CGCG~049-057, we took the $g$-magnitude from SDSS DR9\footnote{\texttt{http://skyserver.sdss3.org/dr9/}} and then converted it to a $B$ magnitude using the method of \citet{windhorst1991}.

\textbf{\textit{$\bullet$ Host galaxy FIR luminosity and SFR}}.   As 10 out of the 17 LIRGs host AGN (see Table~\ref{tab:LIRGsample}), their total FIR luminosity $L_{\rm FIR}$ (defined as their luminosity between 42.4 and 122.5 $\mu$m) will be the summation of FIR luminosity emitted from both the host galaxy and the AGN.  Thus simply using the total FIR luminosity risks overestimating the SFR of the galaxies.  To correct this issue, we disentangle the FIR luminosity emitted by the AGN from the total FIR luminosity using the method of \citet{mullaney2011}. In brief, {\it IRAS\/} flux densities at 4 individual IR wavelengths (taken from \citealt{sanders2003}) were fitted with 5 different host galaxy models (host galaxy IR emission templates) derived by \citet{mullaney2011}. The best fitting models were used
to predict the fraction of AGN emission contributing to the total FIR luminosity, allowing a correction to be made that left the undiluted host galaxy luminosities. Overall, the model predicted a small contribution of AGN emission to the total FIR luminosity ($\sim$ 0 -- 7\%), except for NGC~1068 in which the AGN emission contributes $\sim$27 per cent of the total FIR luminosity. The FIR luminosities of the galaxies corrected for AGN contribution are shown in Table~\ref{tab:galaxy luminosity}.

In addition, to calculate the SFRs, we firstly calculated the IR luminosities in the full 8 -- 1000 $\mu$m regime (${L}_{\text{IR}}$), removing AGN contamination as described above. We then used the same calibration as in \citeauthor{lehmer2010} (2010, their Section 2.2) to convert the total ${L}_{\text{IR}}$ to SFR as:\begin{equation} \label{eq:SFR}
\text{SFR}(M_{\odot} \text{yr}^{-1}) = \gamma9.8 \times 10^{-11}L_{\text{IR}},
\end{equation} where ${L}_{\text{IR}}$ is the infrared luminosity in units of solar bolometric luminosity ($L_{\odot}$ = 3.9 $\times$ 10$^{33}$ erg s$^{-1}$) and $\gamma$ is the correction factor accounting for UV emission emerging from unobscured star-forming regions. The value of the latter parameter for each galaxy was estimated following the method described in Section~2.2 of \citet{lehmer2010}. The SFR of each galaxy is shown in column 8 of Table~\ref{tab:LIRGsample}.

\begin{table}
      \centering
      \caption{Blue and FIR LIRG luminosities}\label{tab:galaxy luminosity}
      \smallskip
      \begin{threeparttable}
          \begin{tabular}{lcc}
             \hline
Galaxy & $L_{\rm B}$	&	$L_{\rm FIR}$$^{a}$	\\
   &   (10$^{42}$ ~erg~s$^{-1}$) & (10$^{42}$ ~erg~s$^{-1}$)  \\
\hline

NGC 1068	&$	26.8	\pm	2.5	$&$	159.8	\pm	0.3	$ \\
NGC 1365	&$	30.1	\pm	1.9	$&$	201.0	\pm	1.2	$ \\
NGC 7552	&$	14.2	\pm	1.7	$&$	207.7	\pm	4.5	$ \\
NGC 4418	&$	3.3	\pm	0.0	$&$	205.9	\pm	7.3	$ \\
NGC 4194	&$	10.3	\pm	1.2	$&$	217.7	\pm	5.3	$ \\
IC 5179	&$	33.1	\pm	4.0	$&$	302.6	\pm	9.5	$ \\
ESO 420-G013	&$	9.7	\pm	0.0	$&$	188.0	\pm	4.4	$ \\
Arp 299	&$	36.8	\pm	0.0	$&$	1436.3	\pm	7.2	$ \\
NGC 838	&$	8.8	\pm	0.3	$&$	190.7	\pm	8.7	$ \\
NGC 5135	&$	27.4	\pm	3.3	$&$	318.7	\pm	17.7	$ \\
NGC 5395	&$	39.8	\pm	7.3	$&$	202.9	\pm	8.6	$$^{b}$\\
NGC 5653	&$	23.1	\pm	2.8	$&$	241.2	\pm	12.4	$ \\
NGC 7771	&$	29.2	\pm	1.3	$&$	450.4	\pm	16.4	$ \\
NGC 3221	&$	24.4	\pm	0.0	$&$	199.3	\pm	11.1	$ \\
CGCG 049-057	&$	0.9	\pm	0.0	$&$	452.5	\pm	45.3	$ \\
IC 860	&$	3.1	\pm	0.0	$&$	336.4	\pm	33.6	$ \\
NGC 23	&$	39.3	\pm	6.2	$&$	214.0	\pm	12.3	$ \\
\hline
TOTAL	&$	360.4	\pm	11.9	$&$	5525.2	\pm	67.9	$\\

               \hline
         \end{tabular}
         \begin{tablenotes}
         \item \textbf{Notes.}  The luminosities are defined as per the text.  $^{a}$FIR luminosity, corrected for an AGN contribution  to the total galaxy luminosity (see text). $^{b}$The value may be overestimated for this galaxy because its FIR flux was not spatially resolved from its neighbouring galaxy NGC~5394.
            
         \end{tablenotes}
      \end{threeparttable}
    \end{table}

\subsubsection{Global sample properties}
\label{sec:ulxperlum}

A convenient and simple method to compare between our sample of galaxies and others is to work out the global properties, in terms of number of ULXs ($N_{\rm ULX}$) per unit luminosity in some relevant reference band.  Here, we compute this quantity for the blue and FIR luminosities introduced above.  In this calculation, we also separate our ULX sample into two sub-samples, those that were detected in the AGN host galaxies and those that were detected in the galaxies without AGN. This is predicated upon the assumption that AGN feedback can affect the SFR of galaxies (see \citealt{fabian2012} and references therein).

The calculated values of the total number of ULXs per unit luminosity are shown in Table~\ref{tab:number per luminosity}.\footnote{As Arp 299 contributes a significant fraction of the total $L_{\rm FIR}$ ($\sim 26$ per cent), we have re-calculated these values after excluding this galaxy.  We find that the values in Table~\ref{tab:number per luminosity} do not change significantly, so Arp 299 does not significantly bias our results.}  Firstly, we find no statistical distinction between the number of ULXs per unit luminosity in galaxies with and without AGN -- so the incidence of ULXs appears independent of the presence of an AGN.  We thus ignored this distinction, and proceeded to compare the global properties of our LIRG sample to other samples of galaxies observed by {\it Chandra\/}.  These include the complete sample of `normal' galaxies within 14.5 Mpc from \citet{swartz2011}; a sample of spiral galaxies from \citet{swartz2004}; and samples of both interacting and strongly interacting systems from \citet{smith2012}.  We find that the number of ULXs per unit luminosity does depend on the host galaxy type.  The ratio $N_{\rm ULX}/L_{\rm B}$ appears heightened in the LIRG sample; it is 3 times higher (at the $\sim 2.7\sigma$ significance level) in the full LIRG sample than the sample of normal galaxies, with smaller, marginal significance excesses also seen above the spiral ($\sim 1.5\sigma$) and interacting galaxy ($\sim 1.2\sigma$) samples.  The differences in the $N_{\rm ULX}/L_{\rm FIR}$ ratio between the LIRGs and other samples are more dramatic, with the LIRGs showing significantly suppressed ULX numbers per unit luminosity, by factors $\sim 5 - 8$ (all significant at a $> 5.6\sigma$ level).  Given that both luminosities are, to first order, related to recent star formation, the fact that the ratios go in different directions compared to the other galaxy samples (i.e. an excess of ULXs compared to blue luminosity; a large deficit compared to FIR) again appears puzzling. We will discuss this further in Section~\ref{sec:discussion}.

\begin{table}
      \centering
      \caption{Number of ULXs per unit galaxy luminosity}\label{tab:number per luminosity}
      \smallskip
      \begin{threeparttable}
          \begin{tabular}{lcc}
             \hline
             
~~~~~Sample  & $N_{\rm ULX} / L_{\rm B}$\tnote{a}  &  $N_{\rm ULX} / L_{\rm FIR}$\tnote{b}  \\
\hline
All LIRGs                 & $1471^{+447}_{-364}$ & $ 959	^{+290}_{-236}$ \\
LIRGs (AGN)               &$1478^{+420}_{-284}$& $861^{+242}_{-162}$ \\
LIRGs (no AGN)              &$1459\pm493$& $1161\pm389$ \\
Normal galaxies \tnote{1}  &$480 \pm 50 $   &$5300 \pm 500$ \\
Spiral galaxies\tnote{2}&        $770 \pm 280$    &       $6467^{+726}_{-655}$    \\
Interacting galaxies\tnote{3}  &$990^{+194}_{-169}$ &$7649^{+1315}_{-1147}$ \\

Strongly interacting & \multirow{2}{*}{$1978^{+660}_{-520}$} & \multirow{2}{*}{-} \\
galaxies\tnote{4} & & \\

              \hline
         \end{tabular}
         \begin{tablenotes}
         \item \textbf{Notes.} \tnote{a}Number of ULXs per unit galaxy blue luminosity (units of 10$^{-46}$~(erg~s$^{-1}$)$^{-1}$).  \tnote{b}Number of ULXs per unit galaxy FIR luminosity (10$^{-47}$~(erg~s$^{-1}$)$^{-1}$). These numbers are reported with a 1$\sigma$ error.  The comparator samples are taken from: $^{1}$\citet{swartz2011}; $^{2}$\citet{swartz2004}; $^{3,4}$\citet{smith2012}.  All reference values were calculated and reported by \citet{smith2012}.

         \end{tablenotes}
      \end{threeparttable}
    \end{table}


\subsubsection{Correlation between ULXs and their host galaxy luminosities}

\begin{figure*}
\begin{center}

\includegraphics[width=6cm]{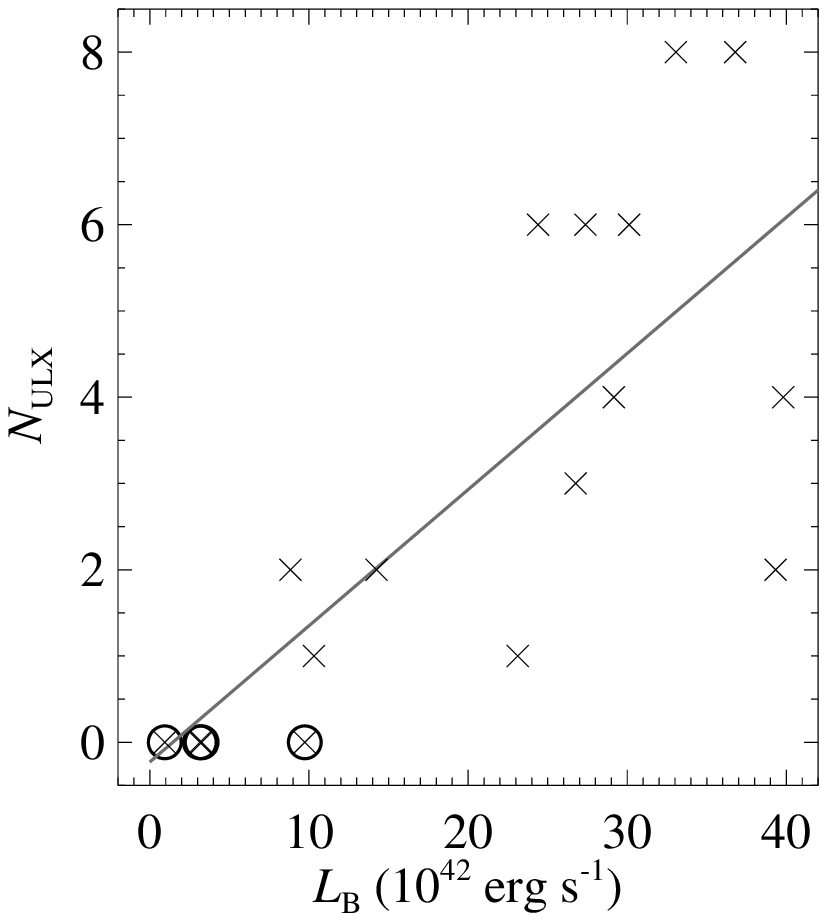}\includegraphics[width=6cm]{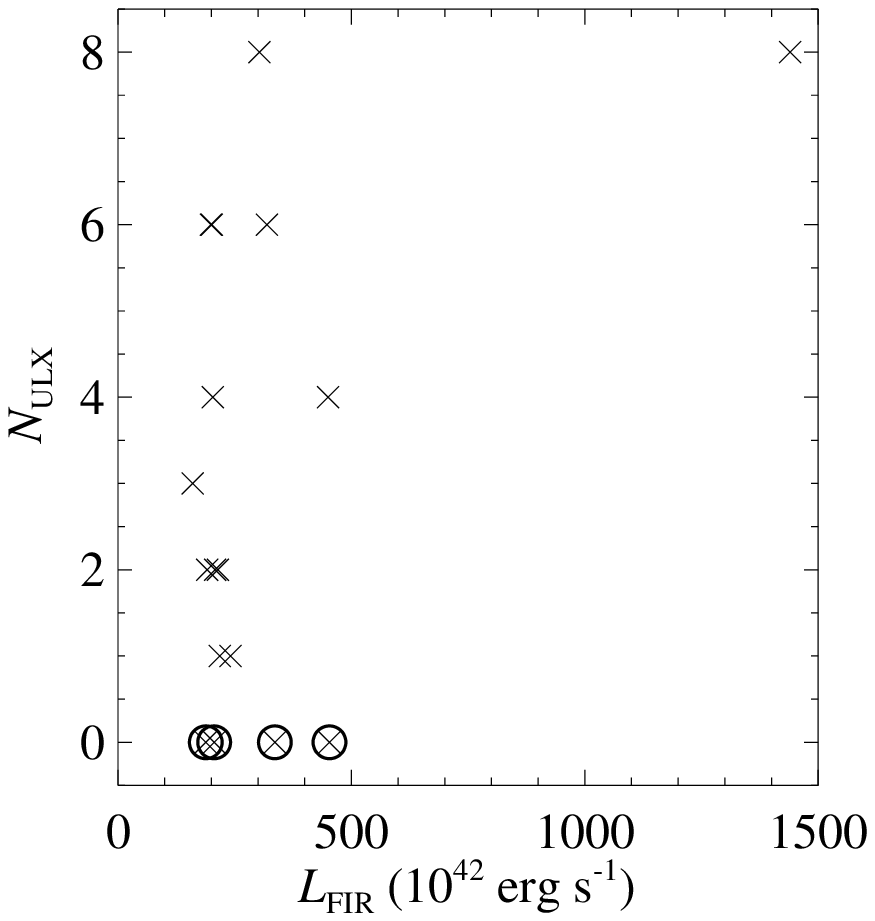}
\includegraphics[width=6cm]{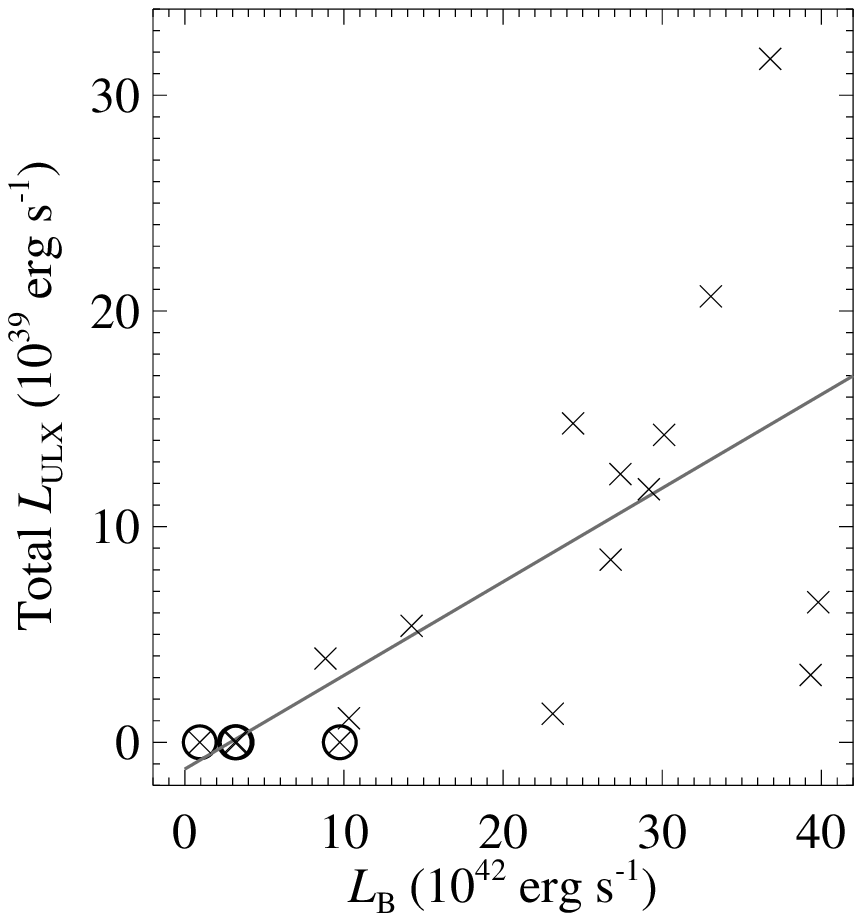}\includegraphics[width=6cm]{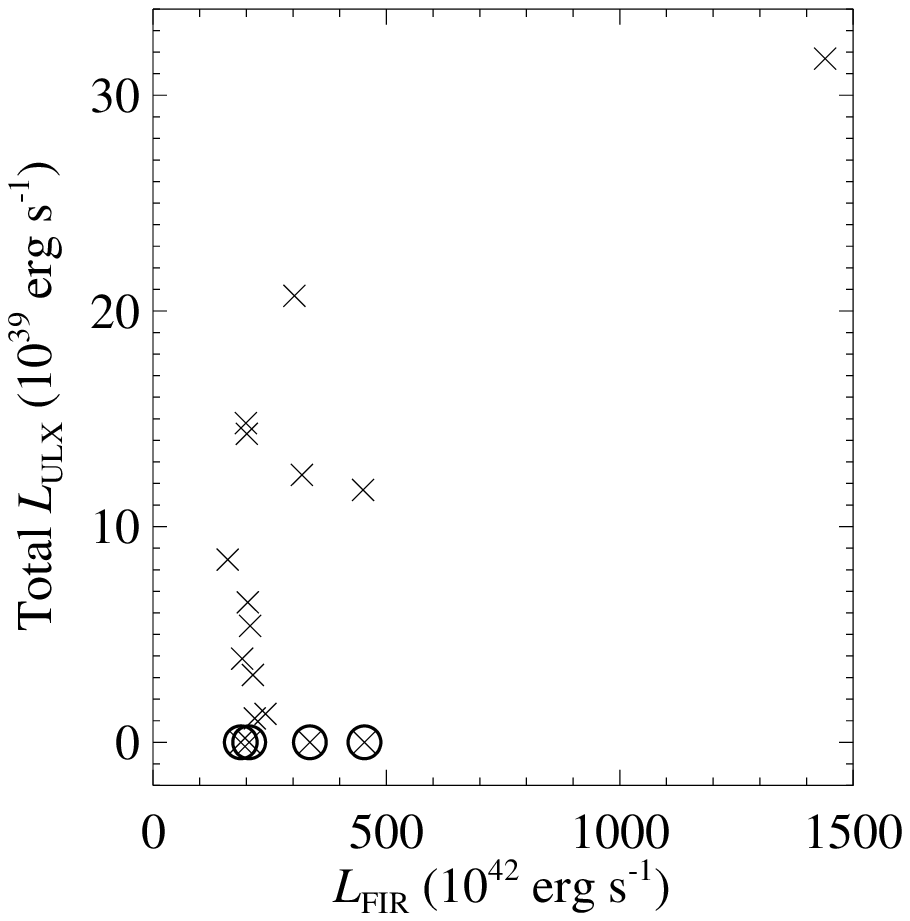}

\caption[]{Scatterplots showing the ULX population properties against the host galaxy luminosities for the LIRG sample.  The panels show: {\it top left} -- number of detected ULXs versus host galaxy blue luminosity; {\it top right} -- number of detected ULXs versus host galaxy FIR luminosity; {\it bottom left} -- summed ULX luminosity versus host galaxy blue luminosity; {\it bottom right} -- summed ULX luminosity versus host galaxy FIR luminosity.  The linear best fit is also shown as a solid straight line in the plots in which we found a correlation (see text). The circles highlight the four galaxies in which no ULXs were detected: NGC~4418, ESO~420-G013, CGCG~049-057 and IC~860.}
\label{fig:ULX-lum_correlation}
\end{center}
\end{figure*}

We also investigated whether there was any relationship between $N_{\rm ULX}$ or the cumulative luminosity of the ULXs in each individual galaxy, $L_{\rm ULX}$, and the luminosity of that galaxy in both blue and FIR light.  We show scatterplots for all four possible relationships in Fig.~\ref{fig:ULX-lum_correlation}, with the best fitting linear relationship shown.  Clearly, a possible relationship exists between $L_{\rm B}$ and both $N_{\rm ULX}$ and $L_{\rm ULX}$, albeit with large scatter; however the ULX population appears not to scale with $L_{\rm FIR}$.  We tested this statistically using the Pearson correlation coefficient ($r$), and find evidence of a positive correlation for both $N_{\rm ULX}$ and $L_{\rm ULX}$ with $L_{\rm B}$ ($r = 0.74$ and 0.65, respectively -- corresponding to significance level of $\ga$ 99.5$\%$); however no evidence for a linear relationship is forthcoming for the possible relationships with $L_{\rm FIR}$ ($r \sim 0$ in both cases).\footnote{We confirmed that the result was not biased by the selected range of IR luminosity we used by examining the relationship between the $N_{\rm ULX}$, $L_{\rm ULX}$ and IR luminosity in the full band (8 -- 1000 $\mu$m). The same result was found: no correlation between $N_{\rm ULX}$, $L_{\rm ULX}$ and the IR luminosity.}

\subsection{X-ray luminosity functions}
\label{sec:xlf}

\begin{figure*}
\begin{center}
\includegraphics[width=8.5cm]{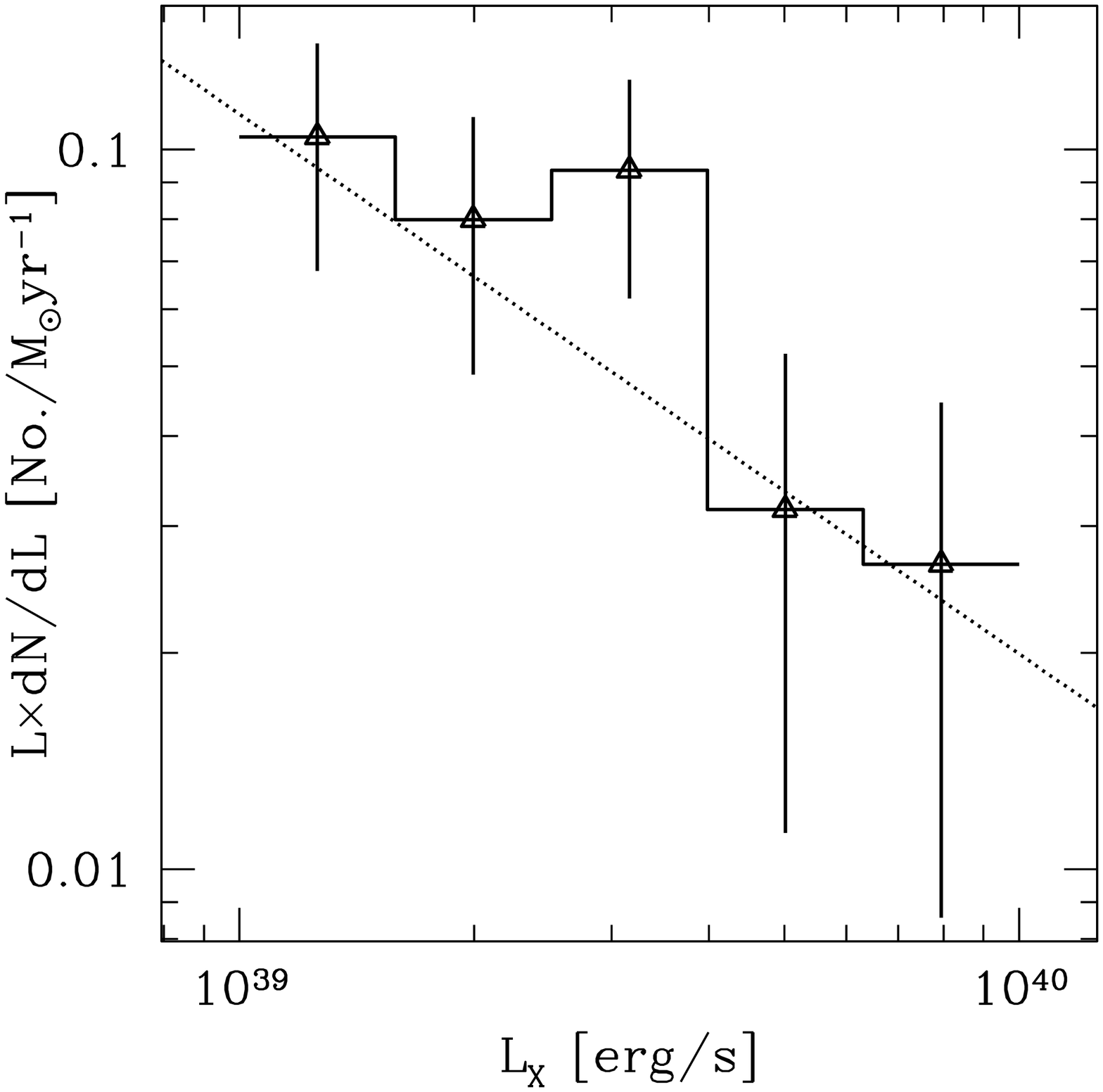}
\includegraphics[width=8.5cm]{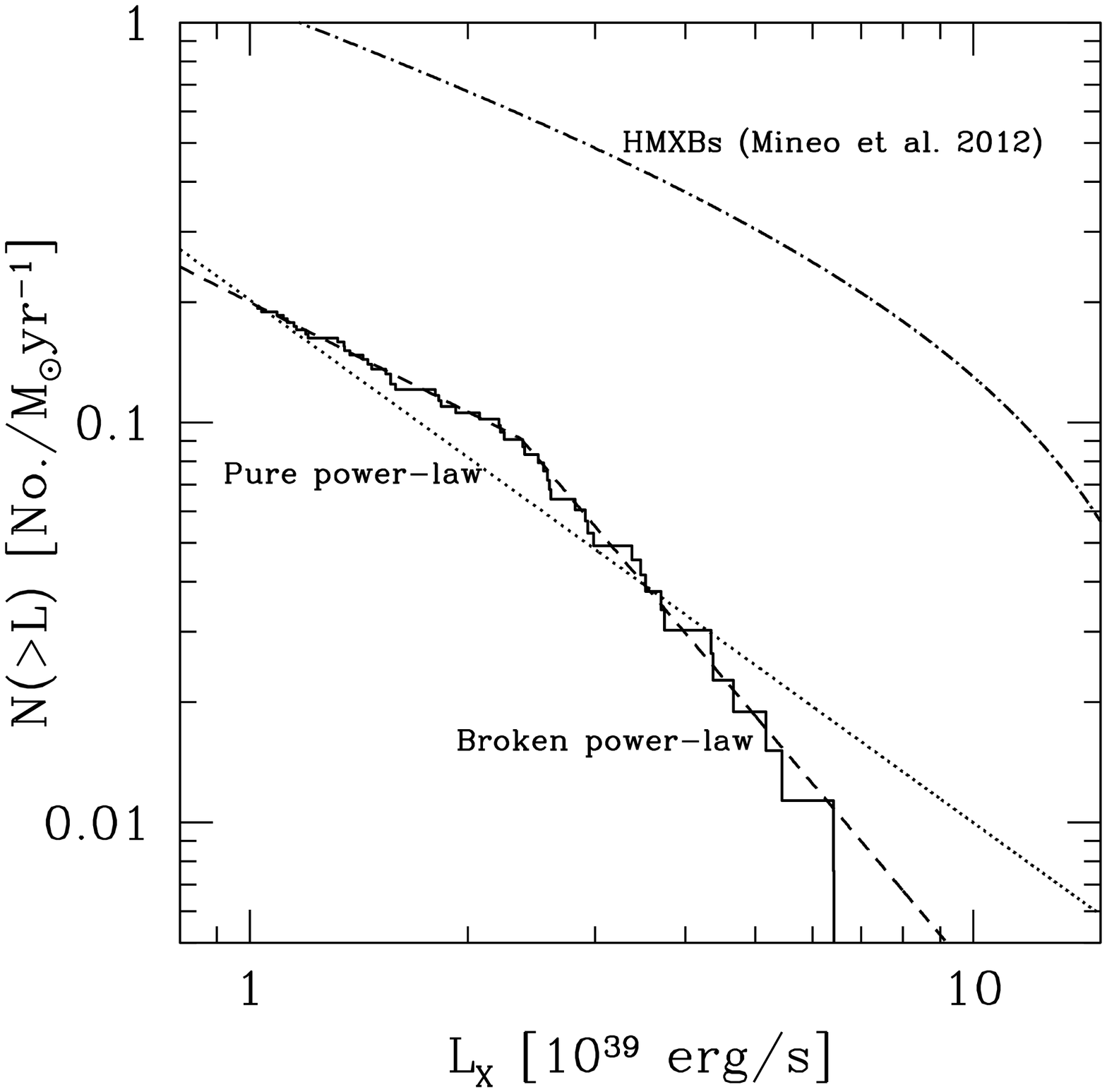}
\caption[]{SFR normalised XLFs for the 53 ULXs in our sample of LIRGs.  {\it Left panel\/}: differential XLF.  
We plot $1\sigma$ errors on each data point. The best-fitting pure power-law model obtained from the ML by is 
overlaid in red. {\it Right panel\/}: cumulative XLF. The best fitting pure power-law model (similar for ML and C-
stat) is plotted in red, while the broken power-law model is in green. For comparison, \citeauthor{mineo2012a} (2012a) HMXB XLF is plotted in blue. All luminosities are in the 0.3--10\,keV 
band. See Section~\ref{sec:xlf} for details.}
\label{fig:XLF}
\end{center}
\end{figure*}

\begin{table}
\centering
\caption{X-ray luminosity function: fitting results from maximum likelihood method}\label{tab:XLFfit2}
\smallskip
\begin{threeparttable}
\begin{tabular}{ccc}
\hline
XLF form & \multicolumn{2}{c}{Pure power-law model}  \\
\cline {2-3}
 &&\\
 & $\gamma$ &  \textit{A}  \\
\hline
&&\\
Differential & $1.75\pm0.25 $ & $0.11\pm0.02$ \\
&&\\
Cumulative & $1.31^{+0.19}_{-0.17} $ & $0.20$ \\
&&\\
\hline
\end{tabular}
\begin{tablenotes}
\item \textbf{Notes.} The best fitting results from the maximum likelihood method. The definitions of the models and their parameters are given in Equations~\ref{eq:powerlaw_diff} and~\ref{eq:brokpowerlaw_diff}, and the text.  All best fitting parameter values are reported with 1$\sigma$ errors.

\end{tablenotes}
\end{threeparttable}
\end{table}

\begin{table*}
\centering
\caption{X-ray luminosity function: fitting results from Sherpa}\label{tab:XLFfit1}
\smallskip
\begin{threeparttable}
\begin{tabular}{cccccccccc}
\hline
XLF form & \multicolumn{3}{c}{Pure power-law model} & & \multicolumn{5}{c}{Broken power-law model} \\   
\cline {2-4}  \cline {6-10}
&&&&&&&&&\\
   & $\gamma$ &  \textit{A} & Stat./\textit{d.o.f.}$^{a}$ & & $\gamma_{1}$ & $\gamma_{2}$ &  \textit{A} & $L_{\rm b}$ 
   &Stat./\textit{d.o.f.}$^{a}$ \\
\hline
&&&&&&&&&\\

Differential & $1.84^{+0.31}_{-0.27} $ & $0.15^{+0.05}_{-0.04}$ & 1.64/3 & & - & - & - & - & - \\
&&&&&&&&&\\
Cumulative & $1.29\pm0.06 $ & 0.20 & 25.05/50 & & $0.91^{+0.13}_{-0.14}$ & $2.14^{+0.29}_{-0.24}$ & 0.28  
& $2.37^{+0.36}_{-0.23} $ & 3.38/48 \\
&&&&&&&&&\\
\hline
\end{tabular}
\begin{tablenotes}
\item \textbf{Notes.} The best fitting results from SHERPA. The definitions of the models and their parameters are given in Equations~\ref{eq:powerlaw_diff} and~\ref{eq:brokpowerlaw_diff}, and the text. The broken power-law model was fitted to the cumulative XLF only. $^{a}$Value of minimised statistic/degrees of freedom; we use the $\chi^2$ statistic ({\it chi2gehrels}) to fit the differential XLF, and the Cash statistic ({\it cstat}: \citealt{cash1979}) to fit the cumulative XLF. All best fitting parameter values are reported with 1$\sigma$ errors.

\end{tablenotes}
\end{threeparttable}
\end{table*}

A further means of examining the ULX population of the LIRGs is by constructing their X-ray luminosity function (XLF).  We constructed the XLF in both cumulative and differential form for the full sample of 53 ULXs, combining the data for all galaxies.

To obtain the XLF in differential form, we binned the sources into logarithmically spaced luminosity bins, 
taking into account the correction for incompleteness. Since we use only sources with luminosity above 
$10^{39}\,\rm{erg}\,\rm{s}^{-1}$, the latter correction is small (see Table~\ref{tab:detectedULX}). We subtracted the 
expected contribution of background AGNs from each bin, using the $\log N-\log S$ from \citet{georgakakis2008}, based on 
        the sum of the areas within the $R_{20}$ of all the LIRGs.\footnote{Due to the fact that different galaxies have different point-source detection sensitivity, only galaxies which $K(\rm L)$ $\geq$ decided completeness limit should contribute to the given luminosity bin (see \citeauthor{mineo2012a} 2012a). However since the LIRG sample has $L_{\rm comp}$ $\la$ 10$^{39}$ erg s$^{-1}$, in this case all galaxy sample contribute to every given bin.} The number of 
background AGNs in each luminosity bin was computed after having converted the luminosity at the distance 
of each galaxy into flux. The resulting incompleteness-corrected, AGN-subtracted number of sources was 
normalised by the sum of the SFR of all the LIRGs.$^{18}$ The differential XLF is shown in the left panel of Fig.~\ref{fig:XLF}. The error bars correspond to 1$\sigma$ uncertainty using Poisson statistics and 
accounting for the uncertainty in the number of both ULXs and background AGNs. The combined XLF in 
cumulative form was constructed by integrating the un-binned differential XLF and is shown in the right-hand 
panel of Fig.~\ref{fig:XLF}. Interestingly, this highlights that there are no ULXs with luminosities above $10^{40} \rm 
~erg~s^{-1}$ in our sample (see Section~\ref{sec:discussion} for further discussion).

Following \citeauthor{mineo2012a} (2012a), the fit of the XLF was performed on un-binned data using a maximum-likelihood 
(ML) method. The predicted contribution of resolved CXB sources was included in the model as 
described above. We began by fitting the XLF with a pure power-law model, defined as follows:
\begin{equation}
\label{eq:powerlaw_diff}
\frac{\rmn{d}N}{\rmn{d}L_{39}} =A\times \rm{SFR} \times L_{39}^{-\gamma}, ~~~~~L_{39}\leq L_{\rmn{cut}}
\end{equation} 
where $L_{39}=L_{\rm X}/10^{39}$ erg s$^{-1}$, $L_{\rmn{cut}}= 10^{42}\,\rm{erg}\,\rm{s}^{-1}$, $A$ is the 
normalisation. Using this model, we obtained the best-fitting values for the XLF parameters: $
\gamma=1.75\pm0.25$ and $A=0.11\pm0.02$ (see Table~\ref{tab:XLFfit2}). The slope is in agreement with the typical value for star forming 
galaxies (\citealt{grimm2003, swartz2011}; \citeauthor{mineo2012a} 2012a). However, its normalisation at $10^{39} 
\rm ~erg~s^{-1}$ is substantially lower than for these other XLFs, for example it is a factor $\sim 5$ below the 
normalisation found by \citeauthor{mineo2012a} (2012a).\footnote{Our SFR estimator assumes a \citet{kroupa2001} initial mass function (IMF) whereas a Salpeter IMF was used by \citeauthor{mineo2012a} (2012a). Using this form of the IMF leads to a corrected value of the normalisation of the \citeauthor{mineo2012a} (2012a) XLF of 0.59.} This again points to a relative deficit of ULXs in the LIRG sample. 

Next, We fitted the cumulative XLF with a pure power-law model (Equation~\ref{eq:powerlaw_diff}) using the ML method described above (see Table~\ref{tab:XLFfit2} for the fitting result). However, the shape of the cumulative XLF (right panel of Fig.~\ref{fig:XLF}), induced us to search for a possible break by repeating the ML fit, using a broken 
power-law model:
 \begin{equation}
\label{eq:brokpowerlaw_diff}
\frac{\rmn{d}N}{\rmn{d}L_{39}} = A\times \rmn{SFR}\times 
\begin{cases}
L_{39}^{-\gamma_{1}}, & \quad L_{39}<L_{b}\\
&\\
L_{b}^{\gamma_{2}-\gamma_{1}}\,L_{39}^{-\gamma_{2}}, & \quad L_{b} \leq L_{39}\leq L_{\rmn{cut}}\\ 
\end{cases}
\end{equation}
where $L_{b}$ is the break luminosity. We did not find convergence for a break luminosity, although we 
explored a very broad range of parameters searching for the ML.

We repeated the fitting procedure described above using version 4.4 of the {\sc sherpa}\footnote{\texttt{http://
cxc.harvard.edu/sherpa4.4/}} package. We first analyzed the cumulative XLF. Given the un-binned 
nature of the cumulative data we used the {\it cstat\/} function \citep{cash1979} in {\sc sherpa} to optimise the 
goodness of fit for the power-law model. We accounted for Poisson uncertainties on the cumulative number of 
sources. In this case, the broken power-law model provided an estimate for 
the break luminosity, $L_{\rm b}=2.37 \times 10^{39}$ erg s$^{-1}$, with a substantial improvement to the fit 
compared to the pure power-law model ($\Delta\rm{C}$-stat of 22 for 2 additional degrees of freedom). It also 
appears to better describe the data by eye. 
However, the pure power-law model provided a statistically good fit to the 
data too (see Table~\ref{tab:XLFfit1}), therefore we cannot reject one model in favour of the other. 

The statistical significance of the break is low ($\sim$$0.5-0.8\sigma$), which is the likely explanation for the ML 
procedure for a broken power-law model not converging.  Moreover, none of the more extensive population studies of the last decade (\citealt{grimm2003, swartz2011}; \citeauthor{mineo2012a} 2012a) reported on features at a similar luminosity. Even if the break is present, another possibility is that a feature resembling a break may be artificially caused by the considerable ($\sigma/\langle N(>L)\rangle \sim1.7$) dispersion in the normalisation of the individual XLFs, which is comparable with that observed by \citeauthor{mineo2012a} (2012a) at the same luminosity threshold ($\sigma/\langle N(>L)\rangle \sim1.5$, see discussion in Section 7.1 of \citeauthor{mineo2012a} 2012a).  A second means of artificially inducing a break could be by over-estimating the completeness of the sample, i.e. incompleteness could act to turn down the XLF slope at lower luminosities if we have not accounted for it properly (see the discussion in Section 3 of \citealt{fabbiano2006}).  However, this is eminently testable in our data as several of our galaxies are complete well below the putative break (cf. Table 3); hence the number of sources above and below the break in these galaxies can be compared to the numbers in the galaxies with sensitivity limits closer to the break luminosity.  We find that 6/11 ULXs are below the putative break in the five nearest host galaxies (NGC 1068, NGC 1365, NGC 7552, NGC 4418 \& NGC 4194); this compares to 21/42 in the more distant LIRGs.  Given the small numbers of objects involved these numbers appear consistent, hence there is no reason to suspect that incompleteness is inducing the putative break feature.  If the break is a real feature intrinsic to the ULX population, it immediately becomes very interesting as the steepening of the XLF slope above $L_{b}$ may suggest that much of the deficit in ULXs is originating above this putative break. We will further discuss the possible physical explanation of the very putative break in Section~\ref{sec:ulx population}.

As a check, we also used {\sc sherpa} to fit the differential XLF. In this case we had to assume $
\chi^{2}$ statistics ({\it chi2gehrels}) and fit the binned differential XLF. This is, strictly speaking, not appropriate 
for the number of sources per luminosity bin. The reduced fitting statistic is substantially below unity, primarily 
as a consequence of the small number of bins.  However, we note that the best-fitting parameters for the two 
forms of the XLF obtained with the procedure described above for a pure power-law model (Table~
\ref{tab:XLFfit1}), are in full agreement with those provided by the ML fitting (Table~\ref{tab:XLFfit2}).

\section{Discussion}
\label{sec:discussion}

\subsection{X-ray spectral results}
\label{accretion state in black holes}

In Section~\ref{sec:stack} we presented an analysis of the stacked spectra for three groups of ULXs, segregated by their observed luminosity.  The analysis of stacked spectra is notoriously difficult to interpret, as summing spectra that may have a variety of underlying levels of absorption and intrinsic forms may not necessarily reveal anything about the population of sources contributing to the stack.  However, in the case of the stacked spectra we create, we see one remarkable result; the stacked spectrum of the sources below $\sim 2 \times 10^{39} \rm ~erg~s^{-1}$ appears markedly different to the stacked spectra of the two more luminous groups.  Both the power-law continuum fit and MCD model fit show that the lowest luminosity spectrum is significantly softer than that in the two more luminous groups, which are statistically indistinguishable from each other. 

Further interpretation is difficult due to the uncertainty in the absorption column, which may vary substantially for the individual ULXs given that LIRGs are known to contain substantial quantities of cold, neutral  gas \citep{mirabel1988}.  This would work to distort any underlying spectrum -- for example, if the average spectrum was heavily absorbed, just a handful of non-representative, low absorption sources would `fill-in' the soft end of the spectrum and lead to a relatively low column being measured.  Indeed, this may be part of the reason for the columns on the MCD fits being consistent with zero, although they are substantially higher than the enforced Galactic foreground upper limit for the sample of $< 5 \times 10^{20} \rm ~cm^{-2}$ in the power-law fit.  We therefore do not directly interpret the power-law slopes or disc temperatures in light of Galactic binaries and/or other ULXs.  We do note, however, that previous analyses have suggested transitions in ULX spectra at similar luminosities.  Both \cite{gladstoneroberts2009} and \cite{yoshida2010} examine ULXs in the interacting galaxies NGC 4485/90 (and also M51 in the latter paper), and note a change from power-law-like spectra dominating, to more disc-like spectra at $\sim 2 - 3 \times 10^{39} \rm ~erg~s^{-1}$, that they also reconcile with the objects becoming substantially super-Eddington beyond that luminosity.  Furthermore, the detailed ULX spectroscopy of \cite{gladstone2009} and follow-up work by \citeauthor{sutton2013b} (2013b) show that below $\sim 3 \times 10^{39} \rm ~erg~s^{-1}$ most ULXs have disc-like spectra in the wider {\it XMM-Newton} bandpass; above this luminosity their spectra resolve into two components, with roughly equal numbers dominated by either the hard or soft component.  This is again interpreted as a transition from a $\sim$ Eddington rate phase, to a super-Eddington state.  Based on their work we can speculate why our stacked spectra may harden at higher luminosities -- if there is substantial absorption present in the LIRGs, it will act to extinguish the signal from the soft ULXs more efficiently than the harder objects.  A salient example of this effect is the extreme ULX in NGC 5907, where a high foreground column may hide a large soft contribution to its spectrum (\citeauthor{sutton2013a} 2013a).  This would therefore preferentially hide the softer ULXs at high luminosities, leaving only the hard spectrum objects to contribute to the stacked spectrum.

One object in our sample had sufficient counts in its {\it Chandra\/} X-ray spectrum for us to analyse it separately.  We found the spectrum of CXOU~J024238.9-000055 to be very hard, with a power-law of photon index $\Gamma = 0.75 \pm 0.1$ providing an acceptable fit to the data.  This result is consistent with previous studies, which also found that the photon index is very hard \citep{smith2003,swartz2004,berghea2008}.  \citet{smith2003} interpret the hard spectrum of this ULX as inverse Compton scattering of synchrotron emission in a jet.  However, they also noted the presence of a broad feature that dips below the continuum level at $\sim 1.7$ keV, and suggest that this may be a spurious feature resulting from an artefact in the gain table of the \textit{Chandra\/} ACIS-S instrument.  We have re-analysed the data with the latest {\it Chandra\/} calibration database that includes corrections for such features\footnote{In a correspondence with the {\it Chandra\/} X-ray centre help desk we were informed that this issue was corrected for in calibration database (CALDB) versions 3.0.0 and later.}, but still find the feature to be present.  We therefore characterised it using a single Gaussian absorption line; this produced a good fit to the data, but produced a broad ($\sim 80$ eV width) line centred at 1.68~keV, that we struggle to attribute to any single physical absorption feature.

We therefore considered an alternative solution -- that the feature is the result of absorption in a range of partially covering, partly-ionised material ( a `warm' absorber model).  Interestingly, a similar model was suggested as a way of producing the apparent two-component {\it XMM-Newton\/} spectra of ULXs by \cite{goncalves2006}, and was also used to describe the spectrum of a ULX in NGC 1365 by \cite{soria2007}.  In the latter paper they propose that the ionising material may be in an outflow, with its origins in the formation and subsequent ejection of a Comptonising region above the inner accretion disc.  Interestingly, the best fitting parameters from \cite{soria2007} -- $N_{\rm H}, \Gamma, \xi$ -- broadly agree with our best fitting parameters.  Our value of ionisation parameter $\xi \approx 158 \pm 87$ for CXOU~J024238.9-000055 corresponds to the ionised material being located at a distance $R < L/(\xi N_{\rm H})$, which for the ionising luminosity (i.e. intrinsic $L_{\rm X}$ extrapolated from the model fit between 13.6 eV and 20 keV) gives a maximum radius for the material of $3.5 \times 10^{15}$ cm ($= 1.2 \times 10^9 R_{\rm g}$ where the gravitational radius $R_{\rm g} = 2GM_{\rm BH}/c^2 = 30$ km for a $10 M_{\odot}$ black hole).  Clearly this is located in the proximity of the ULX, but not close to the inner regions of the disc, consistent with material that has been ejected from the system in some sort of outflow.  We note that a massive, likely highly ionised outflow is a prediction of super-Eddington models (e.g. \citealt{poutanen2007}); evidence has recently been sought for it by looking for partially ionised Fe K lines in spectrally hard ULXs (e.g. \citealt{walton2012,walton2013}), with little success.  However, signatures of partial ionisation may be preferentially located in the softer part of the X-ray spectrum \citep{middleton2014}, that is easier to see in objects that are wind-dominated -- here we may be seeing precisely that.

\subsection{Why is there a deficit of ULXs in LIRGs?}
\label{sec:ulx population}

In this paper we have presented evidence for a significant deficit in the number of ULXs detected in this sample of LIRGs, compared to the expectation based on the relationship between the number of ULXs and the SFR in nearby galaxies.  This manifests itself both in the raw numbers -- we expect to detect $\sim 500$ ULXs in this sample if it follows the number of ULXs per unit SFR relation of \cite{swartz2011}, compared to just 53 detections -- and in the characteristics of the XLFs we construct from our detections.  In the latter, the normalisation of the differential XLF is a factor $\sim 5$ lower at $10^{39} \rm ~erg~s^{-1}$ than the most complete previous analysis of the HMXB XLF (\citeauthor{mineo2012a} 2012a), and although the differential slope is consistent within errors with previous work the cumulative form has a slope that is steeper than that found for most star forming systems (c.f. \citealt{fabbiano2006} and references therein), implying fewer sources are present at high luminosities.  This steep slope may be the result of a break in the XLF at $\sim 2 \times 10^{39} \rm ~erg~s^{-1}$; at energies below the break we see a slope similar to that in star forming galaxies ($\gamma_1 \approx 0.9$).  If this shallow slope were extrapolated to luminosities above $10^{40} \rm ~erg~s^{-1}$ we would expect to see $\sim 7$ ULXs in that regime; instead we see none. In this section, the possible explanations for the deficit will be discussed. 

\subsubsection{A real deficit or observational effects?}

A pertinent question we must therefore ask about the presence of this deficit is simply: why?  A first consideration is whether we really detect all the ULXs that are bright in the 0.3--10\,keV regime in the host galaxies. In this regard we are certainly using the best instrument available for this study; the exquisite sub-arcsecond spatial resolution of {\it Chandra\/} is unparalleled in this respect, allowing us by far the clearest X-ray view of these regions, and minimising any source confusion.  We cannot completely avoid this - for example, the nucleus of M82 hosts at least two extreme ULXs (e.g. \citealt{feng2007}), that we would not resolve if placed at the distance of our furthest LIRGs.  However, we will miss very few ULXs ($\la$10\%) due to sensitivity issues in these galaxies (see Section~\ref{sec:completeness}), with the sample essentially complete in terms of sensitivity to ULX detection.  

But while we may have sufficient sensitivity to detect ULXs over the projected area of the galaxies, this could be compromised locally by extended diffuse X-ray emission.  In particular, the extended source detections that we reject from our point source detection analysis might physically comprise a mixture of extended and faint point-like X-ray emitters, and so could hide ULXs that otherwise would have been detected.  To determine whether this has a large influence on our sample we consider that the diffuse emission is likely to be spectrally soft in SF-dominated galaxies, and so dominant only below 2~keV (cf. \citealt{pietsch2001}; \citealt{franceschini2003}; \citealt{jenkins2004}; \citeauthor{mineo2012b} 2012b); hence the hidden ULXs would be visible as point-like sources at higher energies (above 2~keV).  However, we face the problem that we do not have high photon statistics in this regime, and our sensitivity is relatively poor.  In fact, the best way to proceed is to re-visit our initial source lists for each galaxy where we already have a complete record of the detectable objects above 2~keV as our hard band detections.  We therefore determined which of the hard band detections did not make it into our final source catalogue, and investigated why.

In total, there are 17 objects that have full band luminosities in excess of the ULX threshold of $10^{39} \rm ~erg~s^{-1}$ that are detected as hard band sources but do not make the final source list.  For each of these objects we revisited their hard band data and ran the \textsc{srcextent} algorithm to determine whether they appeared point-like or extended, although we caution that in several cases the photon statistics were limited and so the results may not be conclusive.  We found that a total of 4 sources were spatially extended above 2~keV; the remaining 13 were consistent with point-like objects.  However, 9 of these point-like objects were located with 5 arcseconds of the nuclear position of their host galaxy and so are rejected from consideration.  Of the four remaining objects, two that lie in Arp 299 are questionable ULX detections as they are coincident with plausible contaminants in the host galaxy (a supernova remnant, and the secondary nucleus in this merging system). Hence only two plausible ULX candidates remain, one each in NGC 1068 and Arp 299.  Both objects were originally rejected as extended full band sources, but are resolved to be point-like sources in the hard band.  Given that we can only do this for two objects, this indicates that we are not missing a significant population of ULXs by rejecting extended sources.

Although the nuclei of the galaxies cover a relatively small region of their surface area, we may miss a proportion of the ULX population by excluding this region -- see \cite{lehmer2013} for an example that would be excluded from the nearby starburst galaxy NGC~253.  This is particularly pertinent as the star formation in LIRGs can be very centrally peaked, so by excluding the nucleus we may be rejecting a region containing a large fraction of the galaxy's star formation, and so a proportional number of ULXs.  There will, however, be observational difficulties with detecting such a population -- as Fig.~\ref{fig:detectedULX} shows, the central regions tend to have far higher surface brightness in X-rays and possible spatial complexity compared to the outer regions.  But, given the high spatial resolution of the \textit{Chandra} telescope mirrors (0.5 arcsecond on-axis resolution), we should have the imaging power to resolve the nuclear region of each galaxy, and detect the brightest point-like sources if they are abundant in that region.  Yet, we only find 3 examples of galaxies with more than a single source detection in the nuclear region, and in one of these objects (in NGC 1365) the source furthest from the nuclear position is relatively faint ($\sim 10^{38} \rm ~erg~s^{-1}$).  Conversely, 15 out of 17 galaxies have at least one source detection (in two cases -- NGC 4418 and NGC 5653 -- there are two detections) in the nuclear region with a luminosity in the ULX regime, although seven of these are spatially extended sources in the full band (of which 5 are resolved to point sources above 2 keV).  Eight of these are galaxies with a suspected/confirmed AGN\footnote{Interestingly, the two galaxies without a point source detection in their nucleus -- ESO 420-G013 and IC 860 -- have some evidence for the presence of an AGN; looking at Fig.~\ref{fig:detectedULX} they do have X-ray emission in their nucleus that presumably was too extended for the {\sc wavdetect} algorithm.}; and although we detect a source in each galaxy that we regard as lacking an AGN based on previous evidence, we cannot reject an AGN nature for any of these objects without further observations and detailed analyses that are beyond the scope of this paper.  Thus, the situation is unclear on how many of these detections could be {\it bona fide\/} ULXs in the nuclear region.  However, the total number of detected ULXs in the nuclear regions appears significantly smaller ($< 17$, likely half that amount given the evidence for AGN) than the numbers of ULXs detected outside the nuclei.  So the deficit of ULXs seems to appear significantly worse in the central regions of the galaxies, albeit with some caveats about the difficulty of observing in that region.

To investigate this further we examined the total flux within each 5 arcsecond radius nuclear region, excluding the flux from the detected point sources, in order to place an upper limit on the possible residual flux from ULXs hidden within the diffuse emission.  This analysis was only performed for the seven galaxies in which we have no multi-wavelength evidence for the presence of an AGN, to avoid any complex spatially extended emission that could be related to such an object.  For three galaxies -- NGC~7552,  NGC~838 and NGC~23 -- there are a sufficient number of counts ($\ga$ 300) to extract a spectrum of the emission within the 5 arcsecond nuclear region.  The spectra were fitted with a two component model comprising an optically thin thermal plasma (\textsc{mekal} in \textsc{xspec}) plus a power-law continuum, both subject to the same absorption column. The first component can be used to represent the diffuse emission from the active star forming region, whilst the latter represents the integrated emission from X-ray binary systems in the region.  Both NGC~7552 and NGC~23 displayed similar residual nuclear spectra, with moderate absorption ($5-20 \times 10^{20} \rm ~cm^{-2}$) and similar plasma temperatures and power-law photon indexes ($kT \sim 0.6-0.7$ keV; $\Gamma \sim 2$), whereas NGC~838 appeared somewhat more absorbed, with a cooler diffuse component ($N_{\rm H} \sim 7 \times 10^{21} \rm ~cm^{-2}$; $kT \sim 0.15$ keV; $\Gamma \sim 2.5$).  We estimated the observed flux from X-ray binaries in each galaxy nucleus from the power-law component in each object.  In order to do the same for the galaxies with too few counts for spectral analysis, we extracted the total residual count rates in the nuclear region of each galaxy, and then applied a factor 0.5 multiplier to each total to account for only $\sim 50$ per cent of counts originating in the X-ray binary population of each galaxy (which was estimated from the fitted spectra).  We then converted these counts to fluxes using a typical ULX-like spectrum, consistent with the power-law continua seen in the 3 bright galactic nuclei ($N_{\rm H} \sim 1.5 \times 10^{21} \rm ~cm^{-2}$, $\Gamma \sim 2$).  The estimated observed luminosities in the power-law component for each galaxy are shown in Table~10.

\begin{table}
\centering
\caption{Upper limits on the number of ULXs hidden within the diffuse component of each galaxy nucleus without evidence for an AGN}
\smallskip
\begin{threeparttable}
\begin{tabular}{lccc}
\hline
Galaxy	& $L_{\rm PL}~^a$	& $N_{\rm ULX,hid}~^b$	& ($f_{\rm nuc}/f_{\rm gal})_{\rm IR}~^c$ \\
		& ($10^{39} \rm ~erg~s^{-1}$)		&	& \\\hline
NGC~7552		& 18	& 3.8		& 0.2 \\
IC~5179			& 7.1	& 1.5		& 0.17 \\
NGC~838			& 30	& 6.5		& 0.49 \\
NGC~5653		& 7.9	& 1.7		& 0.2 \\
NGC~3221		& 2.2	& $< 1$	& 0.08 \\
CGCG~049-057	& 1.3	& $< 1$	& 0.81 \\
NGC~23			& 33	& 7.2		& 0.45 \\\hline
\end{tabular}
\begin{tablenotes}
\item {\bf Notes.} $^a$ Total observed luminosity estimate for the hard power-law component in each galaxy nucleus.  $^b$ Estimated upper limit on the number of ULXs hidden within the diffuse emission in this region (see text for details).  $^c$ Fraction of the total 8 $\mu$m flux from each galaxy residing in the nucleus.
\end{tablenotes}
\end{threeparttable}
\end{table}

Assuming that the luminosity distribution of X-ray binary systems in the nuclear region of the LIRGs follows the \citeauthor{mineo2012a} (2012a) XLF, we converted the power-law luminosity into a number of ULXs using that XLF.  In each case we worked out the normalisation required to give the integrated luminosity by integrating under the XLF in the $10^{37}$ -- $10^{40}\rm~erg~s^{-1}$ range, and then using this to calculate the number of ULXs present, that we show in Table~10.  A strict upper limit would simply assume that each ULX has a luminosity of $10^{39} \rm ~erg~s^{-1}$; however a population of ULXs should have a range of luminosities, and integrating under the XLF accounts for this.  What this method does not account for is that the brightest objects should still be detected as point sources, even within bright diffuse emission, by {\it Chandra\/}; but as we saw above the number of such objects detected within the nuclear regions is small.  Given that these nuclei are likely to be far more complex than our simple modelling allows for, and may contain additional components that contaminate the hard band (e.g. hard diffuse emission, extended reflection nebulae from an unseen AGN), we consider our $N_{\rm ULX,hid}$ estimate as a reasonable upper limit on the number of ULXs that may be hidden within these nuclei.  In total we may have up to $\sim 21$ ULXs hidden within these galaxies, or $\sim 3$ per nucleus.  Over the whole sample this would extrapolate to $\sim 50$ ULXs, or a rough doubling of our ULX numbers, and hence an increase in ULX incidence to 0.4 ULX per unit SFR. 

As an independent check of the effect of excluding the nuclear region, we estimated the fraction of the global star formation we are excluding for each galaxy.  We used the FIR emission as mapped by 8 $\mu$m \textit{Spitzer} images\footnote{\texttt{http://sha.ipac.caltech.edu/applications/Spitzer/SHA/}} as a proxy for the SFR, and in each case we corrected for the contribution of an old stellar population to the 8 $\mu$m flux using the 3.6 $\mu$m images and Equation~1 of \citet{calzetti2007}.  Then, we took the ratio between the SFR in the central 5 arcsecond nuclear region and that of the rest of the galaxy (out to the \textit{R}$_{20}$ ellipse), again for the 7 galaxies with no evidence for an AGN.  We found that the nuclear region contribution to the total SFR, $(f_{\rm nuc}/f_{\rm gal})_{\rm IR}$, varied widely, between $\sim 8 - 80$ per cent in the AGN-less galaxies (see Table~10).  However, there was only one case in which it exceeded 50 per cent, so if we take this as the upper limit on the coarse average contribution from the nuclear region and apply that as a correction to the regions outside the nucleus across the whole sample, then we can set an upper limit on our global incidence of ULXs in the LIRG sample of 0.4 ULXs per unit SFR.  Interestingly this upper limit coincidences with the result from the flux analysis demonstrated above.


However, despite the suggested rise in ULX numbers from the upper limits place on the nuclear regions, if we combine the observational effects described above, it is very implausible that they will result in a factor 10 increase in ULX detections.  We therefore regard the deficit as a real effect, and consider physical origins for it. 

\subsubsection{Possible explanations for the deficit}

So, what is the physical origin of the deficit of ULXs in the LIRGs?  Firstly, we would not expect that all ULXs come from young stellar populations.  There are ULXs seen in elliptical galaxies that must be related to their LMXB populations, and we should expect all galaxies with old stellar populations to possess a similar underlying population of LMXBs that scales with the mass of the system \citep{colbert2004,lehmer2010}.  This is not accounted for by \cite{swartz2011}; their number of ULXs per SFR that we base our deficit deduction on will contain ULXs from both populations.  However, this cannot lead to the deficit of ULXs in the LIRGs, for the simple reason that our LIRG sample contains $\sim 4$ times more mass than the combined mass of the nearby galaxies in \cite{swartz2011}, at $\sim 1.3 \times 10^{13} M_{\odot}$ versus $\sim 3.5 \times 10^{12} M_{\odot}$\footnote{The total mass of the LIRGs was taken from Table~1 of \citet{lehmer2010}.}; thus we would expect our sample to possess more LMXB ULXs, that would act to increase the number of ULXs per SFR.  In fact, using the estimate of \cite{feng2011} and making a conservative (factor 2) correction for the high background contamination in the elliptical sample of \cite{walton2011} we might expect 0.5 LMXB ULXs per $10^{11} M_{\odot}$ of galaxy mass.  We would then expect that less than one fifth of the \cite{swartz2011} ULXs ($\sim 17$) are LMXBs; however we would expect more LMXB ULXs in our sample ($\sim 65$) than we observe ULXs in total. Given the high SFR that accompanies the high mass of our sample systems, this serves to emphasise that a real deficit in ULXs is present in the LIRG sample.

Could this deficit be due to some factor inhibiting the formation of ULXs in the environment of the LIRGs?  An obvious property that might suppress the formation of ULXs, given recent studies (e.g. \citealt{mapelli2011,prestwich2013}; \citeauthor{basu-zych2013a} 2013a, 2\citeyear{basu-zych2013b}; \citealt{brorby2014}), could be the relatively high metallicity of the LIRGs.  A good comparator sample for the LIRGs in this respect is the high metallicity galaxies sample of \cite{prestwich2013}, that was demonstrated to show a much lower incidence of ULXs per unit SFR than a sample of extremely metal poor galaxies.  A comparison of our Table~\ref{tab:LIRGsample} and Table 9 of that paper shows that (where available) both samples have a very similar spread of metallicity; and indeed Table 10 of \cite{prestwich2013} quotes a $N_{\rm ULX}$ per SFR of $0.17\pm0.042$ per $M_{\odot} \rm ~yr^{-1}$, identical within errors to our value of $0.2\pm0.05$ per $M_{\odot} \rm ~yr^{-1}$.  However, \cite{prestwich2013} used a different method to calculate SFR (\citealt{calzetti2010}, based on the $H\alpha$ and 24 $\mu$m fluxes); when we use our SFR calculation method on their sample of galaxies we find that the SFR drops by a factor $\sim 6$ to a total of 16.77 $M_{\odot} \rm ~yr^{-1}$ across the whole sample\footnote{There was no {\it IRAS\/} data available for NGC 4625 in the high metallicity sample; we retained the estimate used by \cite{prestwich2013}, but note that this will not adversely affect the results as it had the smallest estimate of SFR in the sample at $0.22 ~M_{\odot} \rm ~yr^{-1}$.}.    Hence the number of ULXs per SFR for this sample increases dramatically, up to $1.04\pm0.26$ per $M_{\odot} \rm ~yr^{-1}$; notably this is lower than the average relation in nearby galaxies, but it is again significantly higher than the number of ULXs per SFR in the LIRGs (and is also significantly higher than the upper limits accounting for the excluded nuclear regions derived in the previous section).  Hence, given the similar metallicities of the two samples\footnote{We caution that one caveat here is that further, systematic measurements of metallicity, that supersede the current disparate methods, are required to confirm that the samples do overlap in metallicity.}, we conclude that the relatively high metallicity of the LIRGs appears insufficient reason on its own to explain their deficit in ULXs.

One reason for suspecting that metallicity affects the incidence of ULXs is the possibility that larger BHs (the MsBH class, with masses between 20 and 100$M_{\odot}$) might form in low metallicity regions.  Conversely, very few such objects should be seen in higher metallicity regions.  If such objects, accreting at $\sim$ Eddington, constitute a large proportion of ULXs up to the HMXB XLF break, then an interesting consequence should be that the XLF break shifts to lower luminosities in high metallicity environments.  This is precisely what is suggested by the very putative break in the cumulative XLF, with the break apparently shifting down by a decade in luminosity to the Eddington limit for a $\sim$16 $M_{\odot}$ BH, compared to the break found in the \citeauthor{mineo2012a} (2012a) XLF.  The lack of ULXs with luminosities above $10^{40} \rm ~erg~s^{-1}$ certainly appears consistent with this scenario.  In addition, this result is supported by the spectral changes seen in this sample, with the change in spectral form either side of the same luminosity as the putative XLF break suggesting the transition between stellar-mass black holes accreting at $\sim$ Eddington, and those at super-Eddington rates.  However, this is not supported by wider spectral studies that see a similar pattern in ULXs located in all galaxy types (\citealt{gladstone2009}: \citeauthor{sutton2013b} 2013b) -- if a large proportion of ULXs were $\sim$ Eddington accretion onto MsBHs, we would see the disc-like spectra reported by \cite{gladstone2009} extend up to the XLF break in many sources, but they instead are mainly seen below $3 \times 10^{39} \rm ~erg~s^{-1}$, seemingly ruling this out.  One interesting scenario remains, though -- the factor 10 drop in the XLF break luminosity could still be due to a drop in the mass of the compact objects, if the objects below the break are predominantly super-Eddington neutron stars (NSs), that reach maximal super-Eddington luminosities at the putative XLF break.  This might be the case if the high metallicities impede the formation of all BHs, not just MsBHs, and the change in spectrum might then be the difference between the super-Eddington NSs and an underlying population of BHs.  Clearly this is highly speculative, and requires a great deal of improvement in data quality in order to investigate this hypothesis.  Firstly, we need to confirm the presence of the break; then we need to identify the source populations either side of it.  This will almost certainly require next generation X-ray observatories that can begin to disentangle BHs and NSs at large distances based on their observational characteristics.

The next possibility we consider that might cause the deficit of ULXs is whether the stellar populations of the LIRGs are too young to allow a population of ULXs to form.  As ULXs are composed of a compact object accreting material from a stellar companion, they will require time to evolve from an initial massive stellar binary system.  This inevitably introduces a lag between the start of star formation in a system, and the appearance of ULXs.  According to \citet{linden2010}, the number of ULXs formed in solar metallicity environments peaks at $\sim$4~--~5 Myr after the beginning of star formation, and a lag of up to $\sim10$ Myr is required to see the peak for ULX populations formed in a sub-solar metallicity medium.  These ages are consistent with the resolved stellar population ages around two nearby ULXs, that are of the order $\la 20$ Myr (\citealt{grise2008}; \citealt{grise2011}).  So the critical question is: how long is it since the star formation began in the LIRGs?  Older studies, based on the correlation between the Bracket-$\gamma$ and far-infrared emission, estimates star formation ages in LIRGs at between 10 -- 1000 Myr \citep{goldader1997}.   However, later work based on the spectral energy distributions of LIRGs and ULIRGs (ultraluminous infrared galaxies) from near IR to radio wavelengths suggests star formation timescale of $\sim$5 -- 100 Myr (\citealt{vega2008}; see also \citealt{clemens2010}). In fact, the star formation ages of a few individual LIRGs have been reported in previous works: for example, 57 Myr for Arp~299 \citep{vega2008,clemens2010}, and 30 Myr for both NGC~7552 and NGC~4194 \citep{tagagi1999}.  Thus it seems unlikely that the LIRGs are too young for their populations of ULXs to switch on.

However, the age calculations above assume a single age for the newly-formed stellar population. This might not be a good assumption because the star formation activity is unlikely to be uniform throughout the galaxies (see e.g. \citealt{mineo2013}); it is possible that star formation may have begun more recently in some regions in the galaxies than in other areas. For example, the spatial study of NGC~5135 by \citet{bedregal2009} suggests that the most recent star forming areas in the galaxy could be $\sim$0.5 -- 1 Myr younger than other regions.  Moreover, they also suggest that NGC~5135 might host two recently-formed stellar populations with different ages: a new population that formed $\sim$6 -- 8 Myr ago, and a second, older stellar population of $\ga$200 Myr age. Thus simply assuming the same stellar age along the whole galaxy might not reflect the real history of star formation in LIRGs.  However, even bearing these caveats in mind, we would realistically require the vast majority of the star formation in the LIRGs to have started within the last 5 Myr for the populations of ULXs to be significantly suppressed; given the ages quoted above this is very unlikely.  Therefore, while this effect may have some influence in the LIRG sample, we regard it as unlikely to be the major factor in the ULX deficit.

So, if the deficit of ULXs in LIRGs is unlikely to be caused by metallicity effects or a very young stellar population, then is there another possibility that may act to reduce the number of detected ULXs?  The answer is potentially revealed in the global $N_{\rm ULX}$ - luminosity relations discussed in Section~\ref{sec:ulxperlum}.  In particular, \cite{smith2012} find similar ratios in a sample of interacting galaxies taken from the Arp catalogue -- they find an enhancement in the $N_{\rm ULX}/L_{\rm B}$ compared to normal galaxy samples, but in their most IR luminous objects they find the $N_{\rm ULX}/L_{\rm FIR}$ ratio significantly suppressed.  They suggest two reasons for this: either the FIR luminosity is significantly enhanced by nuclear activity; or the large columns of gas and dust in these objects are obscuring a large fraction of the ULX population.  In our work we have explicitly removed possible AGN contamination from the FIR fluxes using the method of \cite{mullaney2011}, and then shown that these ratios are the same for LIRGs that both host and do not host AGN; hence we rule out the first scenario.  We therefore favour the second explanation -- that the suppressed $N_{\rm ULX}/L_{\rm FIR}$ ratio in LIRGs could be a result of missing many ULXs due to their signal being extinguished by the high absorption columns in these galaxies due to dust attenuation. Indeed, Lehmer et al. (2010) suggested that an apparent lack of luminosity in X-rays present in the LIRGs, compared to expectation, might be due to dust obscuration. Furthermore, such an effect is seen also in normal star-forming galaxies (Fig.~11 of \citeauthor{mineo2012a} 2012a) as well as in the sample of Local Lyman Break Galaxy Analogues which is shown in the left panel of Fig.~7 of \citeauthor{basu-zych2013b} (2013b). Indeed, that is also consistent with the other relationships we see -- both X-rays and blue light relate to young stellar populations, so explaining the proportionality relationship between the two; but the unusually high number of ULXs per blue luminosity in these galaxies is plausibly due to a faster loss of blue light than X-rays in the absorbing material.  

\begin{figure*}
\begin{center}
\includegraphics[width=6cm]{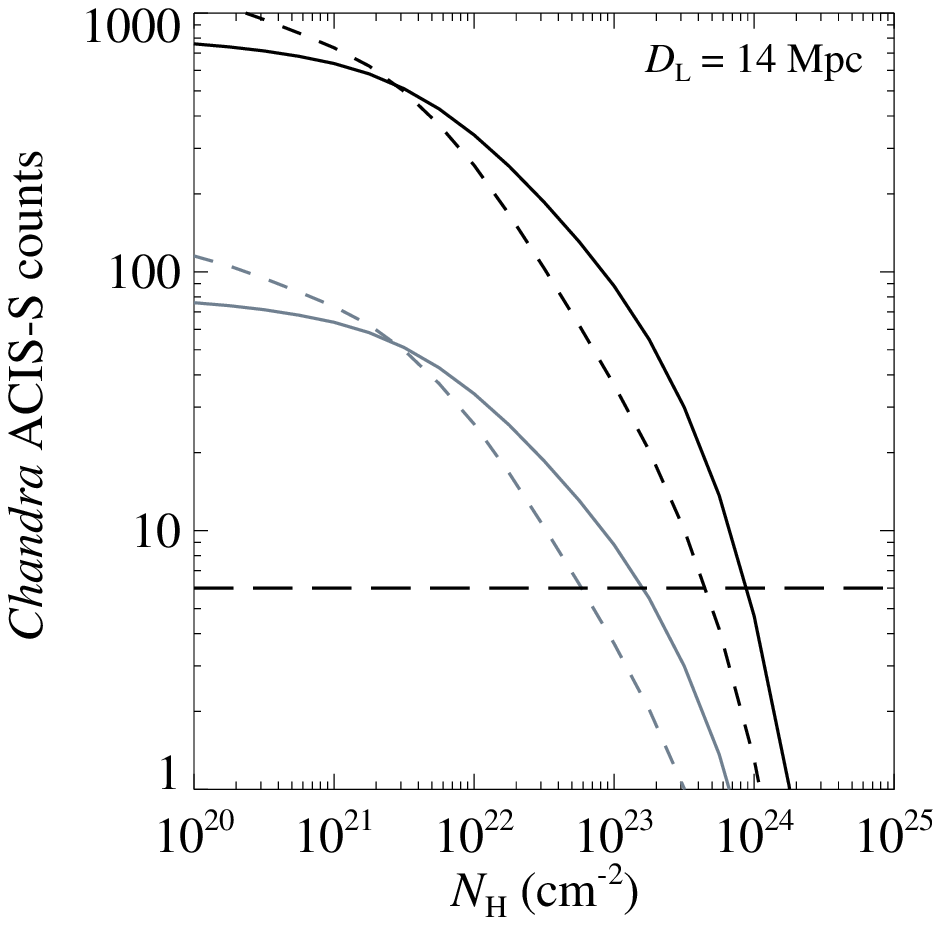}~\includegraphics[width=6cm]{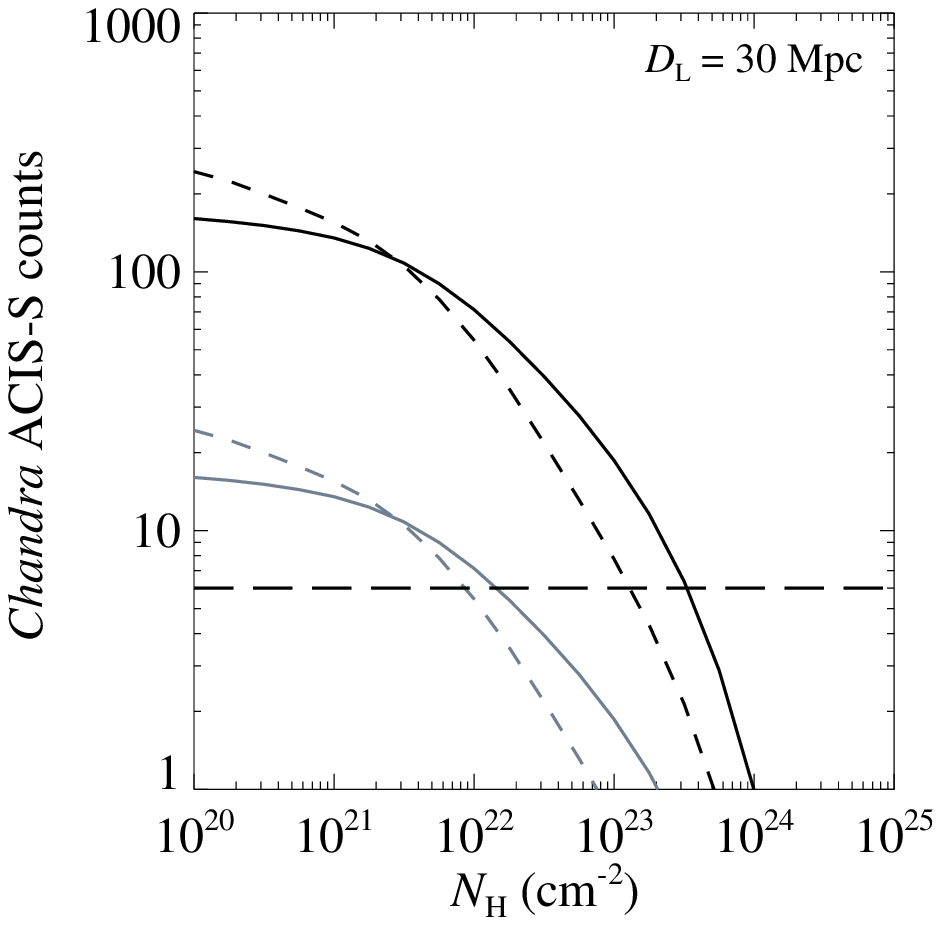}~\includegraphics[width=6cm]{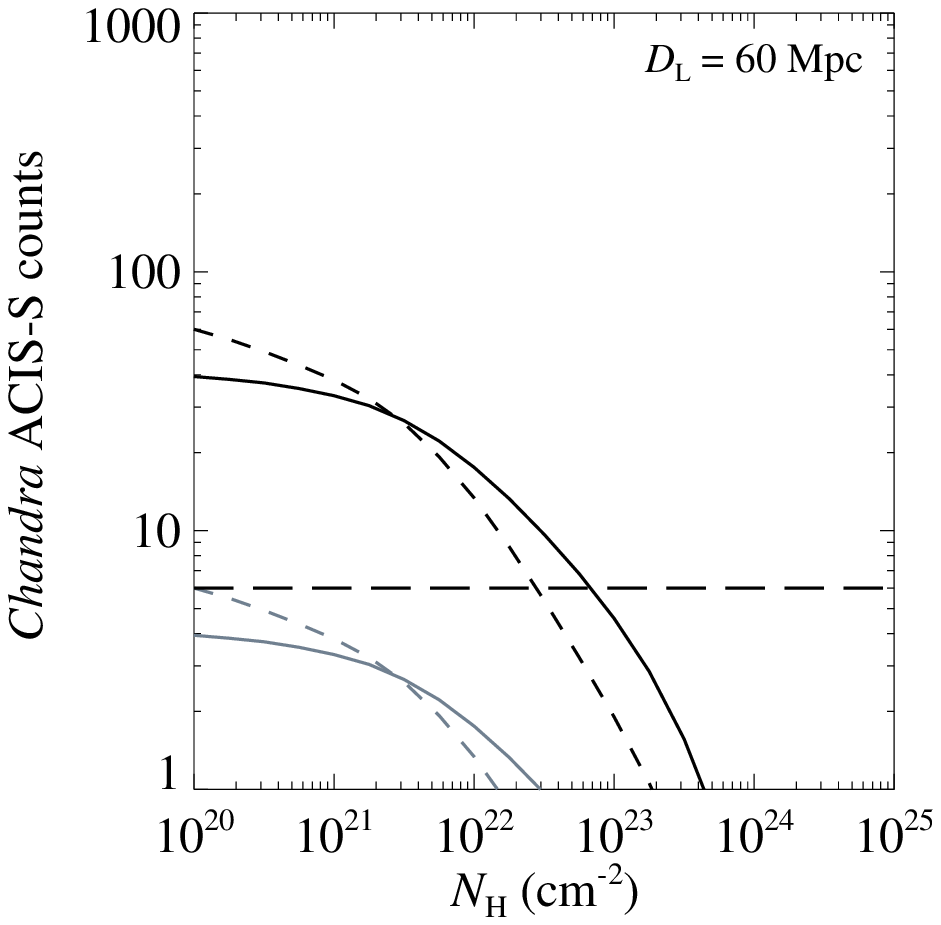}
\caption[]{Simulated \textit{Chandra} ACIS-S counts detected from ULXs as a function of column density ($N_{\rm H}$), calculated at luminosity distances of 14 (left), 30 (middle) and 60 (right) Mpc.  Assuming 20 ks observations, \textsc{webpimms} was used to simulate the number of counts detected from ULXs with photon indexes of 1.5 (solid line) and 2.5 (dashed line). Both calculated values for faint (10$^{39}$ erg s$^{-1}$; grey lines) and bright (10$^{40}$ erg s$^{-1}$; black) ULXs are plotted here. The horizontal dashed line indicates the detection limit -- the minimum number of counts where ULXs are detected as point sources in our observational data by the \textsc{wavdetect} algorithm.}
\label{fig:counts_nh}
\end{center}
\end{figure*}

Hence we have a plausible explanation for the overall deficit of ULXs per SFR in the LIRG sample; we do not see most of them because they are hidden behind the material fuelling the high SFR in these objects. But is this consistent with the columns of material we expect in LIRGs?  We can quantify the column densities required to support this scenario by examining how the detectability of sources with {\it Chandra\/} depends on their intrinsic spectrum, their observed luminosity, their distance and the intervening columns.  If we assume 20 ks exposures (typical for our sample), and a power-law form for our spectra, then we can use \textsc{webpimms} to calculate the dependence of the detected counts on absorption column and distance to the ULXs, which we show in Fig.~\ref{fig:counts_nh}.  Taking a minimum detectability threshold of 6 counts (cf. Table A1) demonstrates that the effective extinction of the ULXs strongly depends on distance to the host galaxies.  In the nearest system (NGC 1068, at 14 Mpc) a faint ULX (at $10^{39} \rm ~erg~s^{-1}$) would be fully extinguished by a column of $\sim 10^{23} \rm ~cm^{-2}$, whilst a brighter object (with luminosity $10^{40} \rm ~erg~s^{-1}$) requires a higher column of $\sim 10^{24} \rm ~cm^{-2}$ to be obscured from our view.  These columns reduce with distance -- at 30 Mpc the equivalent columns are  $\sim 10^{22}$ and a few times $10^{23} \rm ~cm^{-2}$ respectively, while at 60 Mpc even minimal column ($\sim 10^{20} \rm ~cm^{-2}$) will cause a faint ULX to be missed, and $\la 10^{23} \rm ~cm^{-2}$ is required to obscure a bright ULX.  Crucially, these columns are consistent with those seen in LIRGs.  For example, \citet{genzel1998} reported that some ULIRGs have high optical extinctions, $A_{\rm V} \sim 1000$, corresponding to a column density of $\sim$ 2 $\times$ 10$^{24}$ cm$^{-2}$ (using the conversion of \citealt{guver2009} where $N_{\rm H}$ =  2.21 $\times$ 20$^{21}$ cm$^{-2}$ $A_{\rm V}$).  Indeed, a FIR spectral study of NGC~4418 by \textit{Herschel/PACS} revealed a very high value of column density in the nuclear region of the galaxy, $N_{\rm H}$ $\sim$ 10$^{25}$ cm$^{-2}$ \citep{Gonzalez-Alfonso2012}.  Clearly, these high columns are commensurate with completely obscuring ULXs from our view.

One caveat of this conclusion is that it appears unlikely that there is a steady gradation of column as one enters the LIRGs.  In such a situation one might expect to see a range of obscuring columns building up to the high central values, and this would lead to a population of absorbed, and so spectrally hard, but not completely obscured ULXs.  We see no such population in the LIRGs, with just one ULX being a hard band only detection, arguing that the absorbing material is a relatively dense medium enshrouding the ULX population.  We speculate that such a medium might produce the very putative break in the XLF, but only if the brighter ULXs are preferentially buried behind the screen of material\footnote{There is perhaps a hint of the more luminous ULXs being more absorbed in the harder stacked spectra of these objects, although the measured absorption columns do not significantly vary between the three luminosity bins.}.  Interestingly, this is also entirely consistent with the acute lack of ULX detections in the nuclear regions discussed in the previous section; we would expect the innermost regions of each LIRG to be the most deeply enshrouded in dust and gas, and so the most difficult to observe ULXs in.

Finally, interesting supporting evidence for the obscuration scenario is provided by another exotic phenomenon that appears to be under-represented in LIRGs -- core-collapse supernovae (CCSNe).  Optical surveys for these explosive transients reveal that up to 83 per cent of the expected CCSNe (given the calculated SFR) are missing (e.g. \citealt{horiuchi2011,mattila2012}).  Again, this is directly attributable to the very dusty environments of the LIRGs, that obscure and so extinguish the emission from these objects.

\section{Conclusion}

In this paper we have studied the population of ULXs present in a sample of 17 nearby, low foreground absorption LIRGs, based on {\it Chandra\/} observations.  We detect a total of $53^{+16}_{-13}$ ULXs in the galaxies, among a total of 139 point X-ray sources, of which we expect less than one fifth to be background objects.  We consider the sample to be essentially complete in ULX detection.  Our main results can be summarised as follows:

\begin{itemize}
\item
The source spectra were stacked into three groups as a function of source luminosity.  The stacked spectrum of the lowest luminosity sources ($L_{\rm X} < 2 \times 10^{39} \rm ~erg~s^{-1}$) is significantly softer than the two more luminous stacks, that are indistinguishable from one another.  This is consistent with a change in accretion state for $10 M_{\odot}$ BHs as they progress from $\sim $ Eddington rate accretion into a super-Eddington, ultraluminous state.
\item
In one object (CXOU~J024238.9-000055 in NGC 1068) we have sufficient statistics to study its spectrum individually.  Although its spectrum can be acceptably fitted with a power-law continuum, this leaves a large residual at $\sim 1.7$ keV.  We find that this can be modelled by the presence of a partially ionised partial covering medium, located within $10^9 R_{\rm g}$ of the ULX, which might be material expelled from the system in a radiatively-driven outflow, as expected for super-Eddington accretion. 
\item
The LIRGs possess significantly fewer ULXs per unit SFR than nearby `normal' galaxies, by a factor $\sim 10$.  This deficit also manifests itself as a lower normalisation of the differential XLF than for nearby HMXB samples, and a steeper slope to the cumulative XLF than nearby star forming galaxies.  This is very unlikely to be the result of more LMXB ULXs in the normal galaxy sample; nor is it likely to be due to observational effects such as source confusion, or ULXs missed by the detection code due to being embedded in diffuse emission.  Very few of the missing ULXs are detected in the nuclear regions we initially exclude from our analysis; however correcting for those ULXs that may be present based on sources hidden within a diffuse nuclear component, or the SFR of the nuclear region, still presents upper limits significantly below the number of ULXs expected from normal galaxies.
\item
Metallicity may have some influence on ULX numbers. It might suppress the formation of massive stellar remnants, which may be consistent with a very putative break found in the cumulative XLF and the total absence of ULXs above 10$^{40}$ erg s$^{-1}$. However, we regard this effect as unlikely to explain the whole deficit.
\item
A second factor that may have some impact on the deficit is the lag between star formation starting and the appearance of ULXs.  However, given the stellar ages of newly formed populations in LIRGs are typically estimated to be older than 5 Myr, this provides sufficient time to turn ULX populations on, so we are disinclined to regard this as a major contributor to the deficit.
\item
The relatively high ratio of $N_{\rm ULX}/L_{\rm B}$, and very low ratio in $N_{\rm ULX}/L_{\rm FIR}$, of the LIRGs compared to normal galaxy samples point to a very plausible explanation for the main cause of the deficit -- a large part of the ULX population is hidden behind the high columns of gas and dust present in these systems.  We show that the columns of absorbing material present within LIRGs are sufficient to fully obscure a population of ULXs. This is supported by the the detection of far fewer ULXs per unit SFR in the nuclear regions of the LIRGs where we expect the obscuration to be greatest, and by a similar observational deficit in core-collapse supernovae, that should also be more numerous in LIRGs than is observed.
\end{itemize}

If there are large hidden populations of ULXs in the most massive star forming galaxies in the Universe, as is suggested by this sample of LIRGs, then this has interesting implications.  Firstly, given the flat XLFs of star forming systems, it can be shown that ULXs dominate their hard X-ray luminosity.  However, if these objects are hidden from view, then the observed hard X-ray luminosity of these systems will be largely suppressed below its intrinsic value; it cannot then provide a good measure of the star formation rate of the galaxy, invalidating relations derived from lower SFR systems.  This hidden population may also contribute to feedback processes in the galaxies -- a factor 10 more ULXs would boost the radiative feedback by a factor 10, and a similar level of mechanical feedback would also be expected for these systems.  Given that the numbers of massive star forming systems similar to LIRGs increase to peak at redshifts $\sim 1$, this implies that understanding the concealment of X-ray source populations is important for understanding the intrinsic X-ray emission of the galaxies now being picked up in deep surveys, and how the sources that contribute to this emission help shape their host galaxies.  Further work that builds on this study and permits a deeper understanding the X-ray populations of nearby LIRGs is therefore key to understanding our X-ray view of galaxies at the peak of cosmic star formation.

\section*{Acknowledgments}

We thank the anonymous referee for their suggestions, that resulted in improvements to this paper.  WL acknowledges support in the form of funding for a PhD studentship from the Royal Thai Fellowship scheme.  TPR thanks STFC for support in the form of the standard grant ST/G001588/1 and subsequently as part of the consolidated grant ST/K000861/1.  We thank various colleagues for useful conversations, notably Poshak Gandhi for pointing out the CCSNe results and James Mullaney for a helpful discussion in the AGN contribution fitting result. We would also like to thank Steven Willner for a valuable discussion on the effect of dust versus galaxy age.

\bibliography{references}
\bibliographystyle{mn2e}

\appendix

\section[]{A catalogue of point sources detected in the LIRG sample}
\label{sec:appendixA}

Here we present the catalogue of all X-ray point source detections in the {\it Chandra\/} observations of our sample of 17 LIRGs.  For convenience we have divided the sources into two tables: the ULX catalogue (sources with $L_{X}$ $\geq$ 10$^{39}$ erg s$^{-1}$; Table~\ref{tab:ULXcandidates}); and the catalogue of less luminous X-ray point sources (Table~\ref{tab:other_xray}).


\begin{table*}
      \centering
      \caption{ULX candidates}\label{tab:ULXcandidates}
      \smallskip
      \begin{threeparttable}
          \begin{tabular}{ccccccc}
          
             \hline
	RA	&	Dec &	Host galaxy	& Net counts & Flux & Luminosity & Notes \\
	
		&		& & & (10$^{-14}$ erg cm$^{-1}$ s$^{-1}$) & (10$^{39}$ erg s$^{-1}$) \\
	(1)	&	(2)	&	(3)	&	(4)	&	(5)	&	(6)	&	(7)	\\

\hline

00	09	53	&	+	25	55	48	&	NGC 23	& $	6.62	^{+	3.78	}_{-	2.58	}$ & $	0.27	^{+	0.16	}_{-	0.11	}$ & $	1.20	^{+	0.68	}_{-	0.47	}$ &		\\
00	09	53	&	+	25	55	39	&	NGC 23	& $	10.66	^{+	4.43	}_{-	3.26	}$ & $	0.44	^{+	0.18	}_{-	0.13	}$ & $	1.93	^{+	0.80	}_{-	0.59	}$ &		\\
02	09	38	&	--	10	08	47	&	NGC 838	& $	10.27	^{+	4.43	}_{-	3.26	}$ & $	0.58	^{+	0.25	}_{-	0.19	}$ & $	1.80	^{+	0.78	}_{-	0.57	}$ &		\\
02	09	39	&	--	10	08	44	&	NGC 838	& $	11.83	^{+	4.57	}_{-	3.41	}$ & $	0.67	^{+	0.26	}_{-	0.19	}$ & $	2.08	^{+	0.80	}_{-	0.60	}$ &		\\
02	42	38	&	--	00	01	18	&	NGC 1068	& $	385.72	^{+	20.72	}_{-	19.69	}$ & $	6.44	^{+	0.35	}_{-	0.33	}$ & $	1.46	^{+	0.08	}_{-	0.07	}$ &		\\
02	42	39	&	--	00	00	55	&	NGC 1068	& $	1370.25	^{+	38.09	}_{-	37.08	}$ & $	22.87	^{+	0.64	}_{-	0.62	}$ & $	5.17	\pm 0.14	$ &	1	\\
02	42	40	&	--	00	01	01	&	NGC 1068	& $	487.09	^{+	23.84	}_{-	22.81	}$ & $	8.13	^{+	0.40	}_{-	0.38	}$ & $	1.84	\pm 0.09	$ &		\\
03	33	32	&	--	36	08	09	&	NGC 1365	& $	68.79	^{+	9.35	}_{-	8.28	}$ & $	4.02	^{+	0.55	}_{-	0.48	}$ & $	1.57	^{+	0.21	}_{-	0.19	}$ &		\\
03	33	34	&	--	36	11	03	&	NGC 1365	& $	123.75	^{+	12.17	}_{-	11.12	}$ & $	7.24	^{+	0.71	}_{-	0.65	}$ & $	2.82	^{+	0.28	}_{-	0.25	}$ &		\\
03	33	35	&	--	36	09	37	&	NGC 1365	& $	204.89	^{+	15.34	}_{-	14.30	}$ & $	11.99	^{+	0.90	}_{-	0.84	}$ & $	4.66	^{+	0.35	}_{-	0.33	}$ &		\\
03	33	38	&	--	36	09	35	&	NGC 1365	& $	111.89	^{+	11.62	}_{-	10.57	}$ & $	6.55	^{+	0.68	}_{-	0.62	}$ & $	2.55	^{+	0.26	}_{-	0.24	}$ &		\\
03	33	40	&	--	36	07	27	&	NGC 1365	& $	68.83	^{+	9.35	}_{-	8.28	}$ & $	4.03	^{+	0.55	}_{-	0.48	}$ & $	1.57	^{+	0.21	}_{-	0.19	}$ &		\\
03	33	50	&	--	36	10	26	&	NGC 1365	& $	32.88	^{+	6.81	}_{-	5.71	}$ & $	2.80	^{+	0.58	}_{-	0.49	}$ & $	1.09	^{+	0.23	}_{-	0.19	}$ &		\\
10	22	20	&	+	21	34	02	&	NGC 3221	& $	16.61	^{+	5.21	}_{-	4.08	}$ & $	0.69	^{+	0.22	}_{-	0.17	}$ & $	2.91	^{+	0.91	}_{-	0.71	}$ &		\\
10	22	20	&	+	21	33	29	&	NGC 3221	& $	14.82	^{+	4.97	}_{-	3.83	}$ & $	0.61	^{+	0.21	}_{-	0.16	}$ & $	2.60	^{+	0.87	}_{-	0.67	}$ &		\\
10	22	20	&	+	21	34	50	&	NGC 3221	& $	16.74	^{+	5.21	}_{-	4.08	}$ & $	0.69	^{+	0.22	}_{-	0.17	}$ & $	2.93	^{+	0.91	}_{-	0.71	}$ &		\\
10	22	20	&	+	21	34	05	&	NGC 3221	& $	6.63	^{+	3.78	}_{-	2.58	}$ & $	0.27	^{+	0.16	}_{-	0.11	}$ & $	1.16	^{+	0.66	}_{-	0.45	}$ &		\\
10	22	21	&	+	21	33	36	&	NGC 3221	& $	14.73	^{+	4.97	}_{-	3.83	}$ & $	0.61	^{+	0.21	}_{-	0.16	}$ & $	2.58	^{+	0.87	}_{-	0.67	}$ &		\\
10	22	22	&	+	21	33	25	&	NGC 3221	& $	14.87	^{+	4.97	}_{-	3.83	}$ & $	0.62	^{+	0.21	}_{-	0.16	}$ & $	2.61	^{+	0.87	}_{-	0.67	}$ &		\\
11	28	27	&	+	58	34	07	&	Arp 299	& $	16.81	^{+	5.21	}_{-	4.08	}$ & $	0.80	^{+	0.25	}_{-	0.19	}$ & $	2.22	^{+	0.69	}_{-	0.54	}$ &		\\
11	28	31	&	+	58	33	45	&	Arp 299	& $	62.30	^{+	9.84	}_{-	8.77	}$ & $	2.96	^{+	0.47	}_{-	0.42	}$ & $	8.24	^{+	1.30	}_{-	1.16	}$ &		\\
11	28	31	&	+	58	33	27	&	Arp 299	& $	12.03	^{+	4.71	}_{-	3.56	}$ & $	0.57	^{+	0.22	}_{-	0.17	}$ & $	1.59	^{+	0.62	}_{-	0.47	}$ &		\\
11	28	32	&	+	58	33	18	&	Arp 299	& $	48.52	^{+	8.05	}_{-	6.97	}$ & $	2.30	^{+	0.38	}_{-	0.33	}$ & $	6.42	^{+	1.06	}_{-	0.92	}$ &		\\
11	28	33	&	+	58	33	37	&	Arp 299	& $	28.32	^{+	6.55	}_{-	5.45	}$ & $	1.35	^{+	0.31	}_{-	0.26	}$ & $	3.74	^{+	0.87	}_{-	0.72	}$ &	2	\\
11	28	33	&	+	58	34	03	&	Arp 299	& $	32.87	^{+	6.90	}_{-	5.80	}$ & $	1.56	^{+	0.33	}_{-	0.28	}$ & $	4.34	^{+	0.91	}_{-	0.77	}$ &		\\
11	28	34	&	+	58	33	41	&	Arp 299	& $	28.02	^{+	6.55	}_{-	5.45	}$ & $	1.33	^{+	0.31	}_{-	0.26	}$ & $	3.70	^{+	0.87	}_{-	0.72	}$ &		\\
11	28	37	&	+	58	33	41	&	Arp 299	& $	10.84	^{+	4.43	}_{-	3.26	}$ & $	0.51	^{+	0.21	}_{-	0.15	}$ & $	1.43	^{+	0.59	}_{-	0.43	}$ &		\\
12	14	10	&	+	54	31	27	&	NGC 4194	& $	25.27	^{+	6.56	}_{-	5.46	}$ & $	0.56	^{+	0.15	}_{-	0.12	}$ & $	1.11	^{+	0.29	}_{-	0.24	}$ &		\\
13	25	43	&	--	29	50	06	&	NGC 5135	& $	12.57	^{+	4.71	}_{-	3.55	}$ & $	0.40	^{+	0.15	}_{-	0.11	}$ & $	1.35	^{+	0.51	}_{-	0.38	}$ &		\\
13	25	45	&	--	29	50	04	&	NGC 5135	& $	31.36	^{+	6.81	}_{-	5.72	}$ & $	1.01	^{+	0.22	}_{-	0.18	}$ & $	3.37	^{+	0.73	}_{-	0.62	}$ &		\\
13	25	45	&	--	29	50	14	&	NGC 5135	& $	20.53	^{+	5.66	}_{-	4.54	}$ & $	0.66	^{+	0.18	}_{-	0.15	}$ & $	2.21	^{+	0.61	}_{-	0.49	}$ &		\\
13	25	45	&	--	29	50	12	&	NGC 5135	& $	9.64	^{+	4.28	}_{-	3.10	}$ & $	0.31	^{+	0.14	}_{-	0.10	}$ & $	1.04	^{+	0.46	}_{-	0.33	}$ &		\\
13	25	45	&	--	29	50	07	&	NGC 5135	& $	13.69	^{+	4.84	}_{-	3.69	}$ & $	0.44	^{+	0.16	}_{-	0.12	}$ & $	1.47	^{+	0.52	}_{-	0.40	}$ &		\\
13	25	48	&	--	29	49	48	&	NGC 5135	& $	27.77	^{+	6.36	}_{-	5.26	}$ & $	0.89	^{+	0.20	}_{-	0.17	}$ & $	2.99	^{+	0.68	}_{-	0.57	}$ &		\\
13	58	38	&	+	37	25	33	&	NGC 5395	& $	7.70	^{+	3.96	}_{-	2.76	}$ & $	0.39	^{+	0.20	}_{-	0.14	}$ & $	1.35	^{+	0.69	}_{-	0.48	}$ &		\\
13	58	38	&	+	37	24	41	&	NGC 5395	& $	8.80	^{+	4.12	}_{-	2.94	}$ & $	0.44	^{+	0.21	}_{-	0.15	}$ & $	1.54	^{+	0.72	}_{-	0.51	}$ &		\\
13	58	39	&	+	37	25	32	&	NGC 5395	& $	13.68	^{+	4.84	}_{-	3.69	}$ & $	0.69	^{+	0.24	}_{-	0.19	}$ & $	2.40	^{+	0.85	}_{-	0.65	}$ &		\\
13	58	40	&	+	37	26	28	&	NGC 5395	& $	6.87	^{+	3.78	}_{-	2.58	}$ & $	0.34	^{+	0.19	}_{-	0.13	}$ & $	1.20	^{+	0.66	}_{-	0.45	}$ &		\\
14	30	11	&	+	31	13	00	&	NGC 5653	& $	7.54	^{+	3.96	}_{-	2.76	}$ & $	0.36	^{+	0.19	}_{-	0.13	}$ & $	1.32	^{+	0.69	}_{-	0.48	}$ &		\\
22	16	08	&	--	36	50	56	&	IC 5179	& $	13.61	^{+	4.84	}_{-	3.69	}$ & $	0.89	^{+	0.32	}_{-	0.24	}$ & $	2.38	^{+	0.85	}_{-	0.65	}$ &		\\
22	16	08	&	--	36	50	19	&	IC 5179	& $	7.85	^{+	3.96	}_{-	2.76	}$ & $	0.52	^{+	0.26	}_{-	0.18	}$ & $	1.37	^{+	0.69	}_{-	0.48	}$ &		\\
22	16	10	&	--	36	50	40	&	IC 5179	& $	14.30	^{+	4.97	}_{-	3.83	}$ & $	0.94	^{+	0.33	}_{-	0.25	}$ & $	2.51	^{+	0.87	}_{-	0.67	}$ &		\\
22	16	10	&	--	36	50	34	&	IC 5179	& $	6.43	^{+	3.78	}_{-	2.58	}$ & $	0.42	^{+	0.25	}_{-	0.17	}$ & $	1.13	^{+	0.66	}_{-	0.45	}$ &		\\
22	16	10	&	--	36	50	20	&	IC 5179	& $	36.62	^{+	7.14	}_{-	6.05	}$ & $	2.41	^{+	0.47	}_{-	0.40	}$ & $	6.42	^{+	1.25	}_{-	1.06	}$ &		\\
22	16	10	&	--	36	50	24	&	IC 5179	& $	6.57	^{+	3.78	}_{-	2.58	}$ & $	0.43	^{+	0.25	}_{-	0.17	}$ & $	1.15	^{+	0.66	}_{-	0.45	}$ &		\\
22	16	12	&	--	36	50	24	&	IC 5179	& $	12.83	^{+	4.71	}_{-	3.55	}$ & $	0.84	^{+	0.31	}_{-	0.23	}$ & $	2.25	^{+	0.83	}_{-	0.62	}$ &		\\
22	16	12	&	--	36	50	09	&	IC 5179	& $	19.81	^{+	5.56	}_{-	4.43	}$ & $	1.30	^{+	0.36	}_{-	0.29	}$ & $	3.47	^{+	0.97	}_{-	0.78	}$ &		\\
23	16	14	&	--	42	35	02	&	NGC 7552	& $	50.89	^{+	8.19	}_{-	7.12	}$ & $	7.87	^{+	1.27	}_{-	1.10	}$ & $	4.37	^{+	0.70	}_{-	0.61	}$ &		\\
23	16	16	&	--	42	35	18	&	NGC 7552	& $	11.94	^{+	4.57	}_{-	3.41	}$ & $	1.85	^{+	0.71	}_{-	0.53	}$ & $	1.03	^{+	0.39	}_{-	0.29	}$ &		\\

               \hline
         \end{tabular}
         \begin{tablenotes}
         \item
         \end{tablenotes}
      \end{threeparttable}
    \end{table*}

\begin{table*}
      \centering
      \contcaption{ULX candidates and their properties.}
      \smallskip
      \begin{threeparttable}
          \begin{tabular}{ccccccc}
          
             \hline
	RA	&	Dec &	Host galaxy	& Net counts & Flux & Luminosity & Notes \\
	
		&		& & & (10$^{-14}$ erg cm$^{-1}$ s$^{-1}$) & (10$^{39}$ erg s$^{-1}$) \\
	(1)	&	(2)	&	(3)	&	(4)	&	(5)	&	(6)	&	(7)	\\

\hline

23	51	22	&	+	20	06	39	&	NGC 7771	& $	28.82	^{+	6.45	}_{-	5.35	}$ & $	1.35	^{+	0.30	}_{-	0.25	}$ & $	5.44	^{+	1.22	}_{-	1.01	}$ &		\\
23	51	24	&	+	20	06	38	&	NGC 7771	& $	9.67	^{+	4.28	}_{-	3.10	}$ & $	0.45	^{+	0.20	}_{-	0.15	}$ & $	1.82	^{+	0.81	}_{-	0.59	}$ &		\\
23	51	24	&	+	20	06	43	&	NGC 7771	& $	18.68	^{+	5.44	}_{-	4.32	}$ & $	0.88	^{+	0.26	}_{-	0.20	}$ & $	3.53	^{+	1.03	}_{-	0.81	}$ &		\\
23	51	28	&	+	20	06	52	&	NGC 7771	& $	34.72	^{+	6.98	}_{-	5.89	}$ & $	1.63	^{+	0.33	}_{-	0.28	}$ & $	6.55	^{+	1.32	}_{-	1.11	}$ &		\\

               \hline
         \end{tabular}
         \begin{tablenotes}
         \item \textbf{Notes.} The candidate ULX detections, ordered by right ascension (RA). Column 1 and 2: right ascension and declination, at epoch J2000, respectively. Column 3: host galaxy. Column 4: net photon counts. Column 5: source flux. Column 6: source luminosity, assuming the source is at the distance of the host galaxy.  Columns 4 -- 6 are all quoted from data in the 0.3--10\,keV energy band.  Column 7: additional notes for the ULXs:  (1) CXOU~J024238.9-000055 \citep{smith2003};  (2) This ULX is coincident with a radio source detection.  This may be related to the ULX; it may be an unrelated radio supernova in the host galaxy; or it might be a background AGN \citep{ulvestad2009,neff2004,huang1990}.

         \end{tablenotes}
      \end{threeparttable}
    \end{table*}



\begin{table*}
      \centering
      \caption{Less luminous X-ray point sources}\label{tab:other_xray}
      \smallskip
      \begin{threeparttable}
          \begin{tabular}{ccccccc}
          
             \hline
	RA	&	Dec &	Host galaxy	& Net counts & Flux & Luminosity & Notes \\
	
		&		& & & (10$^{-14}$ erg cm$^{-1}$ s$^{-1}$) & (10$^{39}$ erg s$^{-1}$) \\
	(1)	&	(2)	&	(3)	&	(4)	&	(5)	&	(6)	&	(7)	\\

\hline

00	09	53	&	+	25	54	55	&	NGC 23	& $	3.78	^{+	3.18	}_{-	1.91	}$ & $	0.16	^{+	0.13	}_{-	0.08	}$ & $	0.68	^{+	0.57	}_{-	0.34	}$ &		\\
00	09	55	&	+	25	55	25	&	NGC 23	& $	5.29	^{+	3.60	}_{-	2.38	}$ & $	0.22	^{+	0.15	}_{-	0.10	}$ & $	0.96	^{+	0.65	}_{-	0.43	}$ &		\\
02	09	39	&	--	10	08	38	&	NGC 838	& $	4.73	^{+	3.40	}_{-	2.15	}$ & $	0.27	^{+	0.19	}_{-	0.12	}$ & $	0.83	^{+	0.60	}_{-	0.38	}$ &		\\
02	42	33	&	--	00	01	05	&	NGC 1068	& $	58.24	^{+	8.73	}_{-	7.66	}$ & $	0.97	^{+	0.15	}_{-	0.13	}$ & $	0.22	\pm	0.03	$ &		\\
02	42	33	&	--	00	01	30	&	NGC 1068	& $	4.35	^{+	3.40	}_{-	2.15	}$ & $	0.07	^{+	0.06	}_{-	0.04	}$ & $	0.02	\pm	0.01$ &		\\
02	42	37	&	+	00	00	13	&	NGC 1068	& $	10.25	^{+	4.43	}_{-	3.26	}$ & $	0.17	^{+	0.07	}_{-	0.05	}$ & $	0.04	^{+	0.02	}_{-	0.01	}$ &		\\
02	42	37	&	--	00	00	28	&	NGC 1068	& $	12.05	^{+	4.71	}_{-	3.56	}$ & $	0.20	^{+	0.08	}_{-	0.06	}$ & $	0.05	^{+	0.02	}_{-	0.01	}$ &		\\
02	42	37	&	--	00	01	09	&	NGC 1068	& $	38.92	^{+	7.38	}_{-	6.30	}$ & $	0.65	^{+	0.12	}_{-	0.11	}$ & $	0.15	^{+	0.03	}_{-	0.02	}$ &		\\
02	42	37	&	--	00	00	19	&	NGC 1068	& $	8.06	^{+	4.12	}_{-	2.94	}$ & $	0.13	^{+	0.07	}_{-	0.05	}$ & $	0.03	^{+	0.02	}_{-	0.01	}$ &		\\
02	42	38	&	--	00	00	29	&	NGC 1068	& $	49.79	^{+	8.19	}_{-	7.12	}$ & $	0.83	^{+	0.14	}_{-	0.12	}$ & $	0.19	\pm	0.03 $ &		\\
02	42	38	&	--	00	01	43	&	NGC 1068	& $	29.50	^{+	6.64	}_{-	5.54	}$ & $	0.49	^{+	0.11	}_{-	0.09	}$ & $	0.11	^{+	0.03	}_{-	0.02	}$ &		\\
02	42	38	&	--	00	01	37	&	NGC 1068	& $	8.59	^{+	4.58	}_{-	3.42	}$ & $	0.14	^{+	0.08	}_{-	0.06	}$ & $	0.03	^{+	0.02	}_{-	0.01	}$ &		\\
02	42	39	&	--	00	00	50	&	NGC 1068	& $	34.66	^{+	7.56	}_{-	6.47	}$ & $	0.58	^{+	0.13	}_{-	0.11	}$ & $	0.13	^{+	0.03	}_{-	0.02	}$ &		\\
02	42	39	&	--	00	00	52	&	NGC 1068	& $	23.56	^{+	6.77	}_{-	5.67	}$ & $	0.39	^{+	0.11	}_{-	0.09	}$ & $	0.09	^{+	0.03	}_{-	0.02	}$ &		\\
02	42	39	&	--	00	01	23	&	NGC 1068	& $	36.77	^{+	7.55	}_{-	6.47	}$ & $	0.61	^{+	0.13	}_{-	0.11	}$ & $	0.14	^{+	0.03	}_{-	0.02	}$ &		\\
02	42	39	&	--	00	00	57	&	NGC 1068	& $	38.83	^{+	8.03	}_{-	6.95	}$ & $	0.65	^{+	0.13	}_{-	0.12	}$ & $	0.15	\pm	0.03	$ &		\\
02	42	39	&	--	00	01	31	&	NGC 1068	& $	12.27	^{+	5.10	}_{-	3.97	}$ & $	0.20	^{+	0.09	}_{-	0.07	}$ & $	0.05	^{+	0.02	}_{-	0.01	}$ &		\\
02	42	39	&	--	00	00	35	&	NGC 1068	& $	70.92	^{+	9.77	}_{-	8.70	}$ & $	1.18	^{+	0.16	}_{-	0.15	}$ & $	0.27	^{+	0.04	}_{-	0.03	}$ &		\\
02	42	39	&	--	00	01	04	&	NGC 1068	& $	57.03	^{+	8.93	}_{-	7.86	}$ & $	0.95	^{+	0.15	}_{-	0.13	}$ & $	0.22	\pm	0.03	$ &		\\
02	42	40	&	--	00	00	28	&	NGC 1068	& $	183.48	^{+	14.67	}_{-	13.63	}$ & $	3.06	^{+	0.24	}_{-	0.23	}$ & $	0.69	^{+	0.06	}_{-	0.05	}$ &		\\
02	42	40	&	--	00	00	38	&	NGC 1068	& $	57.47	^{+	12.80	}_{-	11.76	}$ & $	0.96	^{+	0.21	}_{-	0.20	}$ & $	0.22	^{+	0.05	}_{-	0.04	}$ &		\\
02	42	41	&	--	00	00	37	&	NGC 1068	& $	174.24	^{+	17.01	}_{-	15.98	}$ & $	2.91	^{+	0.28	}_{-	0.27	}$ & $	0.66	\pm	0.06	$ &		\\
02	42	41	&	--	00	01	32	&	NGC 1068	& $	24.69	^{+	6.27	}_{-	5.16	}$ & $	0.41	^{+	0.10	}_{-	0.09	}$ & $	0.09	\pm	0.02	$ &		\\
02	42	41	&	--	00	00	60	&	NGC 1068	& $	100.64	^{+	11.75	}_{-	10.70	}$ & $	1.68	^{+	0.20	}_{-	0.18	}$ & $	0.38	\pm	0.04	$ &		\\
02	42	41	&	--	00	01	44	&	NGC 1068	& $	32.82	^{+	6.90	}_{-	5.80	}$ & $	0.55	^{+	0.12	}_{-	0.10	}$ & $	0.12	^{+	0.03	}_{-	0.02	}$ &		\\
02	42	41	&	--	00	01	25	&	NGC 1068	& $	35.90	^{+	7.31	}_{-	6.22	}$ & $	0.60	^{+	0.12	}_{-	0.10	}$ & $	0.14	^{+	0.03	}_{-	0.02	}$ &		\\
02	42	41	&	--	00	02	15	&	NGC 1068	& $	33.45	^{+	6.90	}_{-	5.80	}$ & $	0.56	^{+	0.12	}_{-	0.10	}$ & $	0.13	^{+	0.03	}_{-	0.02	}$ &		\\
02	42	41	&	--	00	02	53	&	NGC 1068	& $	8.36	^{+	4.12	}_{-	2.94	}$ & $	0.14	^{+	0.07	}_{-	0.05	}$ & $	0.03	^{+	0.02	}_{-	0.01	}$ &		\\
02	42	42	&	--	00	01	22	&	NGC 1068	& $	7.72	^{+	4.28	}_{-	3.11	}$ & $	0.13	^{+	0.07	}_{-	0.05	}$ & $	0.03	^{+	0.02	}_{-	0.01	}$ &		\\
02	42	42	&	--	00	01	09	&	NGC 1068	& $	14.50	^{+	5.34	}_{-	4.21	}$ & $	0.24	^{+	0.09	}_{-	0.07	}$ & $	0.05	\pm	0.02	$ &		\\
02	42	42	&	+	00	00	02	&	NGC 1068	& $	29.16	^{+	6.73	}_{-	5.63	}$ & $	0.49	^{+	0.11	}_{-	0.09	}$ & $	0.11	^{+	0.03	}_{-	0.02	}$ &		\\
02	42	43	&	--	00	00	52	&	NGC 1068	& $	18.12	^{+	5.67	}_{-	4.55	}$ & $	0.30	^{+	0.09	}_{-	0.08	}$ & $	0.07	\pm	0.02	$ &		\\
02	42	43	&	--	00	00	48	&	NGC 1068	& $	19.87	^{+	5.88	}_{-	4.76	}$ & $	0.33	^{+	0.10	}_{-	0.08	}$ & $	0.08	\pm	0.02	$ &		\\
02	42	43	&	+	00	00	23	&	NGC 1068	& $	10.92	^{+	4.71	}_{-	3.56	}$ & $	0.18	^{+	0.08	}_{-	0.06	}$ & $	0.04	^{+	0.02	}_{-	0.01	}$ &		\\
02	42	43	&	--	00	00	04	&	NGC 1068	& $	10.62	^{+	5.59	}_{-	4.47	}$ & $	0.18	^{+	0.09	}_{-	0.07	}$ & $	0.04	\pm	0.02	$ &		\\
02	42	43	&	--	00	02	45	&	NGC 1068	& $	60.48	^{+	8.86	}_{-	7.79	}$ & $	1.01	^{+	0.15	}_{-	0.13	}$ & $	0.23	\pm	0.03	$ &		\\
02	42	43	&	--	00	01	40	&	NGC 1068	& $	110.24	^{+	11.57	}_{-	10.52	}$ & $	1.84	^{+	0.19	}_{-	0.18	}$ & $	0.42	\pm	0.04	$ &		\\
02	42	44	&	--	00	00	35	&	NGC 1068	& $	95.72	^{+	10.94	}_{-	9.88	}$ & $	1.60	^{+	0.18	}_{-	0.16	}$ & $	0.36	\pm	0.04	$ &		\\
02	42	45	&	--	00	00	10	&	NGC 1068	& $	61.82	^{+	8.98	}_{-	7.91	}$ & $	1.03	^{+	0.15	}_{-	0.13	}$ & $	0.23	\pm	0.03	$ &		\\
02	42	45	&	--	00	01	27	&	NGC 1068	& $	7.33	^{+	3.96	}_{-	2.76	}$ & $	0.12	^{+	0.07	}_{-	0.05	}$ & $	0.03	\pm	0.01	$ &		\\
02	42	45	&	--	00	02	19	&	NGC 1068	& $	51.43	^{+	8.26	}_{-	7.19	}$ & $	0.86	^{+	0.14	}_{-	0.12	}$ & $	0.19	\pm	0.03	$ &		\\
02	42	46	&	--	00	00	30	&	NGC 1068	& $	29.05	^{+	6.55	}_{-	5.45	}$ & $	0.49	^{+	0.11	}_{-	0.09	}$ & $	0.11	\pm	0.02	$ &		\\
02	42	46	&	+	00	00	06	&	NGC 1068	& $	17.92	^{+	5.44	}_{-	4.32	}$ & $	0.30	^{+	0.09	}_{-	0.07	}$ & $	0.07	\pm	0.02	$ &		\\
02	42	47	&	--	00	01	13	&	NGC 1068	& $	6.29	^{+	3.78	}_{-	2.58	}$ & $	0.11	^{+	0.06	}_{-	0.04	}$ & $	0.02	\pm	0.01	$ &		\\
02	42	47	&	+	00	00	28	&	NGC 1068	& $	167.85	^{+	14.03	}_{-	12.99	}$ & $	2.80	^{+	0.23	}_{-	0.22	}$ & $	0.63	\pm	0.05	$ &		\\

               \hline
         \end{tabular}
         \begin{tablenotes}
         \item
         \end{tablenotes}
      \end{threeparttable}
    \end{table*}

\begin{table*}
      \centering
      \contcaption{Other X-ray point sources and their properties.}
      \smallskip
      \begin{threeparttable}
          \begin{tabular}{ccccccc}
          
             \hline
	RA 	&	Dec  &	Host galaxy	& Net counts & Flux & Luminosity & Notes \\
	
		&		& & & (10$^{-14}$ erg cm$^{-1}$ s$^{-1}$) & (10$^{39}$ erg s$^{-1}$) \\
	(1)	&	(2)	&	(3)	&	(4)	&	(5)	&	(6)	&	(7)	\\

\hline

03	33	21	&	--	36	08	15	&	NGC 1365	& $	21.48	^{+	5.77	}_{-	4.65	}$ & $	1.26	^{+	0.34	}_{-	0.27	}$ & $	0.49	^{+	0.13	}_{-	0.11	}$ &		\\
03	33	23	&	--	36	07	53	&	NGC 1365	& $	38.58	^{+	7.31	}_{-	6.22	}$ & $	2.26	^{+	0.43	}_{-	0.36	}$ & $	0.88	^{+	0.17	}_{-	0.14	}$ &		\\
03	33	24	&	--	36	10	50	&	NGC 1365	& $	7.42	^{+	3.96	}_{-	2.76	}$ & $	0.43	^{+	0.23	}_{-	0.16	}$ & $	0.17	^{+	0.09	}_{-	0.06	}$ &		\\
03	33	26	&	--	36	08	38	&	NGC 1365	& $	21.74	^{+	5.77	}_{-	4.65	}$ & $	1.27	^{+	0.34	}_{-	0.27	}$ & $	0.49	^{+	0.13	}_{-	0.11	}$ &		\\
03	33	26	&	--	36	08	50	&	NGC 1365	& $	22.78	^{+	5.87	}_{-	4.76	}$ & $	1.33	^{+	0.34	}_{-	0.28	}$ & $	0.52	^{+	0.13	}_{-	0.11	}$ &		\\
03	33	27	&	--	36	08	14	&	NGC 1365	& $	11.76	^{+	4.57	}_{-	3.41	}$ & $	0.69	^{+	0.27	}_{-	0.20	}$ & $	0.27	^{+	0.10	}_{-	0.08	}$ &		\\
03	33	30	&	--	36	07	56	&	NGC 1365	& $	7.77	^{+	3.96	}_{-	2.76	}$ & $	0.45	^{+	0.23	}_{-	0.16	}$ & $	0.18	^{+	0.09	}_{-	0.06	}$ &		\\
03	33	30	&	--	36	08	30	&	NGC 1365	& $	35.75	^{+	7.06	}_{-	5.97	}$ & $	2.09	^{+	0.41	}_{-	0.35	}$ & $	0.81	^{+	0.16	}_{-	0.14	}$ &		\\
03	33	30	&	--	36	08	21	&	NGC 1365	& $	10.73	^{+	4.43	}_{-	3.26	}$ & $	0.63	^{+	0.26	}_{-	0.19	}$ & $	0.24	^{+	0.10	}_{-	0.07	}$ &		\\
03	33	31	&	--	36	11	06	&	NGC 1365	& $	7.71	^{+	3.96	}_{-	2.76	}$ & $	0.45	^{+	0.23	}_{-	0.16	}$ & $	0.18	^{+	0.09	}_{-	0.06	}$ &		\\
03	33	31	&	--	36	08	08	&	NGC 1365	& $	11.78	^{+	4.57	}_{-	3.41	}$ & $	0.69	^{+	0.27	}_{-	0.20	}$ & $	0.27	^{+	0.10	}_{-	0.08	}$ &		\\
03	33	32	&	--	36	06	43	&	NGC 1365	& $	42.65	^{+	7.61	}_{-	6.53	}$ & $	2.50	^{+	0.45	}_{-	0.38	}$ & $	0.97	^{+	0.17	}_{-	0.15	}$ &		\\
03	33	32	&	--	36	09	03	&	NGC 1365	& $	19.87	^{+	5.56	}_{-	4.43	}$ & $	1.16	^{+	0.32	}_{-	0.26	}$ & $	0.45	^{+	0.13	}_{-	0.10	}$ &		\\
03	33	33	&	--	36	07	15	&	NGC 1365	& $	4.81	^{+	3.40	}_{-	2.15	}$ & $	0.28	^{+	0.20	}_{-	0.13	}$ & $	0.11	^{+	0.08	}_{-	0.05	}$ &		\\
03	33	36	&	--	36	08	21	&	NGC 1365	& $	25.57	^{+	6.56	}_{-	5.46	}$ & $	1.50	^{+	0.38	}_{-	0.32	}$ & $	0.58	^{+	0.15	}_{-	0.12	}$ &		\\
03	33	36	&	--	36	05	57	&	NGC 1365	& $	4.62	^{+	3.40	}_{-	2.15	}$ & $	0.27	^{+	0.20	}_{-	0.13	}$ & $	0.11	^{+	0.08	}_{-	0.05	}$ &		\\
03	33	36	&	--	36	09	58	&	NGC 1365	& $	8.91	^{+	4.12	}_{-	2.94	}$ & $	0.52	^{+	0.24	}_{-	0.17	}$ & $	0.20	^{+	0.09	}_{-	0.07	}$ &		\\
03	33	37	&	--	36	10	26	&	NGC 1365	& $	17.80	^{+	5.33	}_{-	4.20	}$ & $	1.04	^{+	0.31	}_{-	0.25	}$ & $	0.40	^{+	0.12	}_{-	0.10	}$ &		\\
03	33	39	&	--	36	08	01	&	NGC 1365	& $	4.80	^{+	3.40	}_{-	2.15	}$ & $	0.28	^{+	0.20	}_{-	0.13	}$ & $	0.11	^{+	0.08	}_{-	0.05	}$ &		\\
03	33	39	&	--	36	05	24	&	NGC 1365	& $	13.43	^{+	4.84	}_{-	3.69	}$ & $	0.79	^{+	0.28	}_{-	0.22	}$ & $	0.31	^{+	0.11	}_{-	0.08	}$ &		\\
03	33	39	&	--	36	10	02	&	NGC 1365	& $	20.82	^{+	5.66	}_{-	4.54	}$ & $	1.22	^{+	0.33	}_{-	0.27	}$ & $	0.47	^{+	0.13	}_{-	0.10	}$ &		\\
03	33	40	&	--	36	10	37	&	NGC 1365	& $	31.85	^{+	6.72	}_{-	5.62	}$ & $	1.86	^{+	0.39	}_{-	0.33	}$ & $	0.72	^{+	0.15	}_{-	0.13	}$ &		\\
03	33	42	&	--	36	08	19	&	NGC 1365	& $	4.87	^{+	3.40	}_{-	2.15	}$ & $	0.29	^{+	0.20	}_{-	0.13	}$ & $	0.11	^{+	0.08	}_{-	0.05	}$ &		\\
03	33	42	&	-	36	07	41	&	NGC 1365	& $	18.79	^{+	5.44	}_{-	4.32	}$ & $	1.10	^{+	0.32	}_{-	0.25	}$ & $	0.43	^{+	0.12	}_{-	0.10	}$ &		\\
11	28	34	&	+	58	33	24	&	Arp 299	& $	6.41	^{+	3.78	}_{-	2.58	}$ & $	0.30	^{+	0.18	}_{-	0.12	}$ & $	0.85	^{+	0.50	}_{-	0.34	}$ &		\\
12	14	09	&	+	54	31	45	&	NGC 4194	& $	13.84	^{+	4.97	}_{-	3.83	}$ & $	0.31	^{+	0.11	}_{-	0.09	}$ & $	0.61	^{+	0.22	}_{-	0.17	}$ &		\\
12	14	10	&	+	54	31	42	&	NGC 4194	& $	11.61	^{+	4.57	}_{-	3.41	}$ & $	0.26	^{+	0.10	}_{-	0.08	}$ & $	0.51	^{+	0.20	}_{-	0.15	}$ &		\\
12	14	13	&	+	54	31	29	&	NGC 4194	& $	5.59	^{+	3.60	}_{-	2.38	}$ & $	0.14	^{+	0.08	}_{-	0.06	}$ & $	0.27	^{+	0.17	}_{-	0.15	}$ &		\\
12	26	54	&	--	00	52	54	&	NGC 4418	& $	7.78	^{+	3.96	}_{-	2.76	}$ & $	0.31	^{+	0.16	}_{-	0.11	}$ & $	0.38	^{+	0.20	}_{-	0.14	}$ &		\\
13	25	44	&	--	29	49	33	&	NGC 5135	& $	7.67	^{+	3.96	}_{-	2.76	}$ & $	0.25	^{+	0.13	}_{-	0.09	}$ & $	0.82	^{+	0.43	}_{-	0.30	}$ &		\\
13	58	40	&	+	37	24	45	&	NGC 5395	& $	2.86	^{+	2.94	}_{-	1.63	}$ & $	0.14	^{+	0.15	}_{-	0.08	}$ & $	0.50	^{+	0.51	}_{-	0.29	}$ &		\\
14	30	10	&	+	31	12	33	&	NGC 5653	& $	4.80	^{+	3.40	}_{-	2.15	}$ & $	0.23	^{+	0.16	}_{-	0.10	}$ & $	0.84	^{+	0.60	}_{-	0.38	}$ &		\\
23	16	08	&	--	42	34	36	&	NGC 7552	& $	3.93	^{+	3.18	}_{-	1.91	}$ & $	0.61	^{+	0.49	}_{-	0.30	}$ & $	0.34	^{+	0.27	}_{-	0.16	}$ &		\\
23	16	10	&	--	42	35	08	&	NGC 7552	& $	3.92	^{+	3.18	}_{-	1.91	}$ & $	0.61	^{+	0.49	}_{-	0.30	}$ & $	0.34	^{+	0.27	}_{-	0.16	}$ &		\\
23	16	11	&	--	42	34	43	&	NGC 7552	& $	8.82	^{+	4.12	}_{-	2.94	}$ & $	1.36	^{+	0.64	}_{-	0.45	}$ & $	0.76	^{+	0.35	}_{-	0.25	}$ &		\\
23	16	14	&	--	42	35	40	&	NGC 7552	& $	10.95	^{+	4.43	}_{-	3.26	}$ & $	1.69	^{+	0.68	}_{-	0.50	}$ & $	0.94	^{+	0.38	}_{-	0.28	}$ &		\\
23	16	17	&	--	42	35	00	&	NGC 7552	& $	7.93	^{+	3.96	}_{-	2.76	}$ & $	1.23	^{+	0.61	}_{-	0.43	}$ & $	0.68	^{+	0.34	}_{-	0.24	}$ &		\\
23	51	24	&	+	20	06	19	&	NGC 7771	& $	2.81	^{+	2.94	}_{-	1.63	}$ & $	0.13	^{+	0.14	}_{-	0.08	}$ & $	0.53	^{+	0.55	}_{-	0.31	}$ &		\\
23	51	26	&	+	20	06	47	&	NGC 7771	& $	4.62	^{+	3.40	}_{-	2.15	}$ & $	0.22	^{+	0.16	}_{-	0.10	}$ & $	0.87	^{+	0.64	}_{-	0.41	}$ &		\\
23	51	26	&	+	20	06	55	&	NGC 7771	& $	3.64	^{+	3.18	}_{-	1.91	}$ & $	0.17	^{+	0.15	}_{-	0.09	}$ & $	0.69	^{+	0.60	}_{-	0.36	}$ &		\\
23	51	27	&	+	20	06	31	&	NGC 7771	& $	4.79	^{+	3.40	}_{-	2.15	}$ & $	0.23	^{+	0.16	}_{-	0.10	}$ & $	0.90	^{+	0.64	}_{-	0.41	}$ &		\\

               \hline
         \end{tabular}
         \begin{tablenotes}
         \item \textbf{Notes.} The columns are as per Table~\ref{tab:ULXcandidates}.

         \end{tablenotes}
      \end{threeparttable}
    \end{table*}


\bsp

\label{lastpage}

\end{document}